\documentclass[12pt]{article} % use larger type; default would be 10pt

\usepackage[utf8]{inputenc} % set input encoding (not needed with XeLaTeX)

\usepackage{authblk}

\usepackage{amsmath}
\usepackage{graphicx,psfrag,epsf}
\usepackage{enumerate}
\usepackage{natbib}
\usepackage{url} % not crucial - just used below for the URL 

\RequirePackage{amsthm,amsmath,amsfonts}
\RequirePackage[colorlinks,citecolor=blue,urlcolor=blue]{hyperref}
\usepackage{multirow}
\usepackage{algorithm2e}
\usepackage{comment}

\usepackage{pgfplots}
\usepackage{pgfplotstable}
\usepackage{here}
\usepackage{float}
\usepackage[font=small]{caption}

\numberwithin{equation}{section}

\newtheorem{prop}{Proposition}

\newtheorem{rem}{Remark}[section]

\def\bSig\mathbf{\Sigma}

\newcommand\E{\mathbb{E}}
\renewcommand\P{\mathbb{P}}

\usepackage{fancybox}

\usepackage{xcolor}
\colorlet{lightblue}{blue!10}
\usepackage{amssymb}
\graphicspath{{.}{./../Figures_Biometrics/}}

\begin{document}

\def\spacingset#1{\renewcommand{\baselinestretch}%
{#1}\small\normalsize} \spacingset{1}

%%%%%%%%%%%%%%%%%%%%%%%%%%%%%%%%%%%%%%%%%%%%%%%%%%%%%%%%%%%%%%%%%%%%%%%%%%%%%%

\title{\bf Testing independence between two random sets for the analysis  of colocalization in bio-imaging}

\author[1,2]{Fr\'ed\'eric Lavancier}
\author[1]{Thierry P\'ecot}
\author[3]{Liu Zengzhen}
\author[1]{Charles Kervrann}

\affil[1]{{\footnotesize Inria, Centre Rennes-Bretagne Atlantique, SERPICO Project Team, 35042 Rennes, France}}
\affil[2]{{\footnotesize  University of Nantes, Laboratoire de Math\'ematiques Jean Leray, 44322 Nantes, France}}
\affil[3]{{\footnotesize Institut Curie, PSL Research University, CNRS UMR 144, Space Time Imaging of Endomembranes Dynamics Team, 75005 Paris, France}}

\date{}
\maketitle

\vspace{-8.5mm}

\begin{abstract}
Colocalization aims at characterizing spatial associations between two fluorescently-tagged biomolecules by quantifying the co-occurrence and correlation between the two channels acquired in fluorescence microscopy. Colocalization is presented either as the degree of overlap between the two channels or the overlays of the red and green images, with areas of yellow indicating colocalization of the molecules. This problem remains an open issue in diffraction-limited microscopy and raises new challenges with the emergence of super-resolution imaging, a microscopic technique awarded by the 2014 Nobel prize in chemistry.
We propose GcoPS, for Geo-coPositioning System, an original method that exploits the random sets structure of the tagged molecules to provide an explicit testing procedure. 
Our simulation study  shows that
GcoPS unequivocally outperforms the best competitive methods in adverse situations (noise, irregularly shaped fluorescent patterns, different optical resolutions). GcoPS is  also much faster, a decisive advantage to face the huge amount of data in super-resolution imaging. We demonstrate the performances of GcoPS on two biological real datasets, obtained by conventional diffraction-limited microscopy technique and by super-resolution technique, respectively. 

\bigskip

\noindent%

\end{abstract}

%\noindent%
%{\it Keywords:}   %Colocalization, Hypothesis Test, Random Sets,
%Quantitative Fluorescence Microscopy; Spatial Statistics; Stochastic Geometry; Super-Resolution Microscopy.
 %3 to 6 keywords, that do not appear in the title
%\vfill

%\spacingset{1.45} % DON'T change the spacing!

\newpage

\section{Introduction}
\label{sec:intro}

\subsection{Biological challenge}\label{sec:bio}

\begin{figure}
\begin{center}
\includegraphics[width=.3\linewidth]{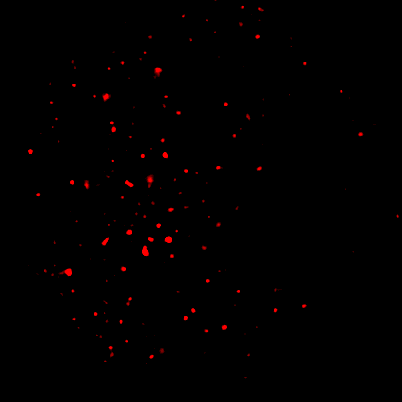}
\includegraphics[width=.3\linewidth]{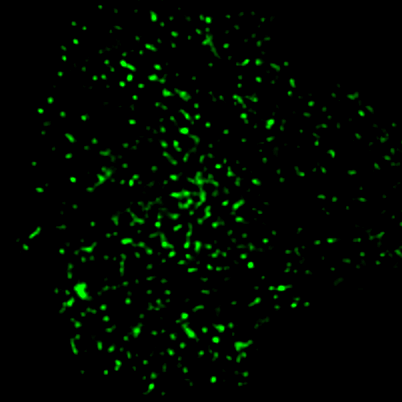}
\includegraphics[width=.61\linewidth]{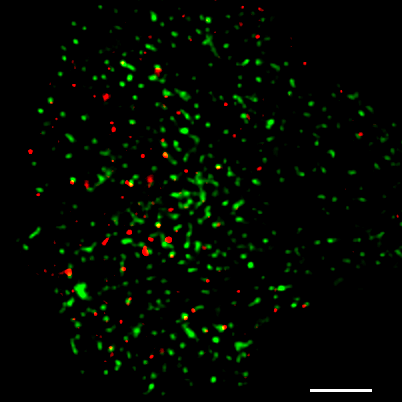}
\end{center}
\vspace{-0.25cm}
\caption{
3D multi-angles TIRFM acquisition of a RPE1 cell \citep{Boulanger2014} expressing m-Cherry Langerin (red tagged molecules) and GFP Rab11 (green tagged molecules). These images are the projection (along the $z$ axis) of the initial 3D $400\times400\times20$ pixels images.  
The image at the bottom is the superposition of the two acquired images displayed at the top. The scale bar corresponds to 10$\mu$m. This figure appears in color in the electronic version of this article, and any mention of color refers to that version.
}
\label{Fig1}
\end{figure}

\begin{figure}
\begin{center}
\includegraphics[width=.43\linewidth]{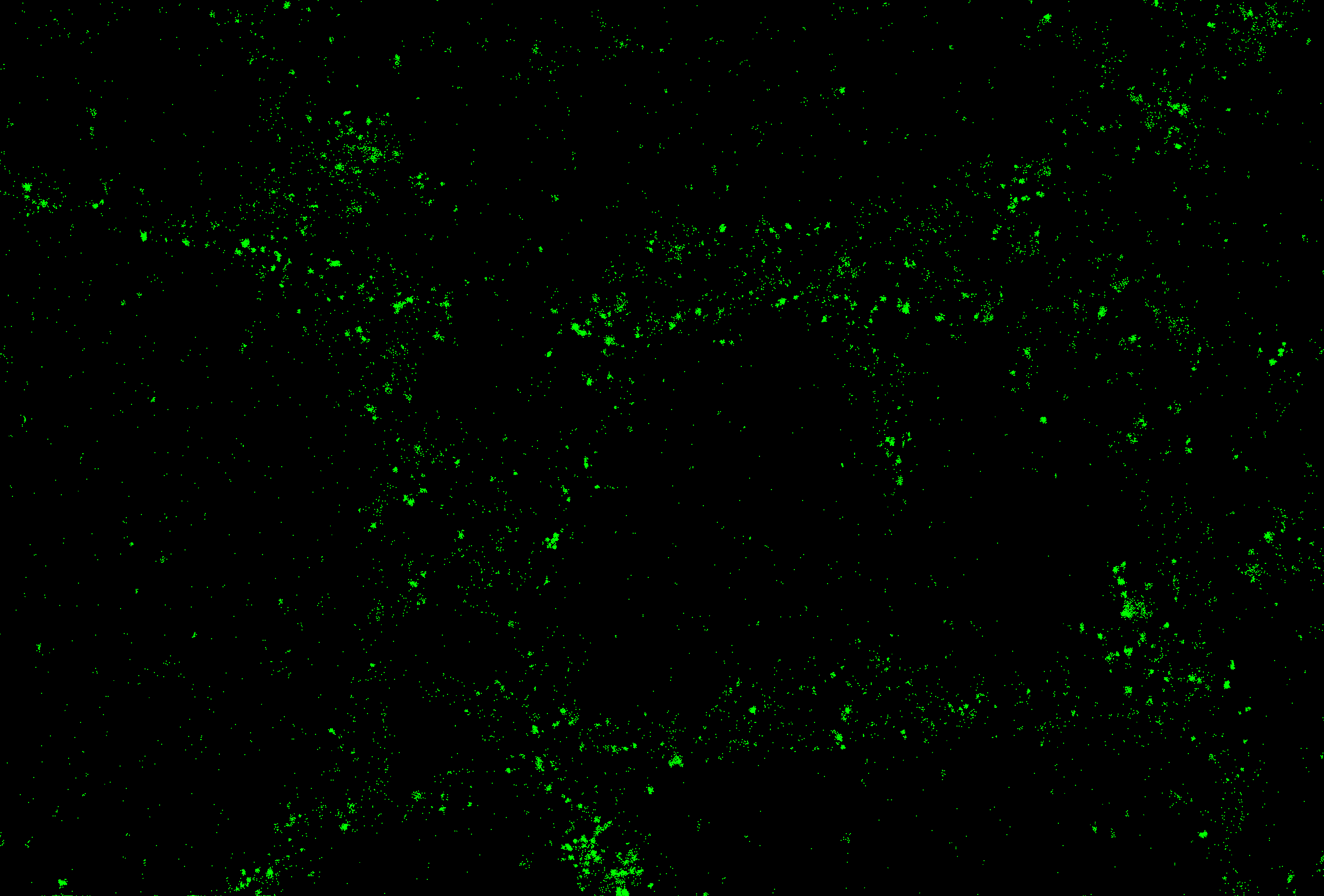}
\includegraphics[width=.43\linewidth]{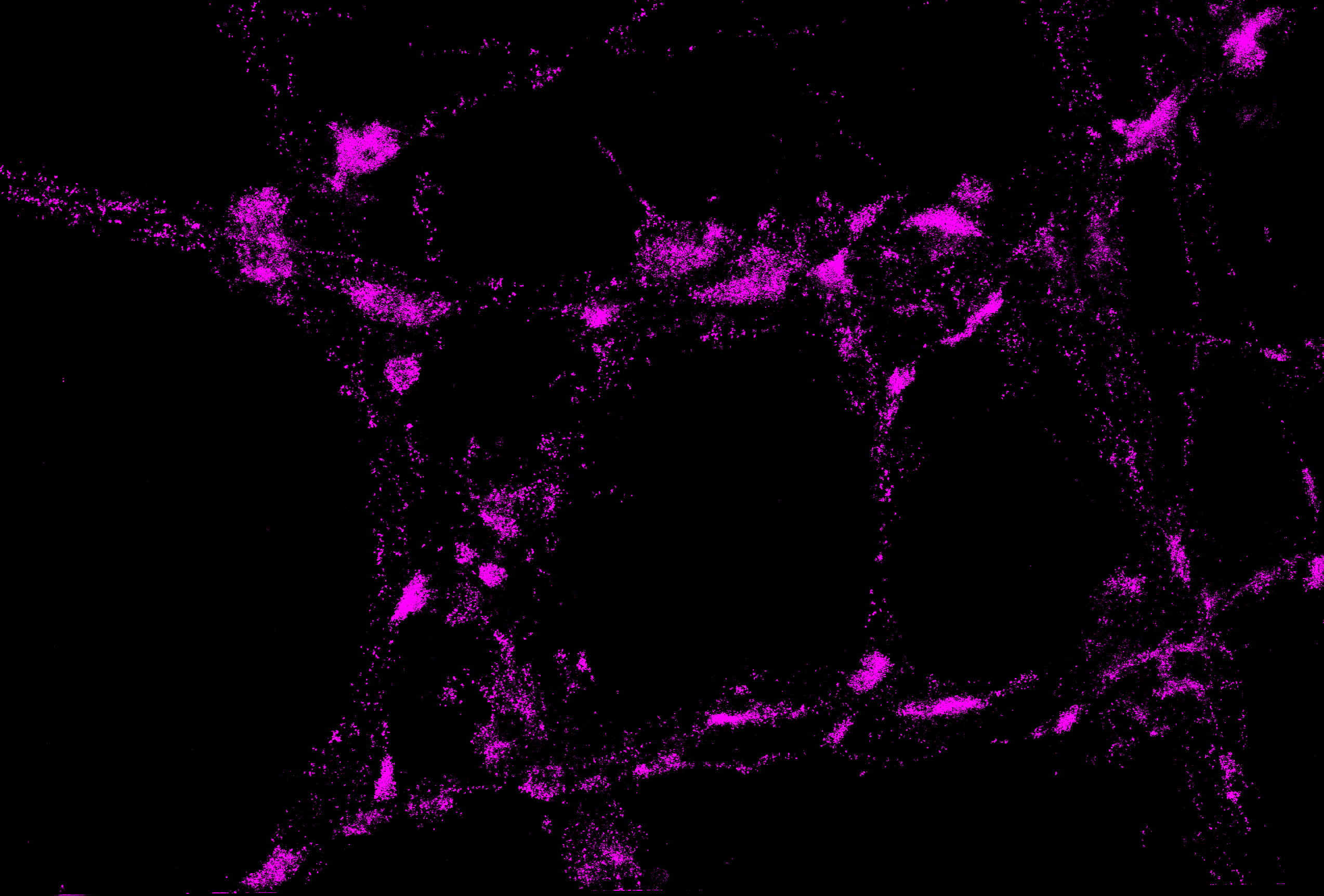}
\includegraphics[width=.87\linewidth]{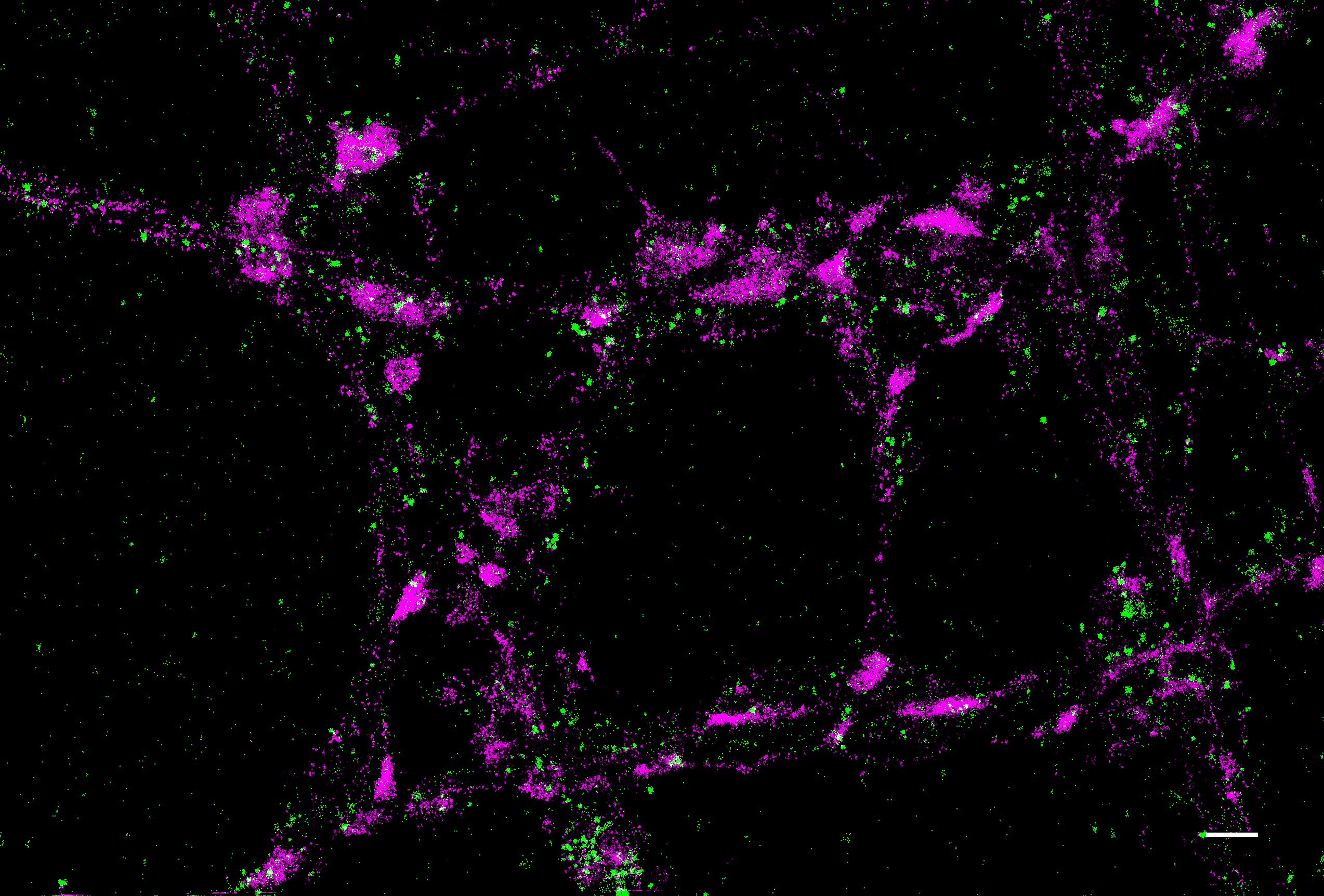}
\end{center}
\vspace{-0.25cm}
\caption{DSTORM acquisition  of a cell from hippocampi of mice \citep{Andreska2014} expressing BDNF proteins (green tagged molecules)  and vGlut (purple tagged molecules). The image at the bottom is the superposition of the two acquired images displayed at the top. The scale bar corresponds to 1$\mu$m and the image size is $ 2547\times 1724$ pixels. This figure appears in color in the electronic version of this article, and any mention of color refers to that version.}
\label{Fig2}
\end{figure}

The characterization of molecular interactions is a major challenge in quantitative microscopy. 
This problem  is usually addressed in living cells by fluorescently labeling  two types of  molecules of interest with spectrally distinct fluorophores, and  simultaneously imaging them. This process provides two images of the same cell, each depicting  one different fluorescently tagged molecule,  both corrupted with diffraction, noise and nuisance background.
As an illustration, Figure~\ref{Fig1} depicts a cell containing Langerin proteins in the red channel,  along with Rab11 proteins in the green channel \citep{Boulanger2014}.  Note that the figures appear in color in the electronic version of this article, and any mention of color refers to that version. 
In the other example of Figure~\ref{Fig2}, BDNF (brain-derived neurotrophic factor) proteins  are visible in green along with vesicle markers for presynapes (vGlut) in purple \citep{Andreska2014}. These two  datasets have been acquired with different microscopy techniques: 3D multi-angle TIRFM (total internal reflection fluorescence microscopy) for the image in  Figure~\ref{Fig1}, and dSTORM (direct stochastic optical reconstruction microscopy) for the image in Figure~\ref{Fig2}. These data are analyzed further in Section~\ref{sec:real} and  the details of experiments are provided in the appendix.

Potential protein-protein interactions inside the cell are determined by the degree of colocalization at the resolution limit of the microscope or, in other words, by the  proportion of interacting proteins co-detected at the same location or in very close proximity \citep{Manders1993,Bolte2006}.  Colocalization often corresponds to co-compartmentalization, implying that two or more molecules bind to the same structure or domain in the cell. 
For this reason, the analysis of colocalization is known to be a critical  step in the analysis of molecular interactions. 
Given the observation of two types of proteins in a cell acquired by fluorescence microscopy, the main questions are whether colocalization occurs, whether it occurs globally in the whole cell or only in certain subregions of the cell, and to spatially quantify colocalization within the cell. 
These problems need to be correctly addressed in presence of the increasing  amount of 2D, 3D, and 3D+time data available in bio-imaging.  
However, for the time being,  there is no definitive solution to colocalization analysis, even for  2D images acquired by conventional diffraction-limited microscopy techniques.

Moreover, the emergence of super-resolution microscopy techniques such as SIM (structured illumination microscopy), dSTORM (direct stochastic optical reconstruction microscopy), PALM (photoactivated localization microscopy)  and STED (stimulated emission depletion) have raised supplementary challenges for the colocalization study. 
These techniques, whose developers Eric Betzig, William Moerner and Stefan Hell were awarded by the 2014 Nobel prize in chemistry  for ``the development of super-resolved fluorescence microscopy'', allow to acquire highly resolved images. This is illustrated by the data in Figure~\ref{Fig2} obtained by dSTORM, in comparison with Figure~\ref{Fig1} where a more conventional TIRFM was used. But the size of these highly resolved images becomes extremely large. Colocalization methods must handle  these very large volumes of data efficiently. 
A more frequently approach is now  to analyze two channels acquired by two different microscopy technologies: a conventional diffraction-limited technique (confocal, TIRFM,...) and a super-resolution technique (e.g. SIM, dSTORM, ...).
This leads to situations where colocalization must be analyzed between two channels with very different resolutions
(of factor 2 to 5 in practice).

\subsection{Statistical formulation and existing methods}

From a statistical perspective, detecting colocalization between two types of proteins in a cell amounts to test whether the two images acquired with fluorescence microscopy are correlated. This might appear at first sight as a basic simple statistical question. However  the objects of interest, the proteins in each image, are not clearly identified. The data are two images of the same cell, corrupted by noise,  nuisance background and diffraction. 
In the review written by \cite{Dunn2011}, the authors conclude that ``the problem of significance testing of colocalization data is one with no simple answer at this point".

 In the literature, two distinct categories of colocalization approaches are generally considered (see  \cite{Bolte2006,Dunn2011} for a review): intensity-based methods and object-based methods. The former analyzes the image contents and focus on  the fluorescence signals, while the latter first detect the objects of interest before analyzing their joint repartition by spatial point process statistics. 
 
 The commonly-used intensity-based technique is to simply compute the Pearson's correlation coefficient between the two images. Nevertheless, this coefficient is sensitive to high intensity backgrounds and provides misinterpretations if the signal-to-noise ratios or the scales of the images are not similar. For these reasons and because each image is spatially autocorrelated, testing the significance of the Pearson correlation is not straightforward. A widely used solution \citep{Costes2004} consists in block-resampling one channel, many times, in order to simulate  by Monte Carlo the distribution of the Pearson's coefficient under the null hypothesis of no-correlation. 
This technique is limited by the choice of the size of the blocks of pixels used for resampling, which strongly influences its result. As a conclusion of previous works \citep{ramirez2010,Dunn2011}, confirmed by our simulation study, this testing procedure, for the choice of blocks  advised in  \cite{Costes2004}, leads to too many false positive colocalized situations. 
 It also suffers from a high computational cost  due to the Monte Carlo step, making this method inappropriate for modern big data. 
Alternative intensity-based coefficients have been proposed  in the bio-imaging community, from the popular Mander's coefficient \citep{Manders1993} to more sophisticated ones \citep{Comeau2008,ramirez2010,Wu2010}. These coefficients suffer from the same basic limitations as the Pearson's coefficient, namely  sensitivity to noise and to nuisance background,  and heavy simulations are required to test their significance.

In the object-based methods, a segmentation procedure is applied in each image to detect the spots (or regions) in cells corresponding to the presence of  proteins of interest. The detected spots are then reduced to points (their centers), providing two point patterns, and the interaction between the two point patterns  is analyzed by spatial point process statistics. 
For this last step,  assuming stationarity of the two point patterns, the most applied strategy is to analyze $K_{12}$, the cross Ripley's  $K$-function between the two point patterns. Roughly, given a point $x$ belonging to the point pattern~1,  $K_{12}(r)$ is the expected number of points in the point pattern~2 located in a ball centered at $x$ with radius  $r$. 
If the two point patterns are independent, $K_{12}(r)$ is simply the volume of the ball with radius $r$ \citep{MW}. Under some conditions, the distribution of the estimate $\hat K_{12}(r)$ can be characterized asymptotically (in the sense that the  spatial domain of observation is large enough) under the null  hypothesis of independence, and this is exploited to construct a testing procedure for colocalization in \cite{Lagache2015}. In \cite{Sherman2011}, the closely related cross pair correlation function is analyzed instead of $K_{12}$, but no testing procedure for significance is considered.  As an alternative, a parametric Gibbs model is  considered in \cite{Helmuth2010} to model the interaction between the two point patterns. The significance of the Gibbs interaction is then tested by Monte Carlo simulations.

Object-based methods are less sensitive to noise and background than intensity-based techniques, though the initial segmentation step to identify the objects can be critical. Transforming each image into a point pattern is particularly well-adapted to images where the objects of interest can be fairly assimilable to points (typically small balls). The statistical analysis then becomes rigorous thanks to the well-established spatial point process techniques. However, this transformation is not suitable in presence of large or anisotropic shaped objects, in which case the reduction of each object to a single point constitutes a dramatic loss of information. For the same reason, these methods can not be applied in subregions of the cell containing only few objects, in which case the associated point patterns would be too sparse to be analyzed. 
Moreover, the computational cost increases with the number of detected points and can become  prohibitive in presence of high-resolved images involving a large number of objects.

 \subsection{GcoPS}

We introduce a new  colocalization method, detailed in Section~\ref{sec:method},  that we name GcoPS for Geo co-Positioning System.  The basic idea is to analyze the two random sets composed of the detected spots obtained after a preliminary segmentation procedure in each image. 
Accordingly, this first step is similar to the object-based methods described before. However we do not reduce the  segmented images  to point patterns but keep them as they are: two binary images, each representing a random set of spots. The analysis of these random sets can then be seen as an intensity-based method, since it basically consists in testing the significance of the Pearson correlation between two binary images, see Section~\ref{sec:method}.  From this perspective, GcoPS can be viewed as a compromise between an object-based method and an intensity-based method. 
Moreover, unlike previous intensity-based methods, our testing procedure exploits a closed formula and no Monte Carlo simulations are needed.

 GcoPS has been implemented within the Icy software \citep{Chaumont2012},  an open platform for bioimage informatics based on Java, and is available as a free downloadable module (more details at {\small \url{http://icy.bioimageanalysis.org/plugin/GcoPS}}). Nonetheless the C++ code is also available in the web supplement of this paper.
  
 The rest of the paper is organized as follows. Section~\ref{sec:method} describes the mathematical foundations of the method and explains the practical choices made for the implementation of GcoPS. Section~\ref{sec:simu} summarizes  the results of our simulation study (detailed in the appendix), which demonstrates the good performances of GcoPS in comparison to the most used intensity-based method of \cite{Costes2004} and to the object-based method of \cite{Lagache2015} that exploits the cross Ripley's  $K$-function. 
In Section~\ref{sec:real}, we apply GcoPS to the analysis of the real datasets presented in Section~\ref{sec:bio}.  A concluding
discussion is made in Section~\ref{sec:discussion}. Finally a supporting information, available in the appendix of this paper, contains the most technical aspects of our contribution, the detailed results of the simulation study, the specificities of data preparation and supplementary figures.

\section{The method}\label{sec:method}

\subsection{Mathematical foundations}
GcoPS applies a test on a binary image pair as explained in this section. The two binary images are obtained by segmenting the input fluorescence images. This splits the set of pixels into a background set and a foreground set, both of which should be non-empty.  The sensitivity to this preliminary step is analyzed in Section~\ref{sec:simu}, where GcoPS is seen to be robust to the choice of the segmentation algorithm.

In what follows, we view the two foreground sets of the two binary images as realizations of two random sets observed through a pixelated image. We refer to \cite{chiu2013} for basic notions on random sets. 
 Formally, let $\Gamma_1\subset\mathbb Z^d$ and $\Gamma_2\subset\mathbb Z^d$ be  two  random sets in $\mathbb Z^d$, where $d$ stands for the dimension (typically $d=2$ or $d=3$ in our examples). Intuitively, $\Gamma_1$ would be the foreground set of the first image if this image was supported on the infinite lattice $\mathbb Z^d$, and similarly for $\Gamma_2$.  We let $\Omega_n\subset\mathbb Z^d$ be the observation region of these random sets, that is the region of interest of the observed images. The index $n$ stands for the number of elements (pixels or voxels) in $\Omega_n$, or in other words its cardinality. For instance when $d=2$, if the region of interest is the whole observed images of size $n_1\times n_2$ pixels, $\Omega_n$ is simply the lattice $\{1,\dots,n_1\}\times \{1,\dots,n_2\}$ with $n=n_1n_2$ pixels. In general yet, the region of interest $\Omega_n$ is a subregion of the observed images, see for instance Figure~\ref{Fig1}. The two observed foreground sets are therefore $\Gamma_1\cap \Omega_n$ and $\Gamma_2\cap \Omega_n$. In the following we develop a procedure to test whether $\Gamma_1$ and $\Gamma_2$ are independent or not (the case of colocalization), based on the observation of $\Gamma_1\cap \Omega_n$ and $\Gamma_2\cap \Omega_n$ and when $n\to\infty$.

Let us consider, for a generic given point $o\in \mathbb Z^d$, the probabilities
\begin{equation}
p_1=\P(o\in\Gamma_1),\; p_2=\P(o\in\Gamma_2),\;  p_{12}=\P(o\in\Gamma_1\cap\Gamma_2).
\end{equation}
If $\Gamma_1$ and $\Gamma_2$ are two  independent random sets, we have $p_{12}=p_1 p_2$.  A natural empirical measure of the departure from independence between $\Gamma_1$ and $\Gamma_2$ is therefore 
\begin{equation}
\hat D_n = \hat p_{12} - \hat p_1 \hat p_2,
\end{equation}
where
 \begin{equation*}
\hat p_1 = \frac{1}{n}|\Gamma_1\cap\Omega_n|,\quad \hat p_2 = \frac{1}{n}|\Gamma_2\cap\Omega_n|, \quad \hat p_{12} = \frac{1}{n}|\Gamma_1\cap\Gamma_2\cap\Omega_n|,\end{equation*}
where $|A|$ denotes the number of elements  in a finite subset $A$ of $\mathbb Z^d$. 
Note that $\hat p_1$ is simply the mean number of ``red'' pixels in the region of interest, while 
$\hat p_2$ is the mean number of ``green'' pixels and $\hat p_{12}$ is the mean number of  ``yellow'' pixels (that are both ``green'' and ``red'').  The following proposition is the basic result for our testing procedure. It provides a central limit theorem for $\hat D_n$  when $\Gamma_1$ and $\Gamma_2$ are independent.

To this end, we assume that $\Gamma_1$ and $\Gamma_2$ are stationary sets.  Denoting by $\Gamma_1-h$ the translation of $\Gamma_1$ by the vector $h\in\mathbb Z^d$, this means that the distribution of $\Gamma_1$ is the same as the distribution of $\Gamma_1-h$ for any $h\in\mathbb Z^d$, and similarly for $\Gamma_2$. We denote by $C_1$ and  $C_2$  the  auto-covariance functions of $\Gamma_1$ and $\Gamma_2$, respectively. Specifically, denoting by $\mathbf 1_{\Gamma_1}(x)$ the indicator function equal to 1 if $x\in\Gamma_1$ and to 0 otherwise, $C_1(h)$ is defined for any $h\in\mathbb Z^d$ and any $x\in\mathbb Z^d$ by
\begin{align*}
C_1(h)&= \E\{\mathbf 1_{\Gamma_1}(x)\mathbf 1_{\Gamma_1}(x+h)\}-\E\{\mathbf 1_{\Gamma_1}(x)\}\E\{\mathbf 1_{\Gamma_1}(x+h)\}\\&=\P\{o\in \Gamma_1\cap(\Gamma_1-h)\} - p_1^2.
\end{align*}
This expression does not depend on $x$ by stationarity of $\Gamma_1$ but involves the probability that two points separated by $h$ belong to $\Gamma_1$. In particular $C_1(0)=p_1(1-p_1)$. The  formulas for $C_2$ are the same where $\Gamma_1$ is replaced by $\Gamma_2$, and $p_1$ by $p_2$. If we assume further that the distribution of $\Gamma_1$ and $\Gamma_2$ are invariant by rotation, then $C_1$ and $C_2$ are isotropic and only depend on the norm $\|h\|$. Our procedure only assumes stationarity but not necessarily isotropy, which makes it adapted to situations in bio-imaging where the observed molecules do not have an isotropic shape. The covariance functions can be estimated by the empirical auto-covariance functions $\hat C_1$ and $\hat C_2$ based on the observation of $\Gamma_1\cap\Omega_n$ and $\Gamma_2\cap\Omega_n$. For $h\in\mathbb Z^d$, let $\Lambda_n(h)=\{x\in\Omega_n, y\in\Omega_n \text{ such that } y=x+h\}$. A standard expression of $\hat C_1(h)$ is given  by
\begin{equation}\label{hatC}
\hat C_1(h) = \frac 1{|\Lambda_n(h)|}\sum_{x\in\Omega_n}\sum_{y\in\Omega_n} \{\mathbf 1_{\Gamma_1}(x) -\hat p_1\}\{\mathbf 1_{\Gamma_1}(y) -\hat  p_1\} \mathbf 1_{x-y=h}, 
\end{equation}
if $\Lambda_n(h)\neq\emptyset$, and $\hat C_1(h) =0$ otherwise, see for instance \cite{Cressie}.  This estimator is a good approximation of $C_1(h)$ whenever $|\Lambda_n(h)|$ is not too small. Robust or tapered estimators of $C_1(h)$ can also be used to reduce the bias \citep{Cressie, guyon}. From a theoretical point of view, we can use any estimator of $C_1(h)$ (and similarly for $C_2(h)$) in Proposition~\ref{prop:asympt}, provided it is consistent for all $h\in\mathbb Z^d$, which is in particular the case of \eqref{hatC} under our assumptions \cite[Theorem~4.1.1]{guyon}.

To state our central limit theorem, we need to assume that the observation domain $\Omega_n$ is a {\it regular} finite subset of $\mathbb Z^d$, which means that $n^{-1}|\partial\Omega_n|\to 0$ as $n\to \infty$ where $\partial \Omega_n$ denotes the boundary of $\Omega_n$. This assumption is not restrictive and holds true if, for instance,  $\Omega_n$ contains a square lattice with side-length $n_0$ and $n_0\to\infty$ as $n\to\infty$.

Moreover, we assume that $\Gamma_1$ and $\Gamma_2$ are two $R$-dependent stationary  random sets. 
We recall that $\Gamma_1$ is $R$-dependent if the events $\mathbf 1_{\Gamma_1}(x)$ and $\mathbf 1_{\Gamma_1}(y)$ are independent whenever $x$ and $y$ are separated by a distance greater than $R$.  In this case, if $\|h\|\geq R$ then  $C_1(h)=0$ and $C_2(h)=0$. This theoretical framework guarantees the  weak spatial dependence of $\Gamma_1$ and $\Gamma_2$, but in practice the value of $R$ does not need to be known, see  Remark~\ref{weak dependence}. This assumption could be weakened to strong mixing conditions as in \cite{bolthausen:82}, but at the cost of slightly more technicalities that we prefer to omit. This would amount to control the rate of convergence to $0$ of $C_1(h)$ and $C_2(h)$ when $\|h\|\to\infty$. Nonetheless, $R$-dependency is a  reasonable assumption for applications in bio-imaging, which provides insightful results. It is for instance fulfilled if $\Gamma_1$ and $\Gamma_2$ are Boolean germ-grain  models with bounded radii~\citep{Gotze1995}.

Under these assumptions, we prove in the following proposition that the variance of $\hat D_n$ is equivalent to $n^{-1}S$ with
\begin{equation}\label{defS} 
S=\sum_{h\in\mathbb Z^d} C_1(h) C_2(h)=\sum_{h\in\mathbb Z^d,\, \|h\|\leq R} C_1(h) C_2(h),\end{equation}
where the last simplification comes from the $R$-dependent assumption we made. 
Our test statistic is finally
\begin{equation}\label{defT}T_n= \sqrt n \frac{\hat D_n}{\sqrt{\hat S_n(\delta)}},\end{equation}
where $\hat S_n(\delta)$  estimates $S$ and is  given for $\delta>0$ by
$$\hat S_n(\delta)=\sum_{h\in\mathbb Z^d,\, \|h\|\leq \delta} \hat C_1(h)\hat C_2(h).$$
Note that if $\hat C_1(h)$ and $\hat C_2(h)$ are consistent estimators, $\hat S_n(\delta)$ is a consistent estimator of 
 $S$ whenever $\delta\geq R$. 
We are now in position to state our main result. Its proof  is postponed to the appendix. 
 
\begin{prop}\label{prop:asympt}
Let $\Gamma_1$ and $\Gamma_2$ be two $R$-dependent stationary  random sets on $\mathbb Z^d$ with autocovariance functions $C_1$ and $C_2$, respectively, satisfying $S>0$ where $S$ is defined in \eqref{defS}.
Assume that $\Omega_n$ is a regular finite subset of $\mathbb Z^d$ with cardinality $n$ and that $\hat C_1(h)$ and $\hat C_2(h)$ are consistent estimators of  $C_1(h)$ and  $C_2(h)$ for any $h\in\mathbb Z^d$. Let $T_n$ be defined by \eqref{defT} with $\delta\geq R$. If
$\Gamma_1$ and $\Gamma_2$ are independent, then $T_n$ tends in distribution to a standard Normal variable as $n\to\infty$.
\end{prop}

\begin{rem}
The statistic $T_n$ can be viewed as a normalized version of the empirical Pearson correlation between the two binary images $(\mathbf 1_{\Gamma_1}(x))_{x\in\Omega_n}$ and $(\mathbf 1_{\Gamma_2}(x))_{x\in\Omega_n}$. From this point of view, our approach shares some similarities with  the  Costes method \citep{Costes2004}, where the correlation between the two raw images is tested by Monte Carlo simulations. The two important differences are that we consider binary images instead of the raw images, which limit the noise and background effects, and that we know the asymptotic distribution of $T_n$, which avoids Monte Carlo simulations. 
\end{rem}

\begin{rem}\label{weak dependence}
We assume that $\Gamma_1$ and $\Gamma_2$ are $R$-dependent but $R$ does not need to be known. This assumption  is a convenient way to capture the spatial weak dependence of  the random sets. Under a more general setting of weak dependence, the statement would be similar although the proof would require more technicalities. A key point is that  $\hat S_n(\delta)$ must be a consistent estimator of $S$. This is in general true for a proper choice of $\delta=\delta_n$ depending on $n$. Specifically, $\delta_n$ needs to tend to $\infty$ but slower than $n$. This choice of $\delta$ corresponds to the choice of the bandwidth (or truncation parameter) for the HAC (heteroskedasticity and autocorrelation consistent) estimator in econometry, see for instance \cite{andrews1991}. For $R$-dependent random sets, $\delta_n\geq R$ is sufficient. The practical way of choosing $\delta$, detailed in the next section, applies for any weak dependent random sets and does not depend on the $R$-dependent assumption. 
\end{rem}

\subsection{Implementation of GcoPS}

The testing procedure for colocalization in $\Omega_n$ boils down to a test of independence between $\Gamma_1\cap\Omega_n$ and $\Gamma_2\cap\Omega_n$, which is  straightforwardly deduced from Proposition~\ref{prop:asympt}:
The null hypothesis of independence is rejected at the asymptotic level $\alpha\in(0,1)$ if $|T_n|>q(\alpha/2)$ where $q(\alpha/2)$ denotes the  upper $\alpha/2$-quantile of the standard normal distribution. The corresponding $p$-value is
\begin{equation}\label{pvaluebi}
p\textrm{-value} = 2(1-\Phi(|T_n|)),\end{equation}
where $\Phi$ denotes the cumulative distribution function of the standard normal distribution.
This bilateral test  can be modified into a unilateral test if the focus for the alternative hypothesis is more specifically colocalization ({\it resp.} anti-colocalization), that is positive ({\it resp.} negative) dependence between $\Gamma_1$ and $\Gamma_2$,  rather than dependence in a general sense. Then a more powerful procedure at the asymptotic level $\alpha$ consists in rejecting the null when $T_n>q(\alpha)$ ({\it resp.} $T_n<-q(\alpha)$), which corresponds to $p\textrm{-value}=1-\Phi(T_n)$ ({\it resp.} $p\textrm{-value}=\Phi(T_n)$).

In the module GcoPS of Icy and in the experiments of the next section, we did the following choices of implementation. 
The empirical covariances $\hat C_1$ and $\hat C_2$ are obtained by an FFT (Fast Fourier Transform) in each image. Concerning the truncation parameter $\delta$ in $\hat S_n(\delta)$, we choose  the larger value of $\|h\|$ such that both $\hat C_1(h)/\hat C_1(0)>0.1$ and  $\hat C_2(h)/\hat C_2(0)>0.1$. In other words, $\delta$ is the maximal range of empirical correlation in $\Gamma_1$ and $\Gamma_2$ beyond which the correlation is less than $0.1$. Beyond this range $\delta$, we view the   values of $\hat C_1(h)\hat C_2(h)$ as a nuisance noise. This  choice of truncation has  also the advantage to speed up the computation in comparison with a larger choice of $\delta$. 
Finally, $\hat p_1$, $\hat p_2$ and $\hat p_{12}$  are immediately obtained from the mean of the binary images $(\mathbf 1_{\Gamma_1}(x))_{x\in\Omega_n}$ and $(\mathbf 1_{\Gamma_2}(x))_{x\in\Omega_n}$, and  from their product. This implementation ensures a very low  computational cost, as attested by  the results discussed in Section~\ref{sec:simu} and reported in the appendix.

\section{Evaluation on synthetic  data sets}\label{sec:simu}

We evaluate in this section the performance of GcoPS on 2D and 3D  data sets in adverse and noisy situations and compare it to the competitive Costes method \citep{Costes2004}, the most used intensity-based method in bio-imaging,  and to the object-based method of \cite{Lagache2015}, which relies on the analysis of  the cross Ripley's  $K$-function. 
The results of our simulations are detailed in the appendix. In summary we have assessed the following aspects:
\begin{enumerate}
\item Computation time in 2D, 2D+time and 3D data;
\item Performances on simulated 2D images, possibly corrupted with noise and a spatial shift;
\item Sensitivity to image segmentation;
\item Sensitivity to the shape and scale of objects;
\item Performances in presence of a different optical resolution in each image;
\item Performances on simulated 3D images.
\end{enumerate}

\begin{figure}
\begin{center}
\begin{tabular}{ccc}
\includegraphics[width=.28\linewidth]{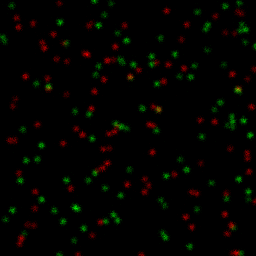}  &
\includegraphics[width=.28\linewidth]{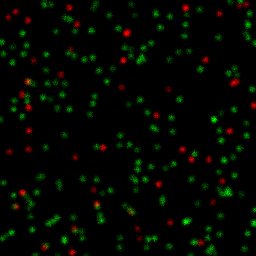} &  
\includegraphics[width=.28\linewidth]{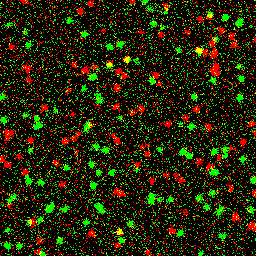} \\
\includegraphics[width=.28\linewidth,height=.28\linewidth]{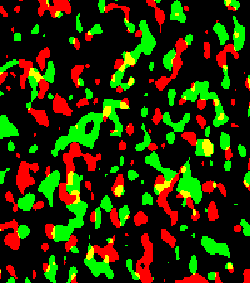} &
\includegraphics[width=.28\linewidth,height=.28\linewidth]{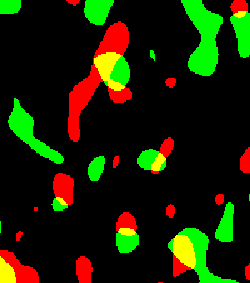} &
\includegraphics[width=.28\linewidth,height=.28\linewidth]{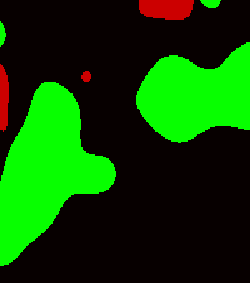}\\
\includegraphics[width=.28\linewidth]{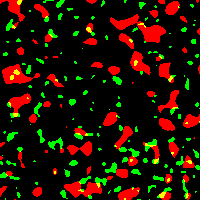}&
\includegraphics[width=.28\linewidth]{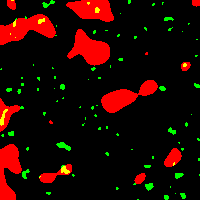} &
\includegraphics[trim=2.8cm 1.55cm 2.8cm 1.55cm,clip,width=.28\linewidth]{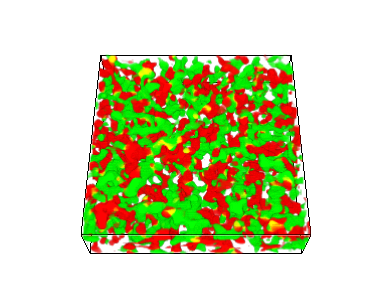} 
\end{tabular}
\end{center}
\caption[bof]{Examples of simulated image pairs used to evaluate the performances of GcoPS. From top left to bottom right: i) 2D image possibly corrupted with noise and a spatial shift; ii) 2D image where the green channel shows 4 times more objects than the red channel; iii) over-segmented image; iv)-vi) various shapes and scales of objects; vii)-viii) different optical resolution in each channel; ix) 3D image. This figure appears in color in the electronic version of this article, and any mention of color refers to that version.}
\label{fig:simus}
\end{figure}

Figure~\ref{fig:simus} shows examples of image pairs that we have evaluated with each method. This figure demonstrates the diversity of the considered situations. Note that the displayed images pairs come from independent channels, but we obviously also considered correlated channels in each situation. In each case, a thousand of image pairs have been simulated to evaluate the methods.  
We refer to the appendix for details about the image generators, more pictures and thorough comments.

In a nutshell, our simulation study reveals that the main drawbacks of the Costes method is the computational time due to its Monte Carlo step, that can be cumbersome for 2D+time and 3D images, along with too many false positive decisions ($12\%$ in average and up to $25\%$, instead of the expected $5\%$ when the level of the test is fixed to 0.05). 
On the other hand, the main weakness of the object-based method  of \cite{Lagache2015} is the lack of sensitivity to colocalization. This clearly appears in presence of large objects (as in the rightmost plot of the middle row of Figure~\ref{fig:simus}), a situation where reducing each object to a point is improper. This method is also very sensitive to the preliminary segmentation step. In particular it completely fails to detect colocalization for over-segmented images (as in the top right plot of Figure~\ref{fig:simus}). 
In turn, GcoPS does not suffer from the aforementioned drawbacks: i) it is fast; ii) it well controls the probability  of false positives; iii) it is very sensitive to colocalization; and iv) it is robust to segmentation, to the shape and the size of the objects, and to a possible different optical resolution in each channel.

An important consequence of the good performance of GcoPS in presence of large objects is that this method can be applied in small windows in the image. Indeed, dividing images into sub-regions acts as enlarging objects, a situation GcoPS handles well. This opens the possibility to accurately localize colocalization.
Moreover, the robustness of GcoPS in presence of a different optical resolution in each image (as in the leftmost plot of the bottom row of Figure~\ref{fig:simus}) shows that  GcoPS is also able to efficiently process images for which a different microscopy technique is used for detecting each type of molecules.

\section{Application to real biological imaging}\label{sec:real}

\subsection{Spatiotemporal colocalization of Langerin and Rab11a in 3D-TIRFM}

\begin{figure}
\begin{center}
\includegraphics[width=.8\linewidth]{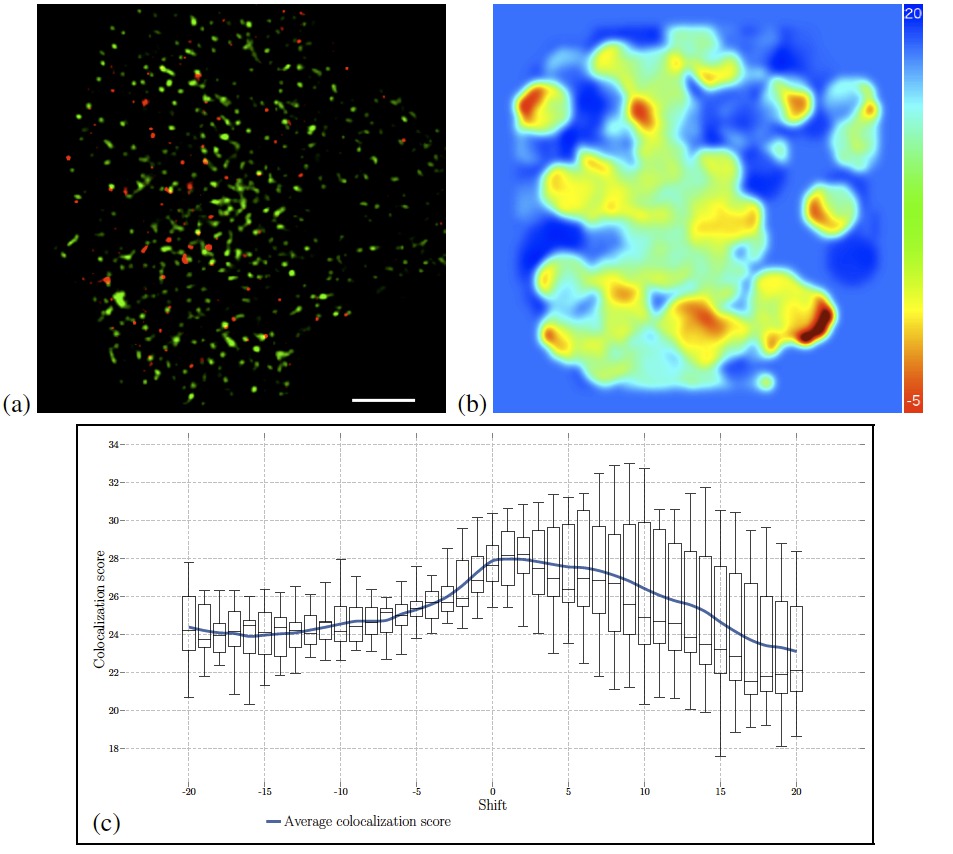}
\end{center}
\vspace{-0.25cm}
\caption[bof]{Spatiotemporal colocalization of Langerin and Rab11a proteins  in a 3D multi-angles TIRF acquisition \citep{Boulanger2014} providing 35 frames of  image pairs with a size equal to $400\times400\times20$ pixels. (a) Average intensity projection (along the $z$ axis)  of a 3D multi-angles TIRF acquisition \citep{Boulanger2014} 
at time $t=0$ showing a RPE1 cell expressing m-Cherry Langerin (red channel) and GFP Rab11 (green channel) on a crossbow shaped micropattern.  The scale bar  corresponds to 10~$\mu$m. (b) Heat map for continuous visualization of colocalization score associated to the left image at time $t=0$. (c) Distribution of colocalization  scores (box plots) and their averages (blue solid line) between pairs of images taken at time $t$ and $t \pm \Delta t$,  for a set of temporal shifts $\Delta t \in \{-20,\dots,20\} $  and computed over the 35 frames ($t \in \{0,\dots,34\}$) as displayed  at the top at $t=0$. This figure appears in color in the electronic version of this article, and any mention of color refers to that version.}
\label{Fig4}
\end{figure}

In the first experiment, a temporal acquisition with 3D-TIRFM was performed \citep{Boulanger2014} using wild-type RPE1 cells transfected with Langerin-mCherry and Rab11a-GFP, resulting in 35 frames of image pairs with a size equal to   $400\times400\times20$ pixels. The details of data preparation are provided in the appendix. Figure~\ref{Fig1} depicts the two images obtained at time $t=0$ (that is the first frame) along with their superposition, once projected along the $z$ axis. 
The same superposition is shown again in Figure~\ref{Fig4}~(a). The value of the test statistic  $T_n$ of Proposition~\ref{prop:asympt} (that we alternatively name colocalization score) computed on the whole cell is very large (about 28), whatever the frame considered in the sequence is, showing a clear colocalization between the two channels. We then processed a dense colocalization map at time $t=0$, see Figure~\ref{Fig4}~(b), by computing the colocalization score $T_n$ every 25 pixels at the medium plane on windows of size $50\times50\times10$ voxels. The spatial density of colocalization scores is  computed by applying a Gaussian kernel density method with a global bandwidth equal to 5. 
This representation has the advantage to discriminate between different levels of colocalization since a more colocalized region will have a higher colocalization score. 

Finally, we applied GcoPS to the whole Langerin-mCherry/Rab11a-GFP image pair sequence, frame by frame, to get the distribution of the colocalization score $T_n$. Next, we shifted the frames between the two channels and applied GcoPS from -20 to +20 frames shift by considering Langerin-mCherry as the reference. The mean colocalization scores along with their distributions are reported for each temporal shift in  Figure~\ref{Fig4}~(c). The slope of the colocalization scores  is steeper for a positive temporal shift than for a negative temporal shift, demonstrating that globally,  Rab11a is visible before Langerin, which is consistent with previous observations \citep{Gidon, Boulanger2014}.

\subsection{Colocalization of BDNF proteins and vGlut in dSTORM}\label{BDNF}
In the second experiment, we evaluated the colocalization between BDNF (brain-derived neurotrophic factor) proteins and vGlut, a vesicle marker for presynapses, on an image acquired with dSTORM \citep{Andreska2014}, see Figure~\ref{Fig2}. While identifying BDNF proteins is straightforward with a standard spot detector, the segmentation of vGlut is more difficult as these markers do not correspond to regular shapes. Consequently, we performed three different segmentations by thresholding the image with three different thresholds, resulting in the three binary images shown in the appendix. The $p$-value obtained with GcoPS is extremely low ($p$-value = 0) for the three segmentations of vGlut, a result consistent with previously published studies \citep{Andreska2014} where  colocalization between BDNF proteins and vGlut was unraveled.
The object-based method of \cite{Lagache2015} needs a segmentation of vGlut with a high threshold to obtain a low $p$-value ($p$-value = 0.005). Segmentations with low thresholds give large objects, leading to a failure for this method ($p$-value = 0.532 and $p$-value = 0.061). The intensity-based method of \cite{Costes2004} provides $p$-values close to 0 whenever the size of blocks for the permutation step is fixed to $2 \times 2$, or $5 \times 5$ or $10 \times 10$ pixels$^2$ in the two channels.  This result is in agreement with the conclusion of GcoPS. However the Costes method needs about 3 minutes to process this pair of images while GcoPS only takes 9 seconds. Nonetheless, it is fair to recall that all these methods assume stationarity of distributions of the proteins in the two channels. This assumption is not satisfied here and one must be careful with the previous conclusions.

\begin{figure}
\begin{center}
\includegraphics[width=\linewidth]{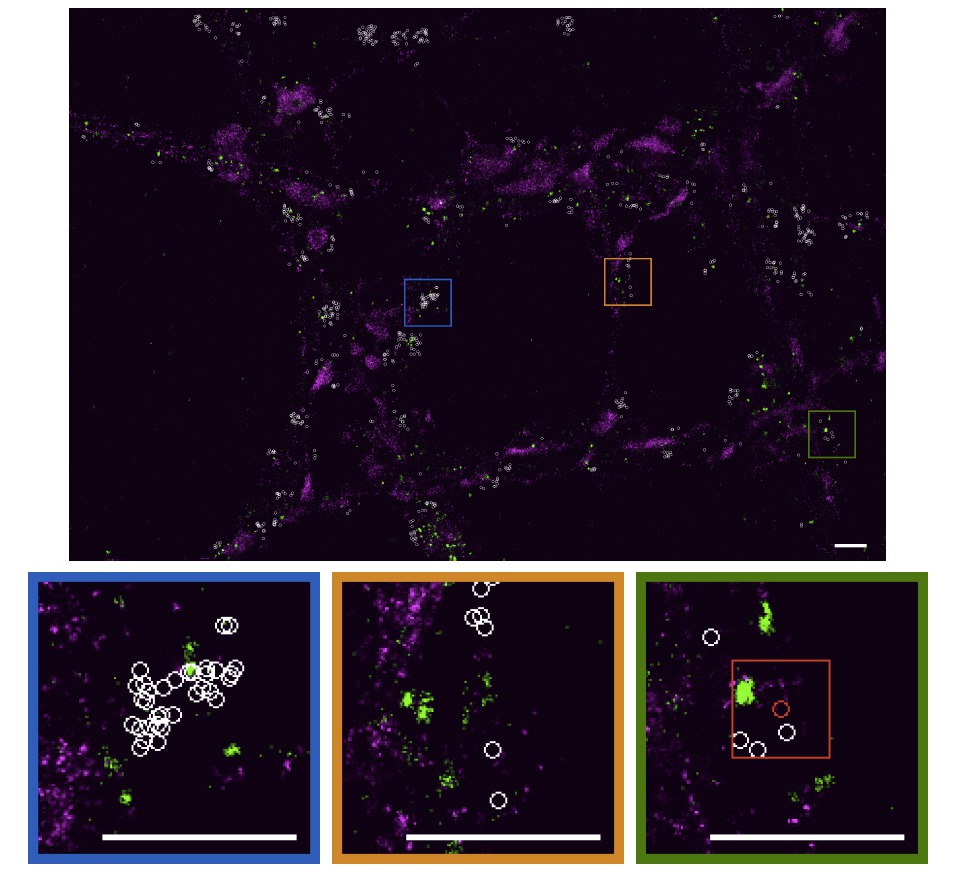}
\end{center}
\vspace{-0.25cm}
\caption[bof]{
  Colocalization of BDNF proteins and vGlut in the  dSTORM images depicted in Figure~\ref{Fig2}. The colocalization regions identified by GcoPS are represented as white circles. Three zoomed-in regions are displayed at the bottom. The red rectangle represents an example of window used to find the hit shown as a red circle. The scale bars correspond to 1$\mu$m. This figure appears in color in the electronic version of this article, and any mention of color refers to that version.
}
\label{Fig3}
\end{figure}

We then applied GcoPS to windows of size $50\times50$ pixels  randomly located in the subregion of the image containing the segmented objects to identify the regions of colocalization  (see the examples at the bottom of Figure~\ref{Fig3}).  The colocalization regions identified by GcoPS are represented as white circles (the centers of the tested windows) in Figure~\ref{Fig3}. The  window size is here chosen sufficiently small to analyze local interactions in detail, while being sufficiently large with respect to the size of the objects to safely apply our testing procedure. Note that the distribution of proteins in these local windows can be more  reasonably considered to be stationary, unlike the distribution in the whole image.  This representation of colocalization hits, based on tests carried out on randomly located windows, is a fastest alternative to the dense colocalization map shown in Figure~\ref{Fig4}, that requires to process the testing procedure on all local windows. The two possibilities to analyze local interactions are included in the GcoPS module of Icy.

\section{Discussion}\label{sec:discussion}

We developed GcoPS, an original, fast and robust approach to test and quantify interactions between molecules.  Given a pair of binary images, obtained by segmentation of the input fluorescence images, GcoPS tests if they are independent or not, at the whole image scale and in image subregions. This testing procedure exploits the fact that these binary images can be viewed as realizations of random sets.

GcoPS can be viewed as a compromise between an object-based method and an intensity-based method. 
As demonstrated by the simulation study summarized in Section~\ref{sec:simu} and detailed in the appendix, it benefits from the advantages of both approaches while avoiding their weaknesses. Indeed, GcoPS is robust to noise and to nuisance background, inheriting the merits of an object-based method. 
But since the objects are not simply reduced to points, GcoPS  is adapted to any kind of object shapes and sizes. For the same reason, it is more powerful to test colocalization in small subregions of the cell than a point pattern approach, that would only be based on a few points. This property opens the possibility to efficiently localize the colocalization between markers within the cell. GcoPS is also able to evaluate the colocalization between large objects and small dots, as in a composite TIRF-PALM experiment. Finally, since the analysis of random sets boils down to the analysis of binary images, GcoPS is very fast, with no dependence on the number of objects per image, unlike spatial point process statistics.

A natural concern may be the sensitivity of the method to the preliminary segmentation step. In fact, GcoPS is not sensitive to the presence of spurious isolated spots, because these ones are typically very small objects that are negligible in volume with respect to the whole detected random set. This contrasts with point-based methods for which each spurious point counts as much as any other detected object in the cell and  falsely influences the statistical analysis. As a consequence, any segmentation algorithm can be used if it provides a biologically reasonable segmentation of the tagged molecules. Our simulation study actually confirms that GcoPS is fairly robust to the choice of segmentation algorithm parameters.

For the practical implementation of GcoPS, it is important to determine the ratio between the size of the tested window 
with respect to the size and the number of objects in the window. In theory, this ratio should be as large as possible so that the GcoPS colocalization score $T_n$ is distributed as a standard Gaussian law, see Proposition~\ref{prop:asympt}. 
The control of the convergence to the  Gaussian distribution, or rather the absence of control of this convergence,  is a common issue in almost all testing procedures.  Nonetheless, 
 when the objects are very large with respect to the size of the windows, as in the rightmost plot in the middle row of Figure~\ref{fig:simus}, GcoPS still behaves satisfyingly, proving that it is very robust to the detection of colocalization in small sub-windows of a real image. On the other hand, GcoPS (as well as the other competitive methods) also assume stationarity of the distribution of the objects inside the tested window. This hypothesis is more easily acceptable in small windows than in large windows, as illustrated in the real dataset analyzed in Section~\ref{BDNF}. This remark confirms the importance to use a procedure that remains effective in small windows, as GcoPS. For a safe decision, we finally recommend to use GcoPS in windows that are at least five times larger than the average size of the objects.  This guarantees that a minimal fluorescence information is available to assess colocalization, while allowing to consider small sub-windows.

%\backmatter

\section*{Acknowledgements}
We thank Jean Salamero, Markus Sauer, Soeren Doose, Sarah Aufmolk, Perrine Paul-Gilloteaux and Fabrice
Cordeli\`eres for assistance with experiments and for helpful insight in the preparation of the manuscript.\vspace*{-8pt}

%  If your paper refers to supplementary web material, then you MUST
%  include this section!!  See Instructions for Authors at the journal
%  website http://www.biometrics.tibs.org

%\section*{Author contributions statement}

%F.L., T.P. and C.K. conceived the study and designed research; F.L., T.P. and C.K. performed research; F.L. and T.P. developed computational software; L.Z. conducted the experiments and acquired experimental data; F.L., T.P., L.Z. and C.K. analyzed data;  F.L., T.P., L.Z. and C.K. wrote the paper. All authors reviewed the manuscript.

% \bibliographystyle{biom} 
%\bibliography{references_gcops}

%\section*{Supporting Information}
%The full simulation study with supplementary figures, the proof of Proposition~\ref{prop:asympt}, and the details of data preparation yielding the images displayed in Figures~\ref{Fig1} and \ref{Fig2} are available with this paper at the Biometrics website on Wiley Online Library. GcoPS is available as a free downloadable module of  the Icy software (see {\small \url{http://icy.bioimageanalysis.org/plugin/GcoPS}}), but as an alternative the web supplement also makes available the C++ code to implement GcoPS.
%\vspace*{-8pt}

\renewcommand{\thefigure}{S\arabic{figure}}

\appendix

\section{Proof of Proposition~1}\label{sec:proof}

Since  $\Gamma_1$ and $\Gamma_2$ are independent $\mathbb E(\hat D_n)=0$, and by stationarity of both random sets 
\begin{align*}
\mbox{Var}(\hat D_n)  &= n^{-2}  \mathbb{E}\left(  \sum_{x\in\Omega_n}  \mathbf 1_{\Gamma_1}(x) \mathbf 1_{\Gamma_2}(x)   - |\Omega_n|^{-1}\sum_{x\in\Omega_n}  \mathbf 1_{\Gamma_1}(x)\sum_{x\in\Omega_n}  \mathbf 1_{\Gamma_2}(x)\right)^2 
  = V_1+V_2+V_3,
\end{align*}
where
\begin{align*}
V_1&=  n^{-2} \sum_{x\in\Omega_n}\sum_{y\in\Omega_n} C_1(x-y) C_2(x-y),\\
V_2&=- 2n^{-3} \sum_{x\in\Omega_n} \left(\sum_{y\in\Omega_n} C_1(x-y) \right)\left(\sum_{y\in\Omega_n} C_2(x-y) \right),\\
V_3& =n^{-4}  \sum_{x\in\Omega_n}\sum_{y\in\Omega_n} C_1(x-y)  \sum_{x\in\Omega_n}\sum_{y\in\Omega_n} C_2(x-y). 
\end{align*}
Since $\Gamma_1$ and $\Gamma_2$ are $R$-dependent, $C_1$ and $C_2$ are summable, implying $V_1=O( n^{-1})$, $V_2=O( n^{-2})$ and $V_3=O(n^{-2})$, as  $n\to\infty$. 
Therefore  $$\mbox{Var}(\hat D_n) \sim V_1\sim  n^{-1} \sum_{h\in \mathbb Z^d,\, \|h\|\leq R} C_1(h) C_2(h), \text{ as }  n\to\infty.$$ 

Concerning the asymptotic normality of $\hat D_n$, let us first consider $\hat p_{12}$. We have 
\[n (\hat p_{12} - p_{12}) = \sum_{x\in\Omega_n} X_x,\]
where $X_x=\mathbf 1_{\Gamma_1\cap\Gamma_2}(x) - p_{12}$. $(X_x)_{x\in\mathbb Z^d}$ is a real valued stationary centered random field on $\mathbb Z^d$, which inherits the  $R$-dependency from $\Gamma_1$ and $\Gamma_2$ and admits moments of any order. A central limit theorem then applies, see for instance \cite{bolthausen:82} where all mixing conditions are verified by the $R$-dependent assumption. We get 
\[n^{-1/2} \sum_{x\in\Omega_n} X_x \overset{d}{\longrightarrow} \mathcal N(0,\sigma^2)\]
as $n\to\infty$, where $$\sigma^2=\sum_{h\in \mathbb Z^d} \mathbb E(X_oX_h)=\sum_{h\in \mathbb Z^d,\, \|h\|\leq R} C_{12}(h),$$ and $C_{12}$ denotes the autocovariance function of the random set $\Gamma_1\cap\Gamma_2$. This gives us the asymptotic normality of $n^{1/2}(\hat p_{12} - p_{12})$. Thanks to the Cramer-Wold device, we can prove similarly the joint  asymptotic normality of $n^{1/2}(\hat p_{12} - p_{12})$, $n^{1/2}(\hat p_{1} - p_{1})$ and $n^{1/2}(\hat p_{2} - p_{2})$. The $\Delta$-method then provides the asymptotic normality of $n^{1/2}\hat D_n$. The asymptotic variance of this term is deduced from our preliminary computations and we obtain that
 \[\sqrt{n} \frac{\hat D_n}{\sqrt{\sum_{h\in \mathbb Z^d,\, \|h\|\leq R} \limits C_1(h) C_2(h)}} \overset{d}{\longrightarrow} \mathcal N(0,1),\]
as $n\to\infty$.  Replacing $C_1(h)$ and $C_2(h)$ by consistent estimators does not change the result, by application of Slutsky's lemma, since the sum above is finite.

\section{Evaluation on synthetic  data sets: details}\label{sec:simuSI}

\begin{figure}
\begin{center}
\includegraphics[width=\linewidth]{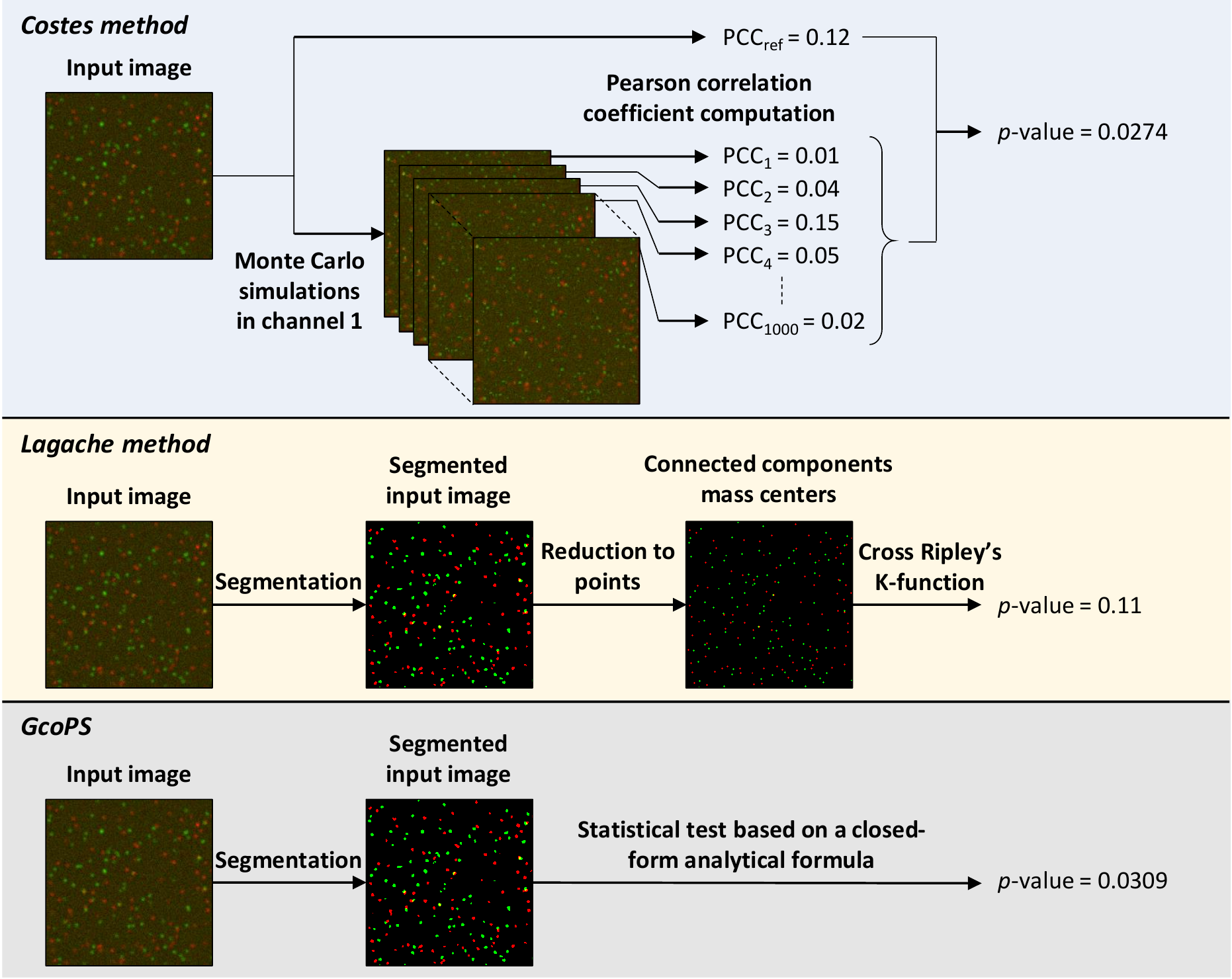}
\end{center}
\caption{Schematic workflow for the intensity-based method of  \cite{Costes2004}, the object-based method of \cite{Lagache2015} and GcoPS, when implementing a colocalization testing procedure from the same raw image.}
\label{workflow}
\end{figure}

This section is a detailed version of Section~3 of the main manuscript. 
The objective is to evaluate the performances of GcoPS, in comparison to the competitive Costes method \citep{Costes2004} and the object-based method of \cite{Lagache2015}, in various adverse situations. 
As a first comparison, Figure~\ref{workflow} depicts the workflow for each testing method, starting from the same raw image. 
For implementation, we use JACoP plugin \citep{Bolte2006} on ImageJ \citep{Schneider2012} to apply the Costes method: i) in the randomization step, $n=1000$ replications are considered; ii) we choose blocks of pixels with a size corresponding to the PSF  (point spread function) for simulated images, as advised in  \cite{Costes2004}, and corresponding to the average size of the objects for segmented images. For the Lagache method, we use the colocalization studio in Icy  \citep{Chaumont2012}.
Moreover, we recall that we have implemented GcoPS within the Icy software, where it is available as a downloadable module (more details at {\small \url{http://icy.bioimageanalysis.org/plugin/GcoPS}}).

\subsection{Time computation}\label{sec:time}
Table~\ref{tab:comp} gives the CPU time for the three methods applied to 2D, 2D+t and 3D images, depending on the number of objects detected after segmentation. As expected, the Costes procedure, which is not an object-based method, is not sensitive to the number of objects. However, it is by far the slowest method because of the Monte Carlo step. On the contrary, the object-based method of \cite{Lagache2015} strongly depends on the number of objects and can therefore be quite slow if this number is large, which is typically the case in 3D.  In contrast, GcoPS is very fast whatever the situation is and is not sensitive to the number of objects. The CPU time of GcoPS is of course minimal when implemented on C++ and is slightly less optimal with the Icy plugin (based on Java), nevertheless keeping much faster than the two alternative methods.

\begin{table}[h]
\centering
{\scriptsize
\begin{tabular}{|c||c|c|c|c|c|c|}
\cline{2-7}
\multicolumn{1}{c||}{} & { \!\!\!\bf 2D image\!\!\!} & { \!\!\!\bf 2D image\!\!\!} & { \!\!\!\bf 2D image\!\!\!} & { \!\!\!\bf 2D+time image \!\!} & { \!\!\!\bf 3D image\!\!\!} & { \!\!\!\bf 3D image\!\!\!} \\
\multicolumn{1}{c||}{} & $256 \times 256$ & $256 \times 256$ & $256 \times 256$  & $256 \times 256 \times 1000$ & $256 \times 256 \times 60$ & $256 \times 256 \times 60$ \\
\multicolumn{1}{c||}{} & { \!\!\! 50 objects \!\!} & { \!\!\! 200 objects \!\!} & {\!\!\!  3500 objects \!\!} & { \!\!\! 100 objects \!\!} & { \!\!\! 1000 objects \!\!} & { \!\!\! 2000 objects \!\!} \\\hline
\hline
{ \bf \cite{Costes2004}} & { 6.1 sec} & { 6.2 sec} & { 6.1 sec} & { 38 min 20 sec} & { 3 min 3 sec} & { 3 min 10 sec} \\
{ ImageJ plugin} & & & & & &\\\hline
{ \bf \cite{Lagache2015}} & { 1 sec} & { 1.96 sec} & { 12.38 sec} & { 12 min 39 sec} & {25 sec} & {60 sec} \\
{ Icy plugin} & & & & & &\\\hline
{ \bf GcoPS} & {\textbf{0.18 sec}} & { \textbf{0.2 sec}} & {\textbf{0.19 sec}} & { \textbf{29.5 sec}} & {\textbf{10 sec}} & {\textbf{9.8 sec}} \\
{ C++ code} & & & & & & \\\hline
{ \bf GcoPS} & \textbf{0.77 s} & \textbf{0.86 sec} & \textbf{0.82 sec} & \textbf{2 min 50 sec } & \textbf{22 sec} & \textbf{21 sec} \\
{ Icy plugin} & & & & & &\\\hline
\end{tabular}
}
\caption[bof]{Comparison of computation time of the intensity-based method of  \cite{Costes2004}, the object-based method of \cite{Lagache2015} and GcoPS, when applied to 2D, 2D+time and 3D images. 
}\label{tab:comp}
\end{table}

\subsection{Evaluation on simulated 2D images}\label{sec:2D}

We evaluate the sensitivity of the methods on synthetic images generated by the simulator described in  \cite{Lagache2015}. This simulator consists in a first step to generate randomly distributed particles (say the red channel), and in a second step to simulate a proportion of green particles nearby red particles  while the rest of green particles are drawn randomly and independently.  Note that the following figures appear in color in the electronic version of this article, and any mention of color refers to that version.
Three scenarios are considered: i) without noise, ii) with noise and iii) with noise and a spatial shift of three pixels between colocalized particles (a 3 pixels shift is more than enough to account for a possible spatial shift due to the experimental device). In each of these scenarios, the two channels were simulated  with a proportion of forced neighbors  of $0\%$ (no colocalization),  $2.5\%$ and $5\%$. In each situation, 1000 pairs of images were generated to evaluate the sensitivity of the three tested  methods.

\begin{figure}
\begin{center}
\begin{tabular}{cc}
\begin{tabular}{c} 
\includegraphics[width=.2\linewidth]{icySimulation_noNoiseNoShift_coloc0} \\
{ 0\% colocalization}\\$\, $\\
\includegraphics[width=.2\linewidth]{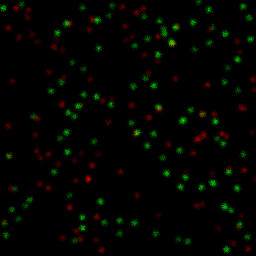}\\ 
{ 2.5\% colocalization}\\$\, $\\
\includegraphics[width=.2\linewidth]{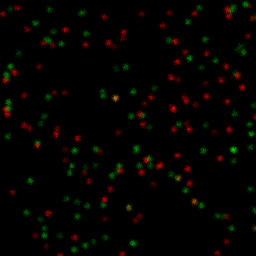}\\
{ 5\% colocalization}
\end{tabular} &
\begin{tabular}{c} \includegraphics[width=0.5\linewidth]{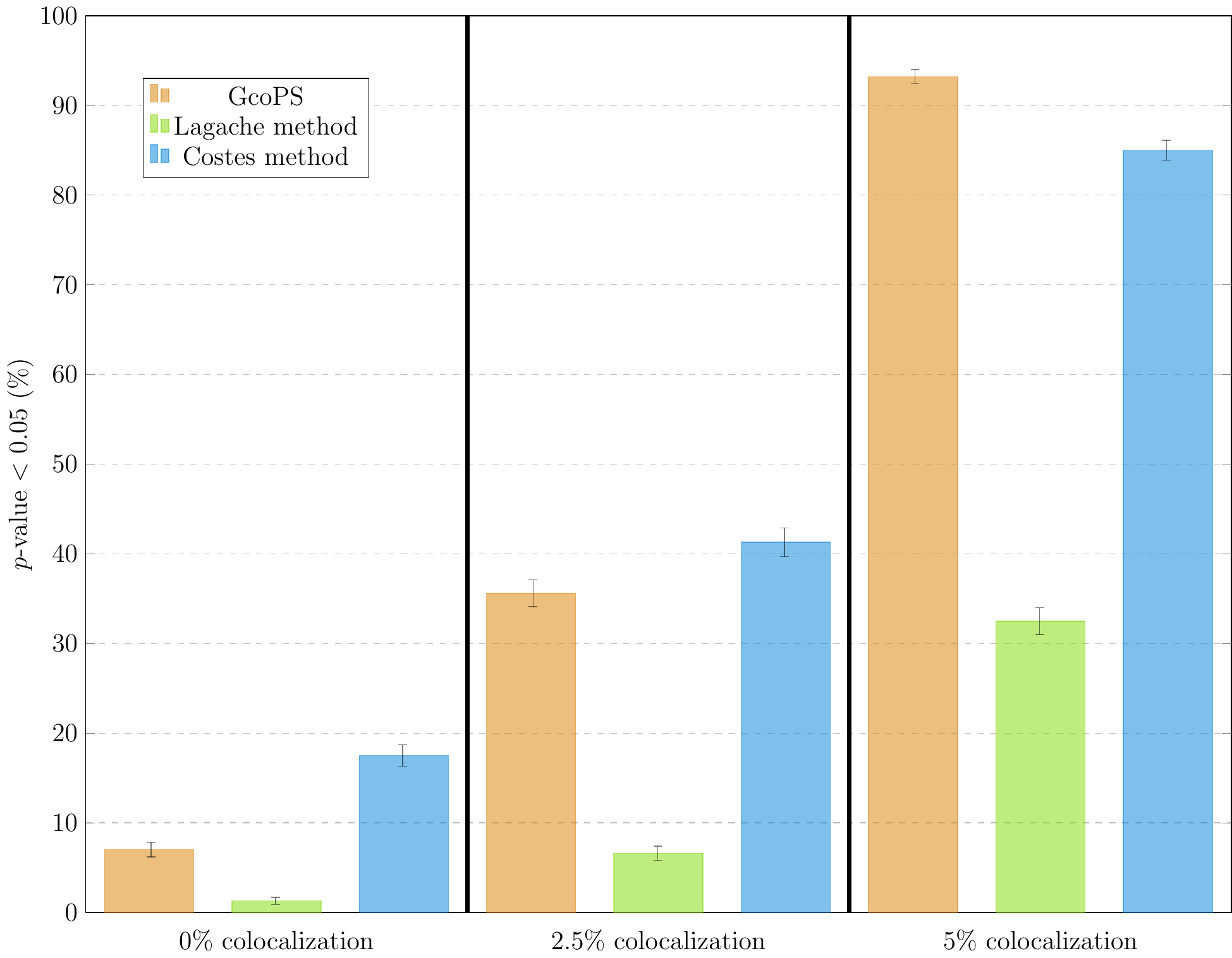} \\
\includegraphics[width=0.5\linewidth]{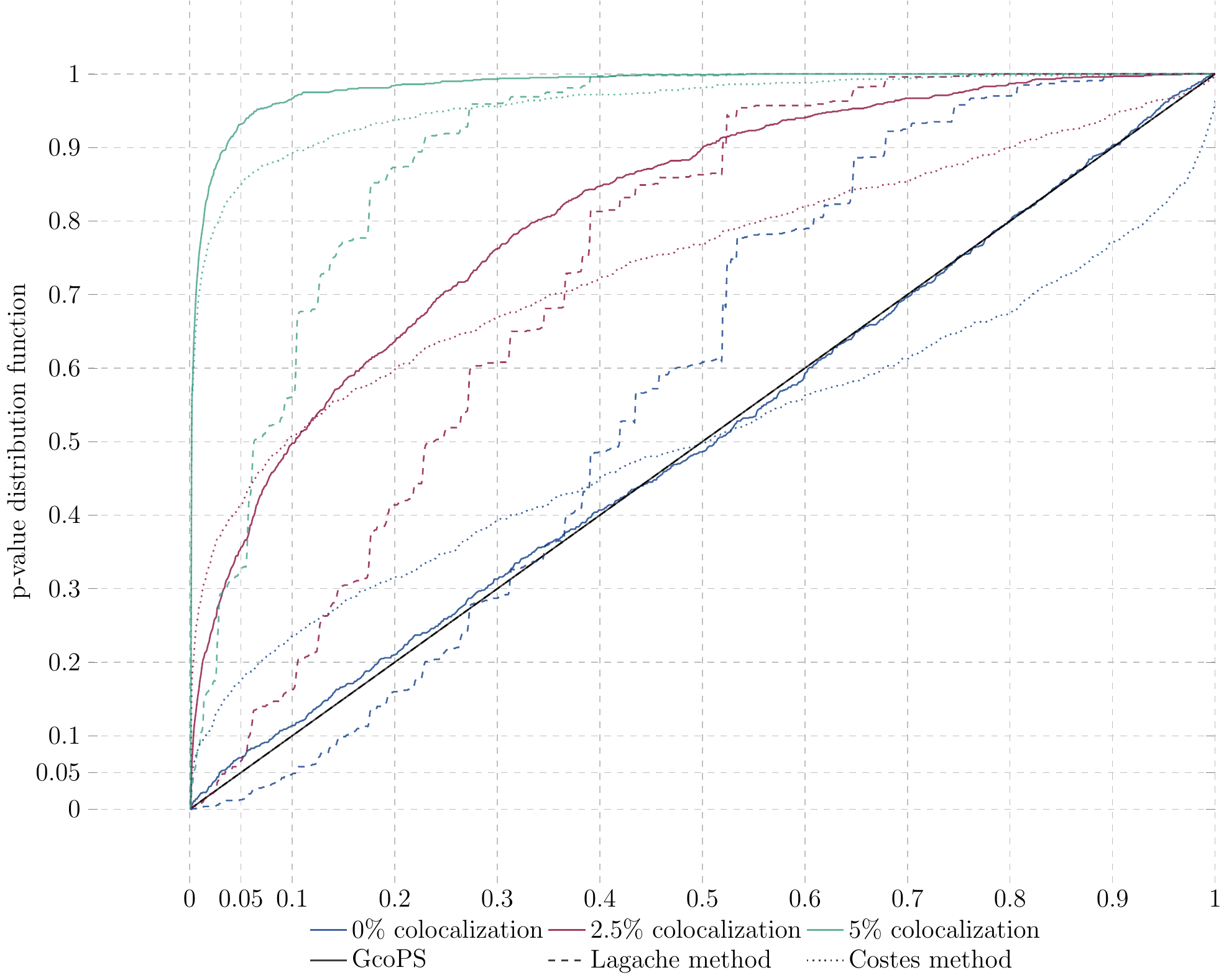} \end{tabular}
\end{tabular}
\end{center}
\caption[bof]{Evaluation of GcoPS on 2D simulated images.  Top right: proportion of $p$-values lower than $0.05$ obtained with GcoPS,  the object-based method of \cite{Lagache2015} and the intensity-based method of  \cite{Costes2004} over 1000 simulated images without noise and without shift for 0\%, 2.5\% and 5\% forced neighbors. The confidence intervals at the top of each bar represent  one standard deviation over all 1000 simulations. Bottom right:  empirical distribution functions  of $p$-values. Left: an example of simulated images. }
\label{fig:2Dnonoisenoshift}
\end{figure}

\begin{figure}[H]
\begin{center}
\begin{tabular}{cc}
\begin{tabular}{c} 
\includegraphics[width=.2\linewidth]{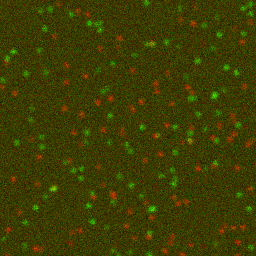} \\
{ 0\% colocalization}\\$\, $\\
\includegraphics[width=.2\linewidth]{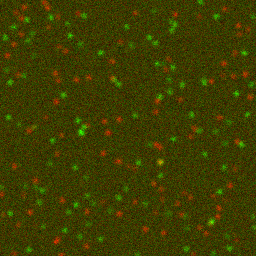}\\ 
{ 2.5\% colocalization}\\$\, $\\
\includegraphics[width=.2\linewidth]{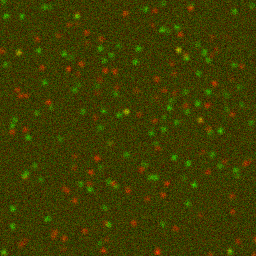}\\
{ 5\% colocalization}
\end{tabular} &
\begin{tabular}{c} \includegraphics[width=0.5\linewidth]{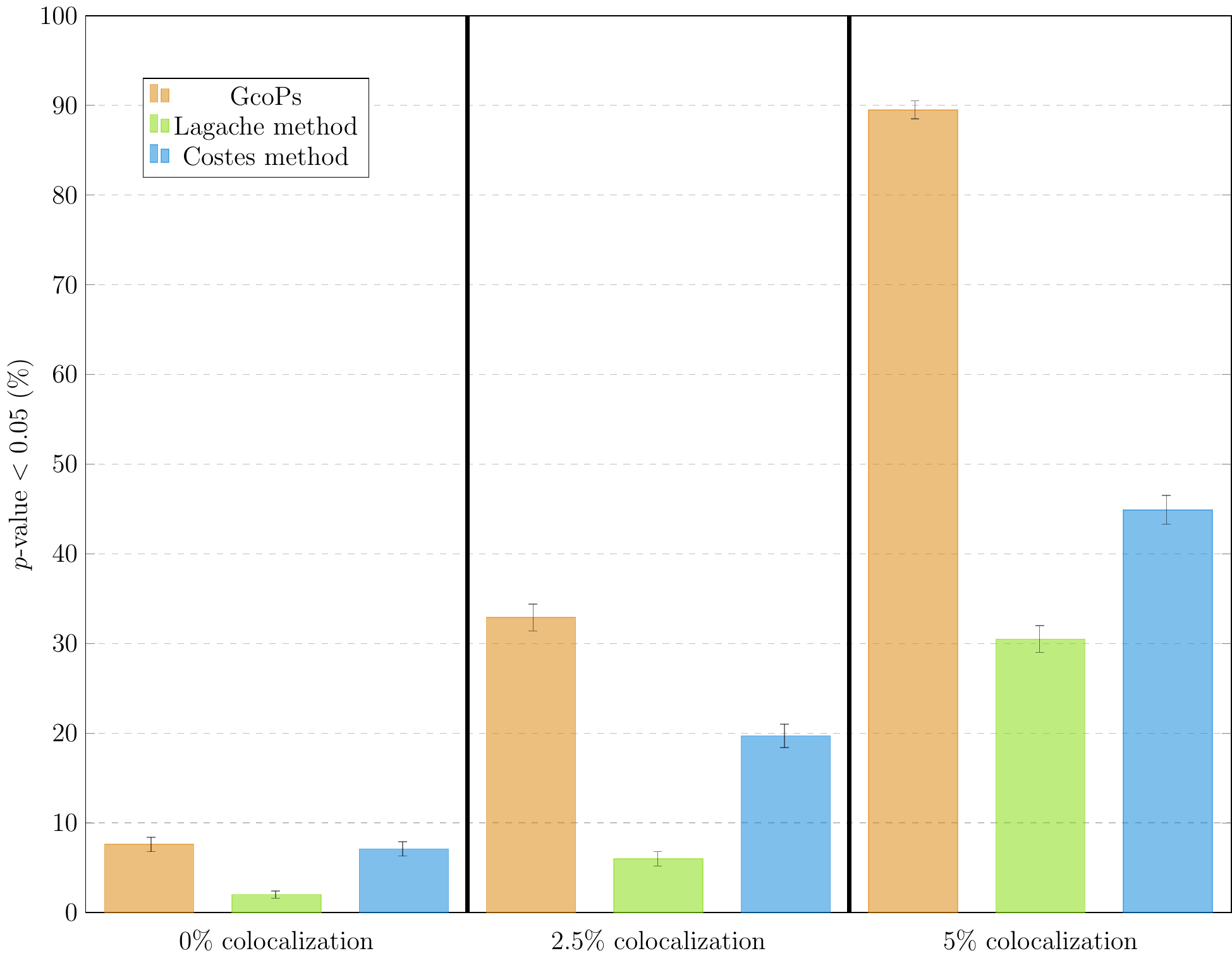} \\
\includegraphics[width=0.5\linewidth]{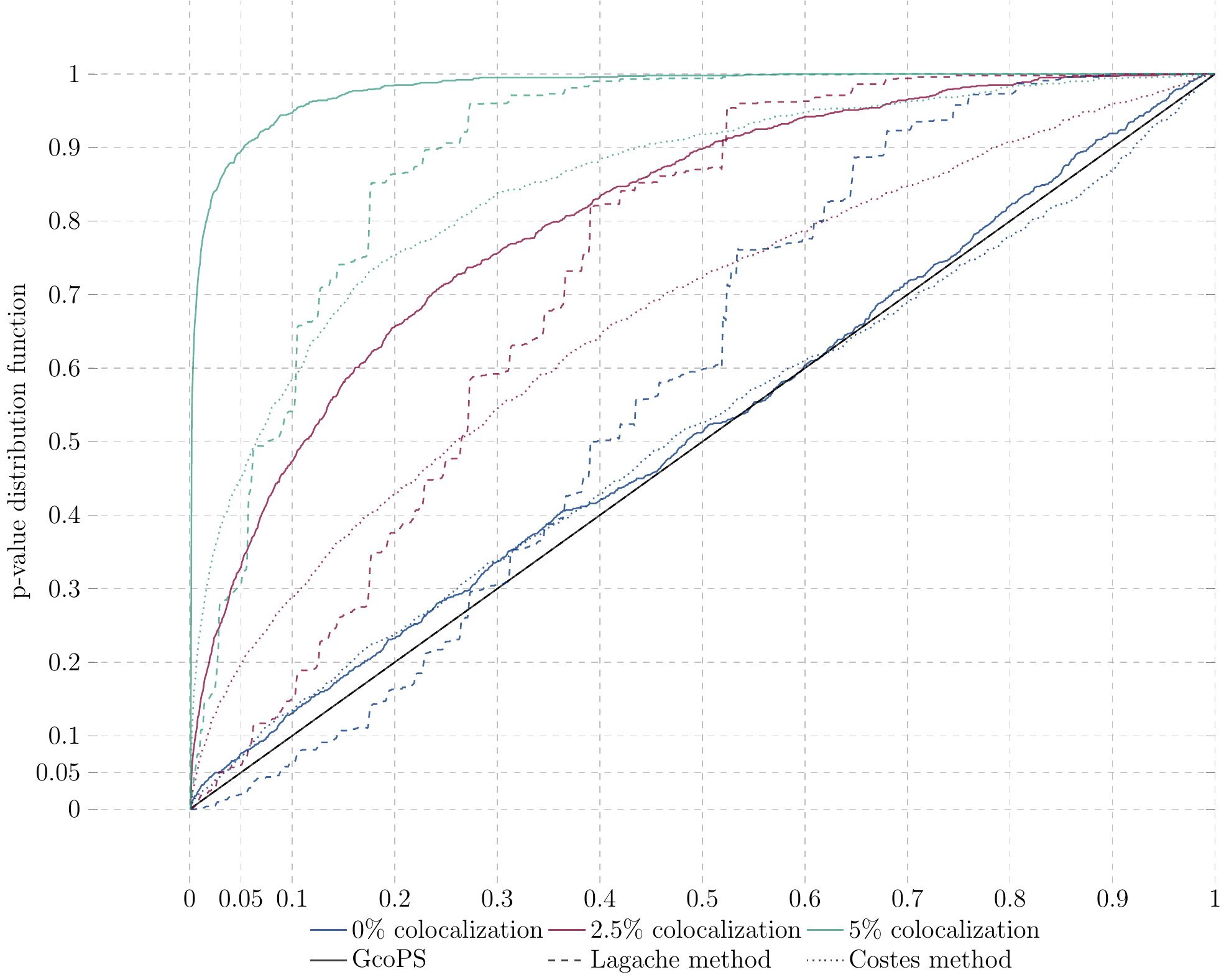} \end{tabular}
\end{tabular}
\end{center}
\caption{Evaluation of GcoPS on 2D simulated images with noise.  Same plots as in Figure~\ref{fig:2Dnonoisenoshift}  except that the simulated images are corrupted with noise.}
\label{fig:2Dnoisenoshift}
\end{figure}

\begin{figure}
\begin{center}
\begin{tabular}{cc}
\begin{tabular}{c} 
\includegraphics[width=.2\linewidth]{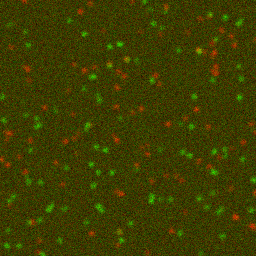} \\
{ 0\% colocalization}\\$\, $\\
\includegraphics[width=.2\linewidth]{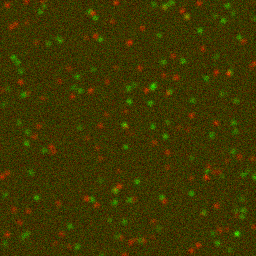}\\ 
{2.5\% colocalization}\\$\, $\\
\includegraphics[width=.2\linewidth]{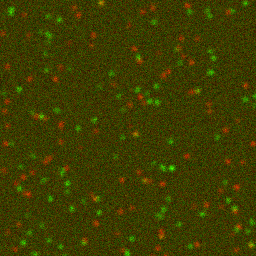}\\
{ 5\% colocalization}
\end{tabular} &
\begin{tabular}{c} \includegraphics[width=0.5\linewidth]{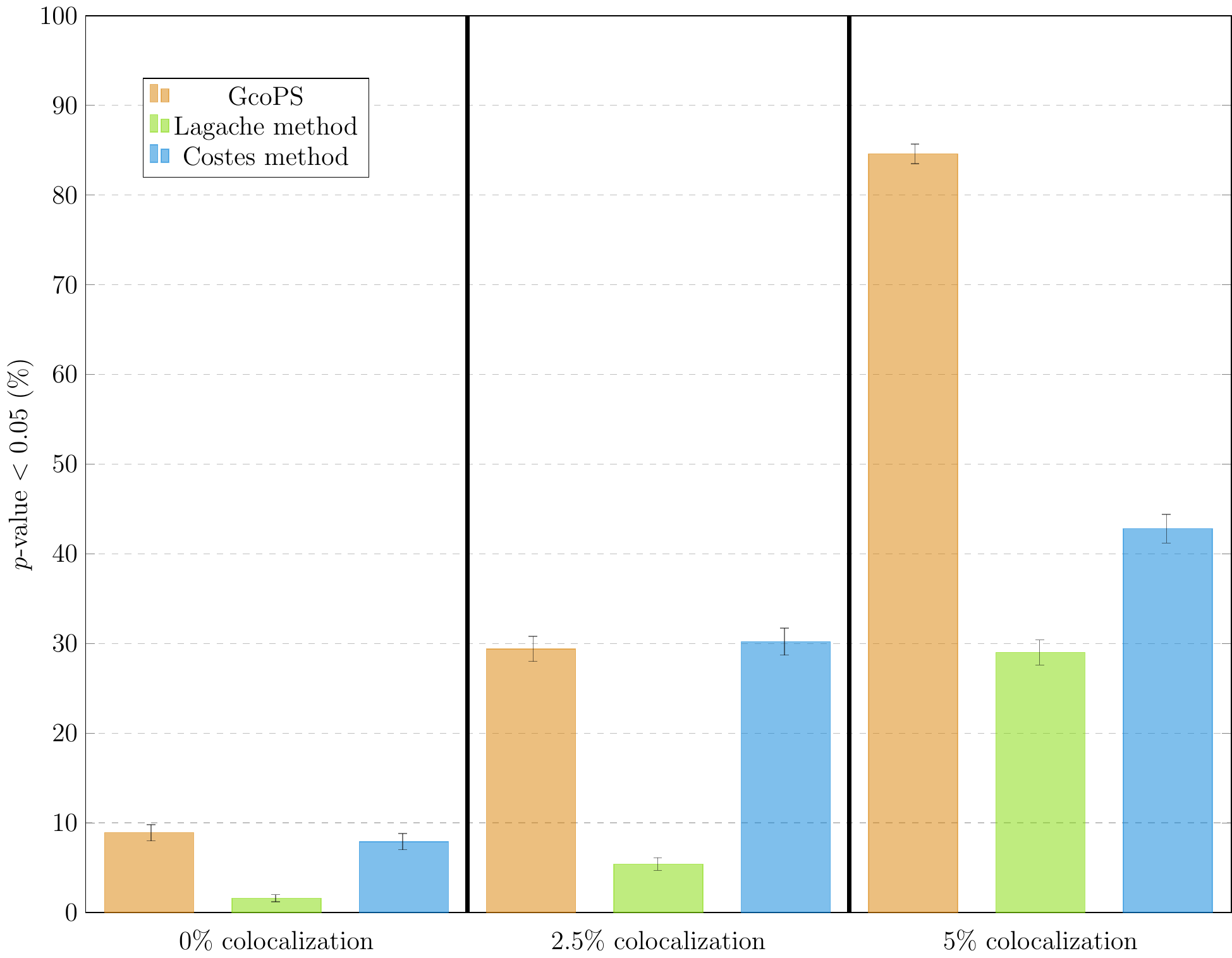} \\
\includegraphics[width=0.5\linewidth]{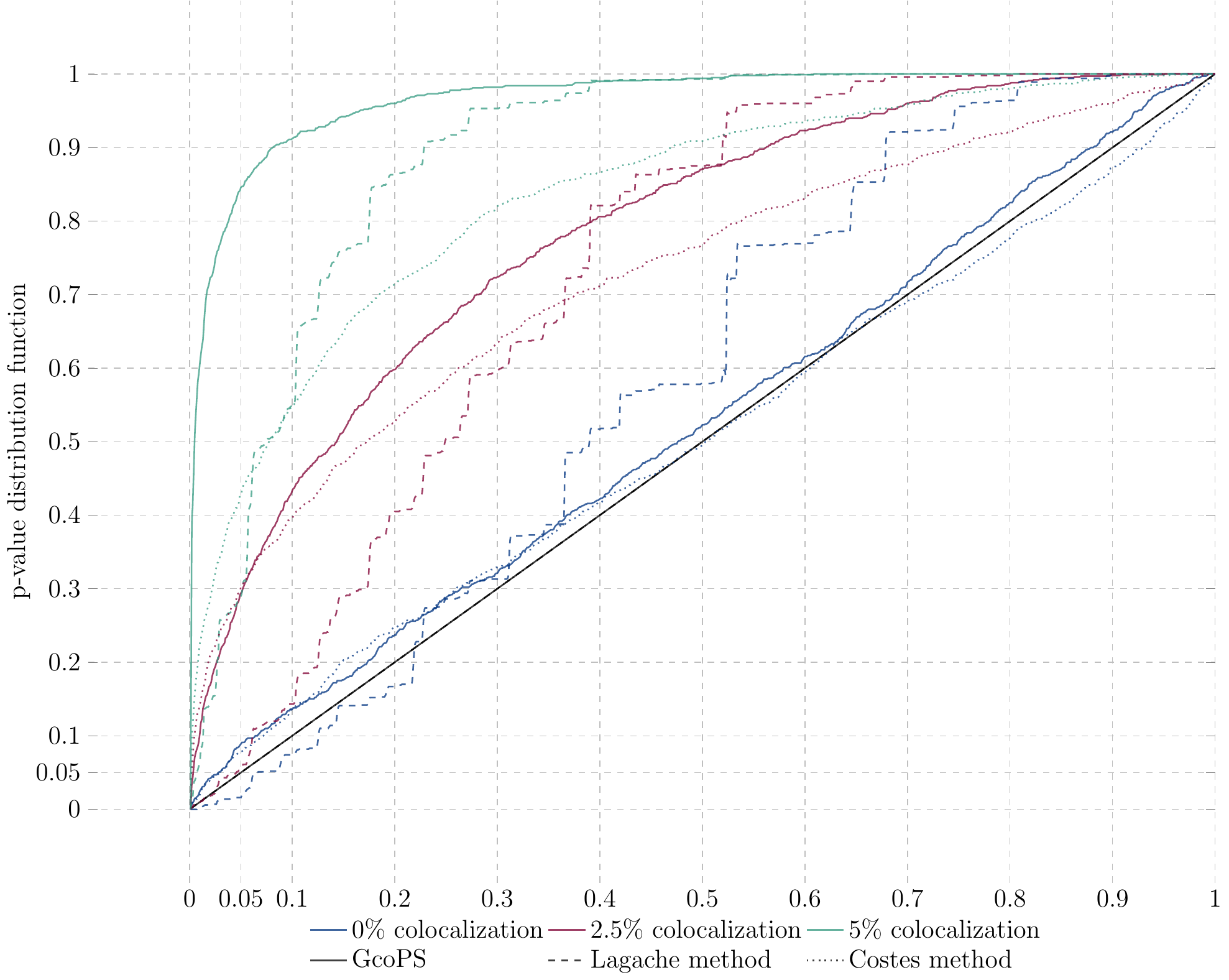} \end{tabular}
\end{tabular}
\end{center}
\caption{Evaluation of GcoPS on 2D simulated images with noise. Same plots as in Figure~\ref{fig:2Dnonoisenoshift} except that the simulated images are corrupted with noise and a shift of three pixels is applied between the two channels.}
\label{fig:2Dnoiseshift}
\end{figure}

\begin{figure}
\begin{center}
\begin{tabular}{cc}
\begin{tabular}{c} 
\includegraphics[width=.2\linewidth]{50vs200_0} \\
{ 0\% colocalization}\\$\, $\\
\includegraphics[width=.2\linewidth]{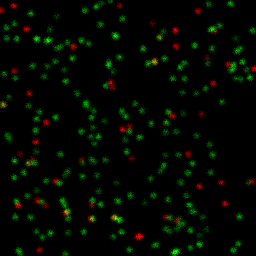}\\ 
{2.5\% colocalization}\\$\, $\\
\includegraphics[width=.2\linewidth]{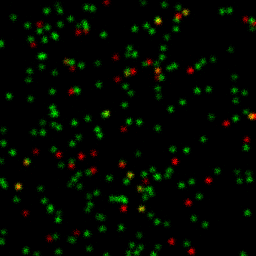}\\
{ 5\% colocalization}
\end{tabular} &
\begin{tabular}{c} \includegraphics[width=0.5\linewidth]{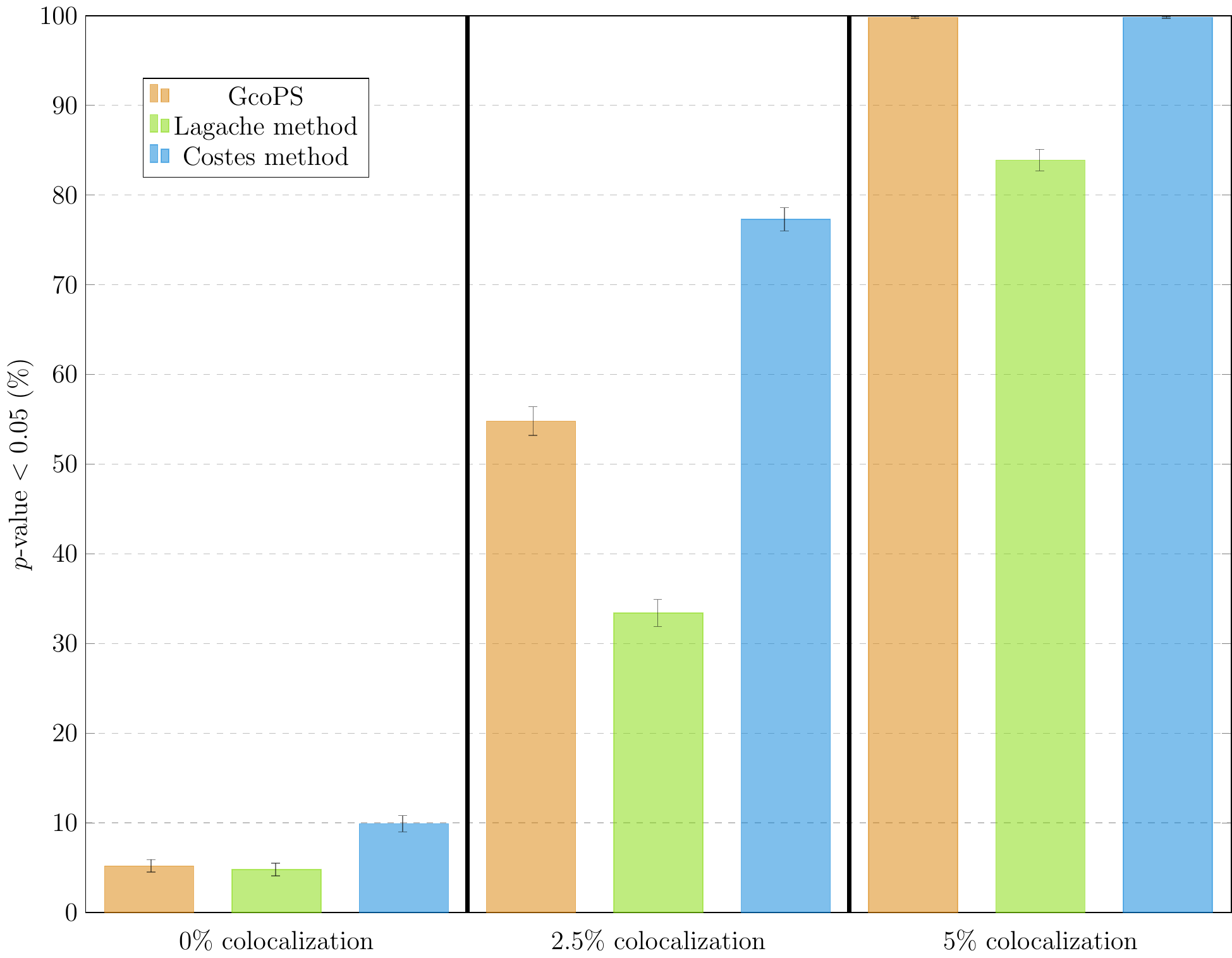} \\
\includegraphics[width=0.5\linewidth]{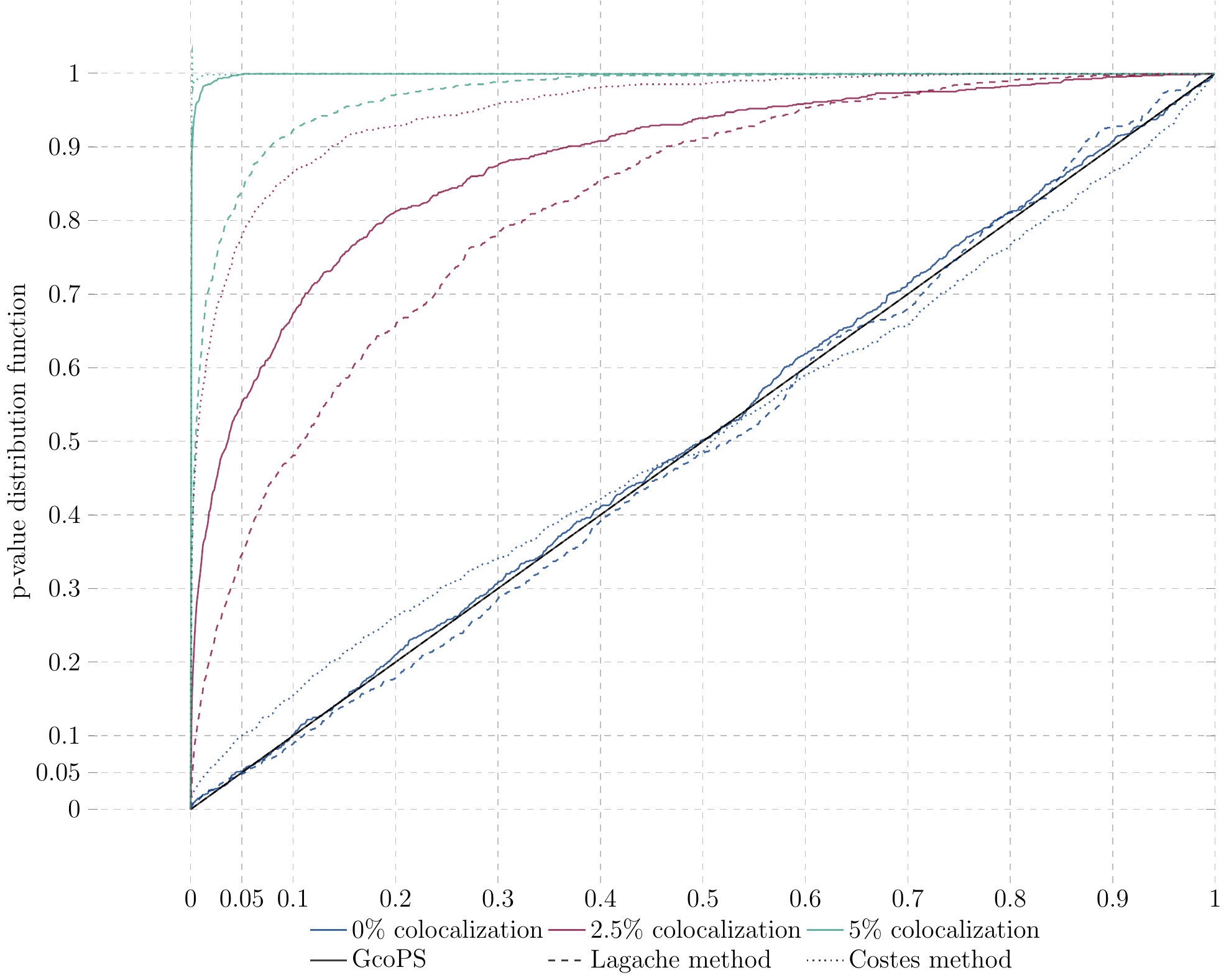} \end{tabular}
\end{tabular}
\end{center}
\caption{Evaluation of GcoPS on 2D simulated images with noise. Same plots as in Figure~\ref{fig:2Dnonoisenoshift} except that the green channel shows 4 times more objects than the red channel.}
\label{fig:2D50vs200}
\end{figure}

Figure~\ref{fig:2Dnonoisenoshift} compares the results of GcoPS, the object-based method of \cite{Lagache2015} and the intensity-based method of  \cite{Costes2004} for the first scenario (no noise, no shift), by reporting the proportion of $p$-values lower than 0.05, along with the empirical distribution function (edf) of the $p$-values. Recall that a perfect testing procedure would result in an edf equal to the first diagonal in absence of colocalization and in an edf which is uniformly equal to 1 in presence of colocalization.
Figure~\ref{fig:2Dnonoisenoshift} reveals that the object-based method is not sufficiently sensitive  (less than $35\%$ of images have a $p$-value inferior to 0.05 with $5\%$ forced neighbors against more than $90\%$ for GcoPS). The Costes method is in turn too sensitive in absence of colocalization (about $18\%$ of images without colocalization have a $p$-value lower than 0.05), which leads to too many false positive decisions. 
These observations are confirmed by the simulations carried out in the other two scenarios, see Figures~\ref{fig:2Dnoisenoshift} and \ref{fig:2Dnoiseshift}, where the power of GcoPS is clearly superior to the other methods (twice more images have a $p$-value lower than 0.05 with $5\%$ forced neighbors in presence of noise and/or shift).
Finally, Figure~\ref{fig:2D50vs200} displays an unbalanced situation where the number of objects in one channel is 4 times superior to the number of objects than in the other channel. The results confirm the previous conclusions.

\subsection{Sensitivity to image segmentation}\label{sec:seg}
\begin{figure}
\centering
\begin{tabular}{cc}
{\bf  Simulated image} & {\bf   Segmented image with ATLAS}\\
\includegraphics[width=.3\linewidth]{icySimulation_noiseShift_coloc0} & \includegraphics[width=.3\linewidth]{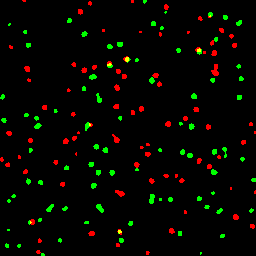} \\\\
{\bf  Thresholded image, $\tau=15$} & {\bf  Thresholded image, $\tau=20$} \\
\includegraphics[width=.3\linewidth]{icySimulation_noiseShift_coloc0_threshold15} &
\includegraphics[width=.3\linewidth]{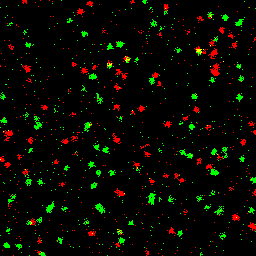} \\
{\bf  Thresholded image, $\tau=25$} & {\bf  Thresholded image, $\tau=30$} \\
\includegraphics[width=.3\linewidth]{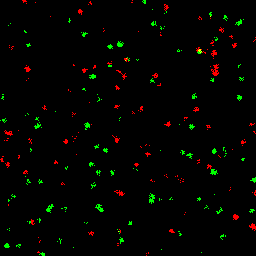} &
\includegraphics[width=.3\linewidth]{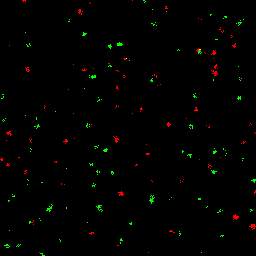} \\
\end{tabular}
\caption{Sensitivity to segmentation. Example of a simulated image with noise and shift and the corresponding segmentations with ATLAS  \citep{Basset2015} and a threshold $\tau$ ranging from $15$ to $30$.}
\label{fig:segexamples}
\end{figure}

\begin{figure}
\begin{tabular}{cc}
{\bf  GcoPS} & {\bf  Lagache method} \\
\includegraphics[width=.48\linewidth]{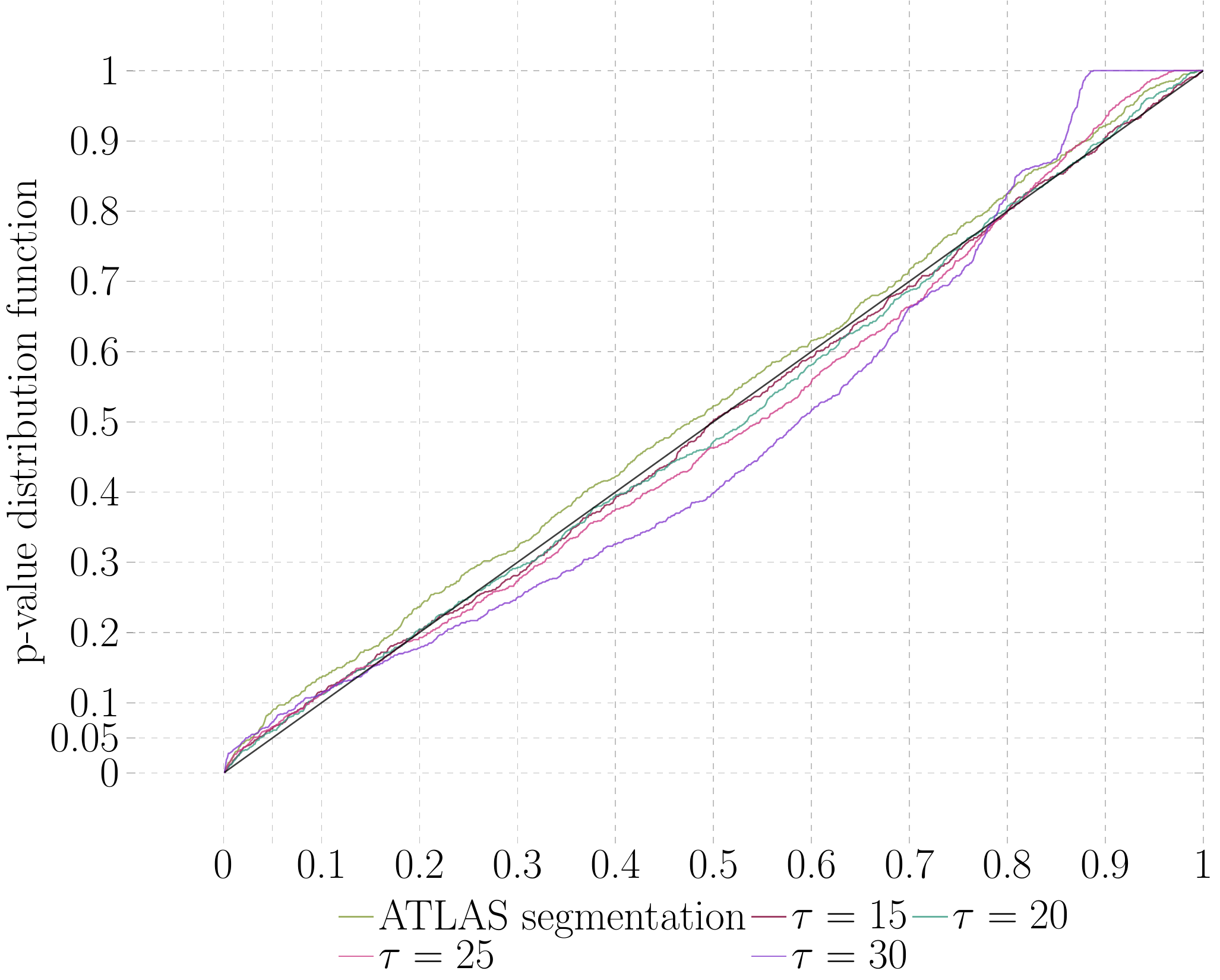} &
\includegraphics[width=.48\linewidth]{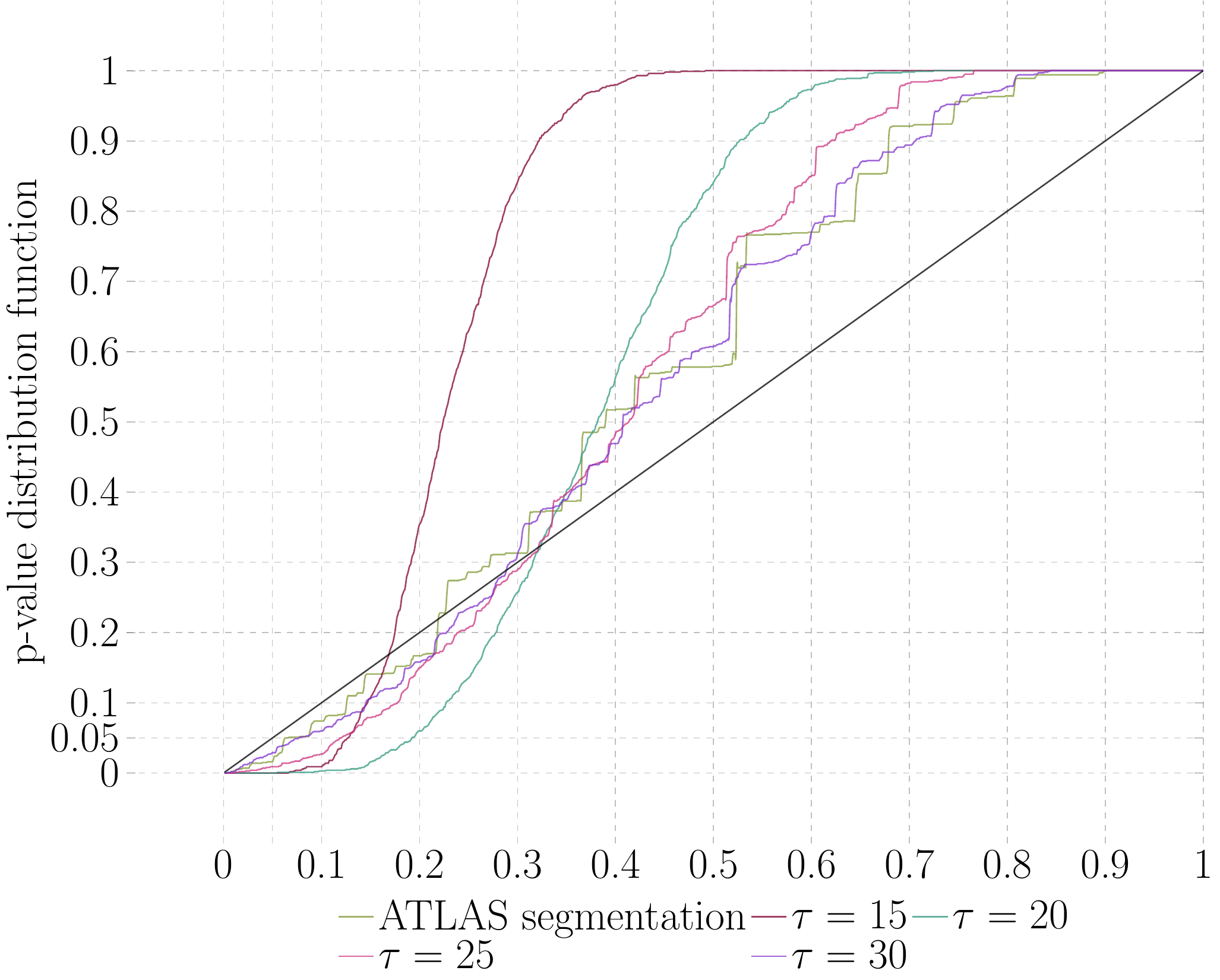}\\
$\,$ & \\
{\bf  Costes method} &  \\
\includegraphics[width=.48\linewidth]{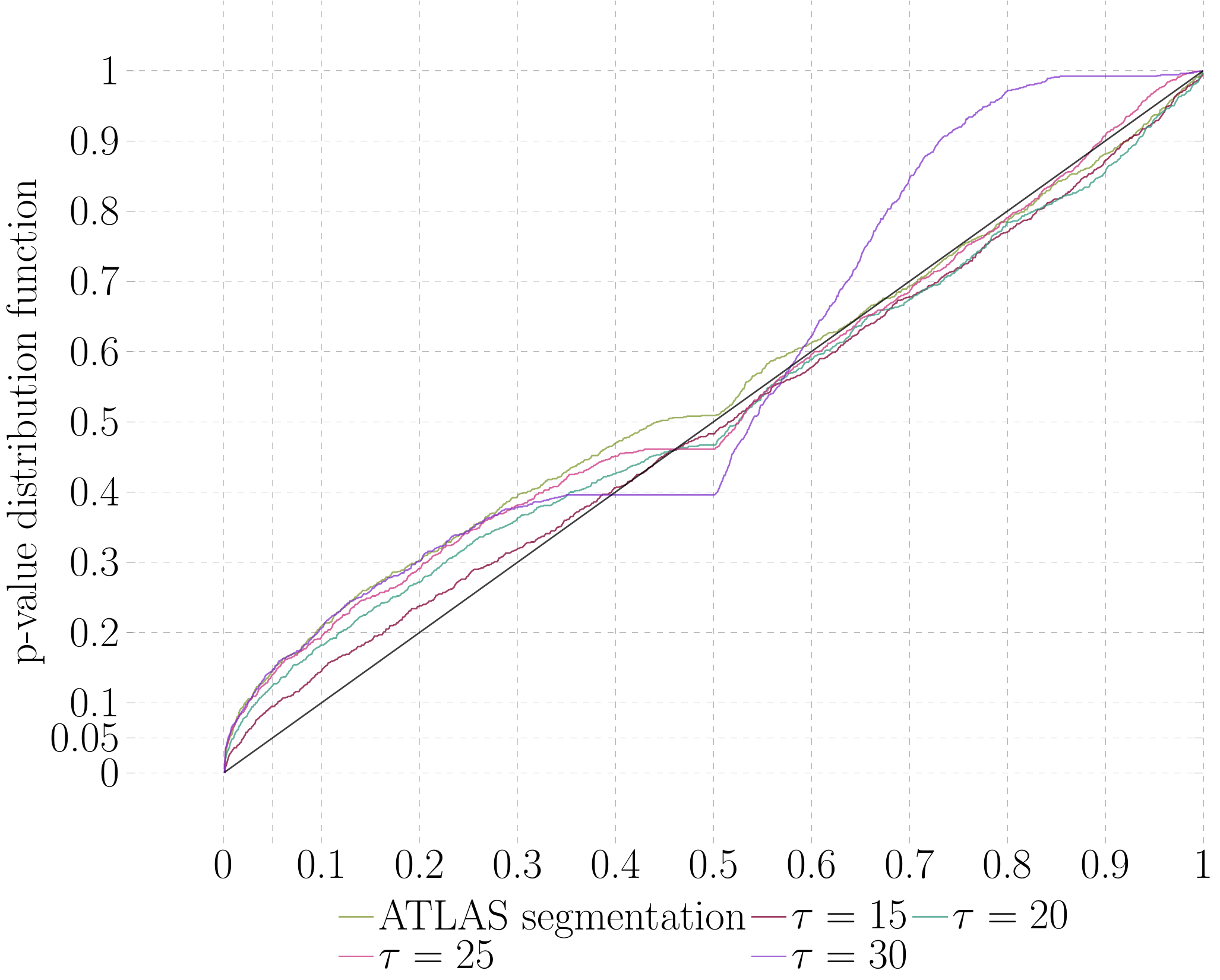} &
\includegraphics[width=.48\linewidth]{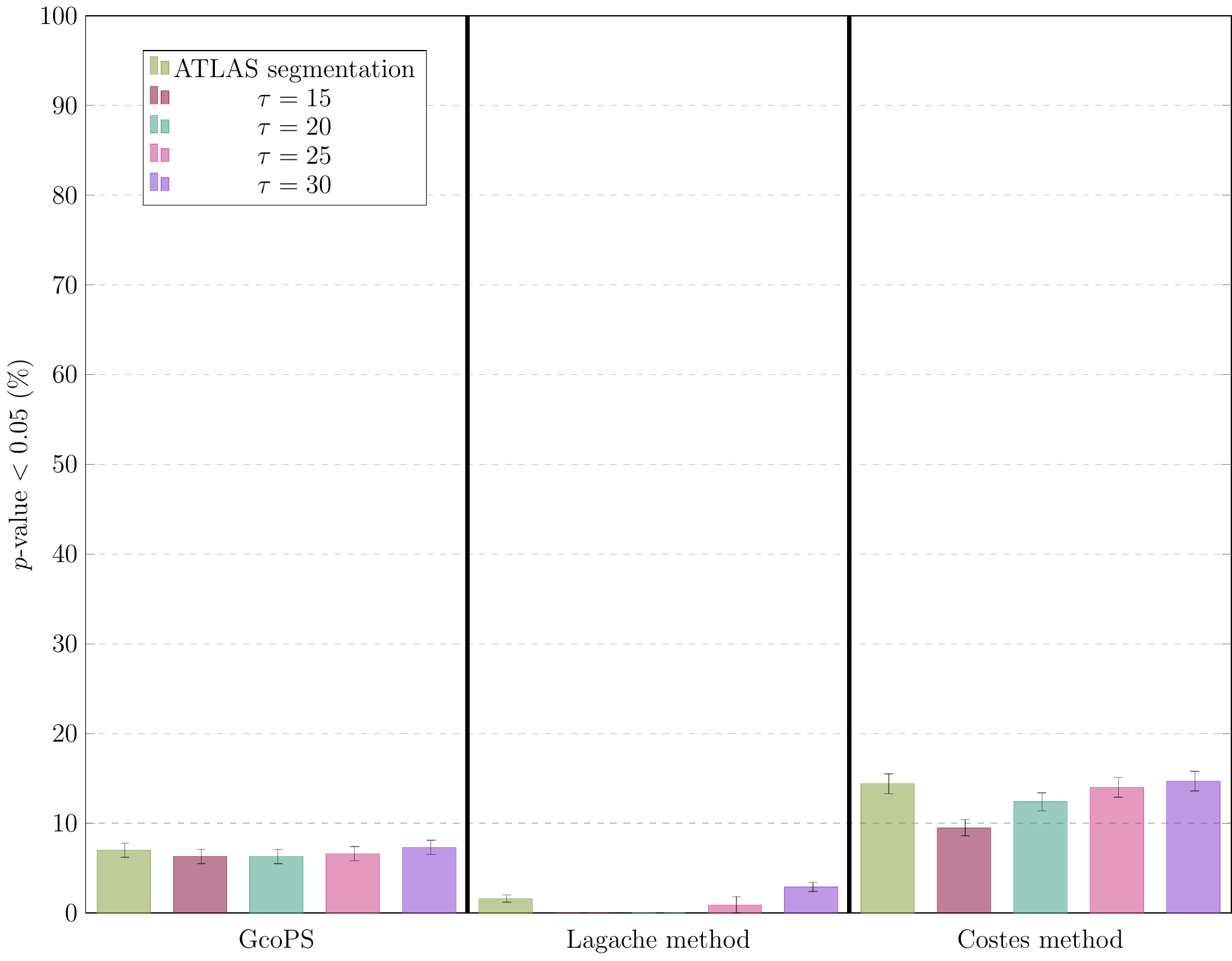} \\
\end{tabular}
\caption{Sensitivity to segmentation. Proportion of $p$-values lower than $0.05$ (bottom right) and  empirical distribution functions of $p$-values obtained with (from top left to bottom left) GcoPS, the object-based method of \cite{Lagache2015} and the intensity-based method of  \cite{Costes2004}   over 1000 simulated images with noise and shift, segmented with ATLAS  \citep{Basset2015} or a threshold $\tau$ ranging from $15$ to $30$, when there is no colocalization.}
\label{fig:seg0}
\end{figure}

\begin{figure}
\begin{tabular}{cc}
{\bf  GcoPS} & {\bf  Lagache method} \\
\includegraphics[width=.48\linewidth]{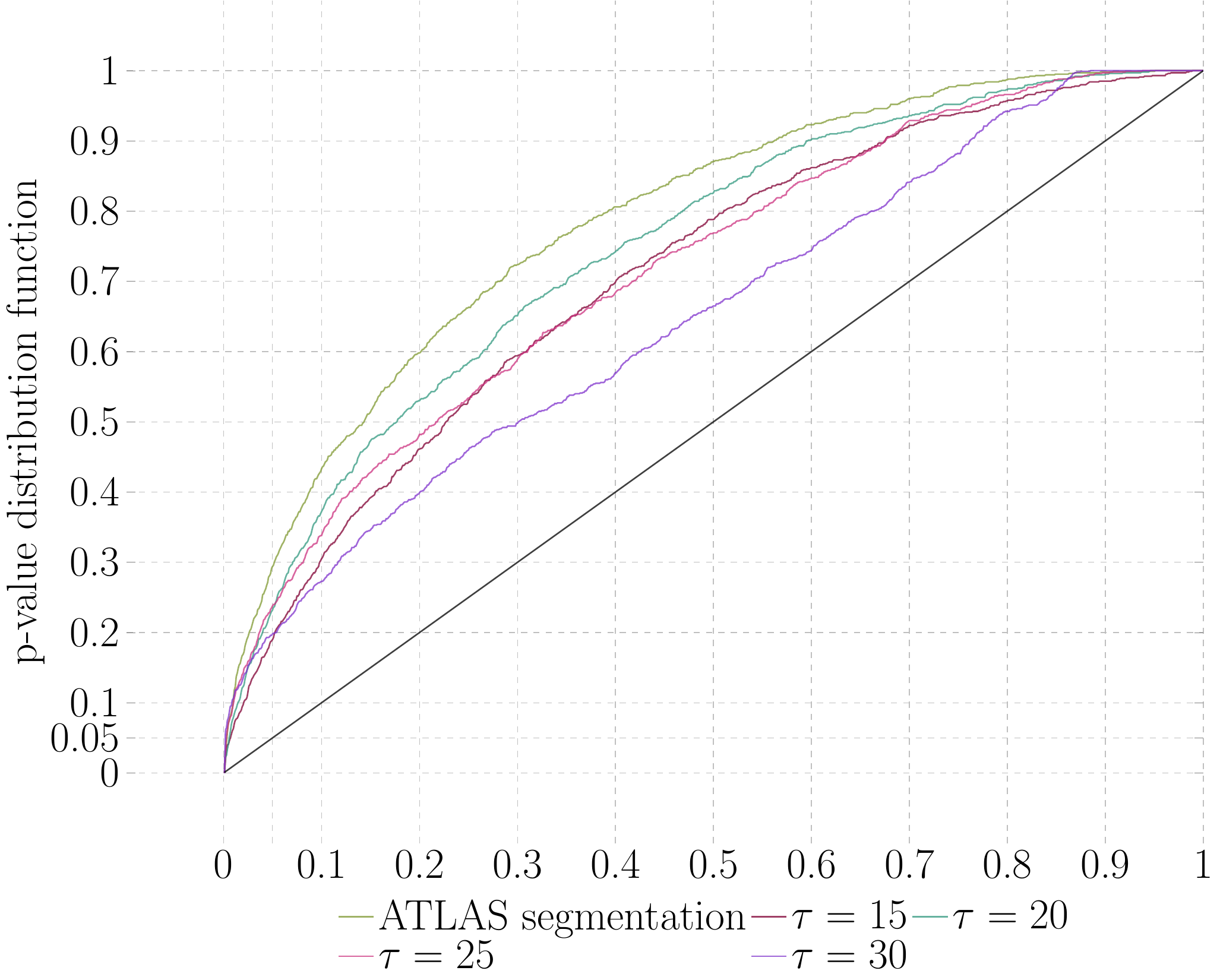} &
\includegraphics[width=.48\linewidth]{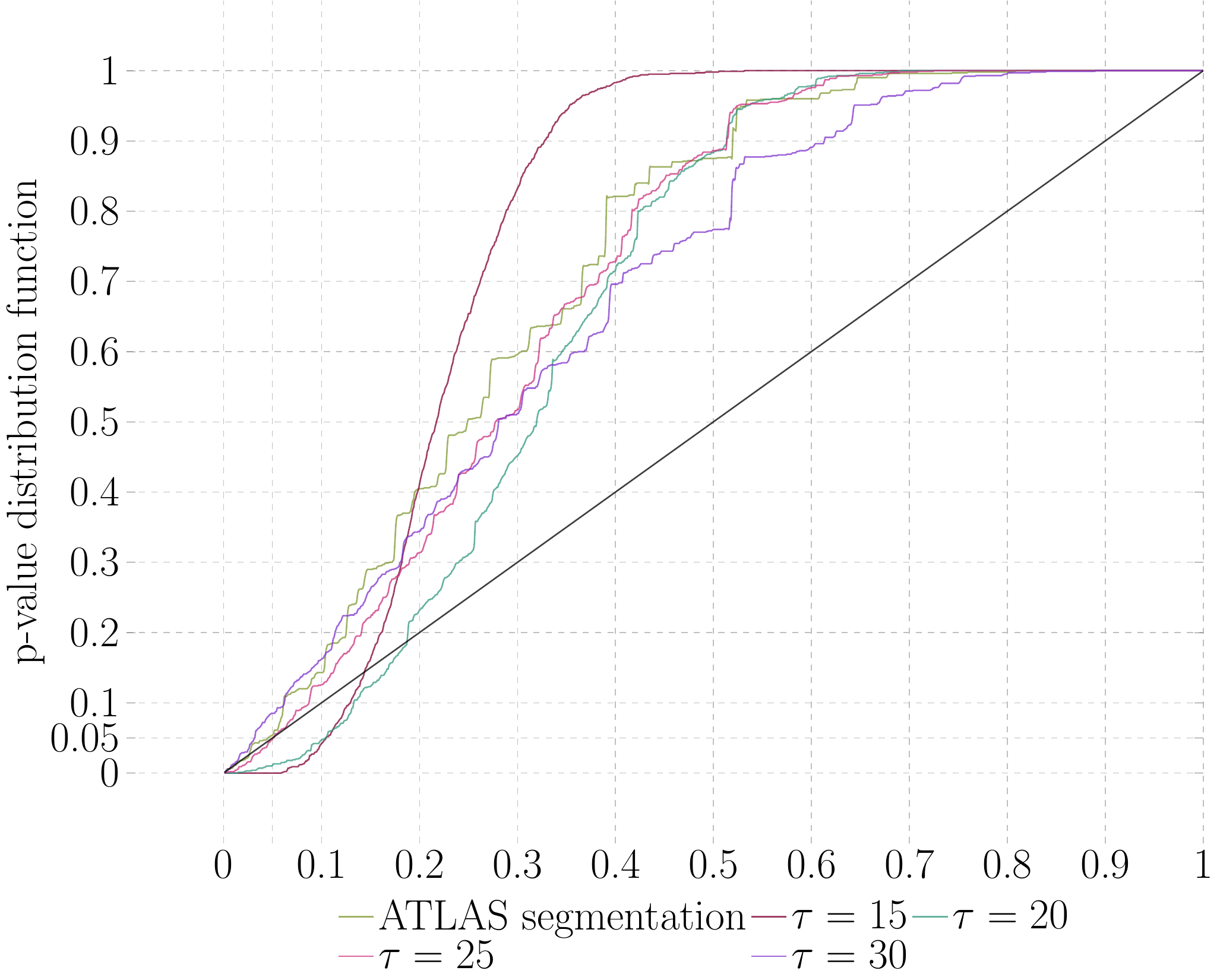}\\
$\,$ & \\
{\bf  Costes method} &  \\
\includegraphics[width=.48\linewidth]{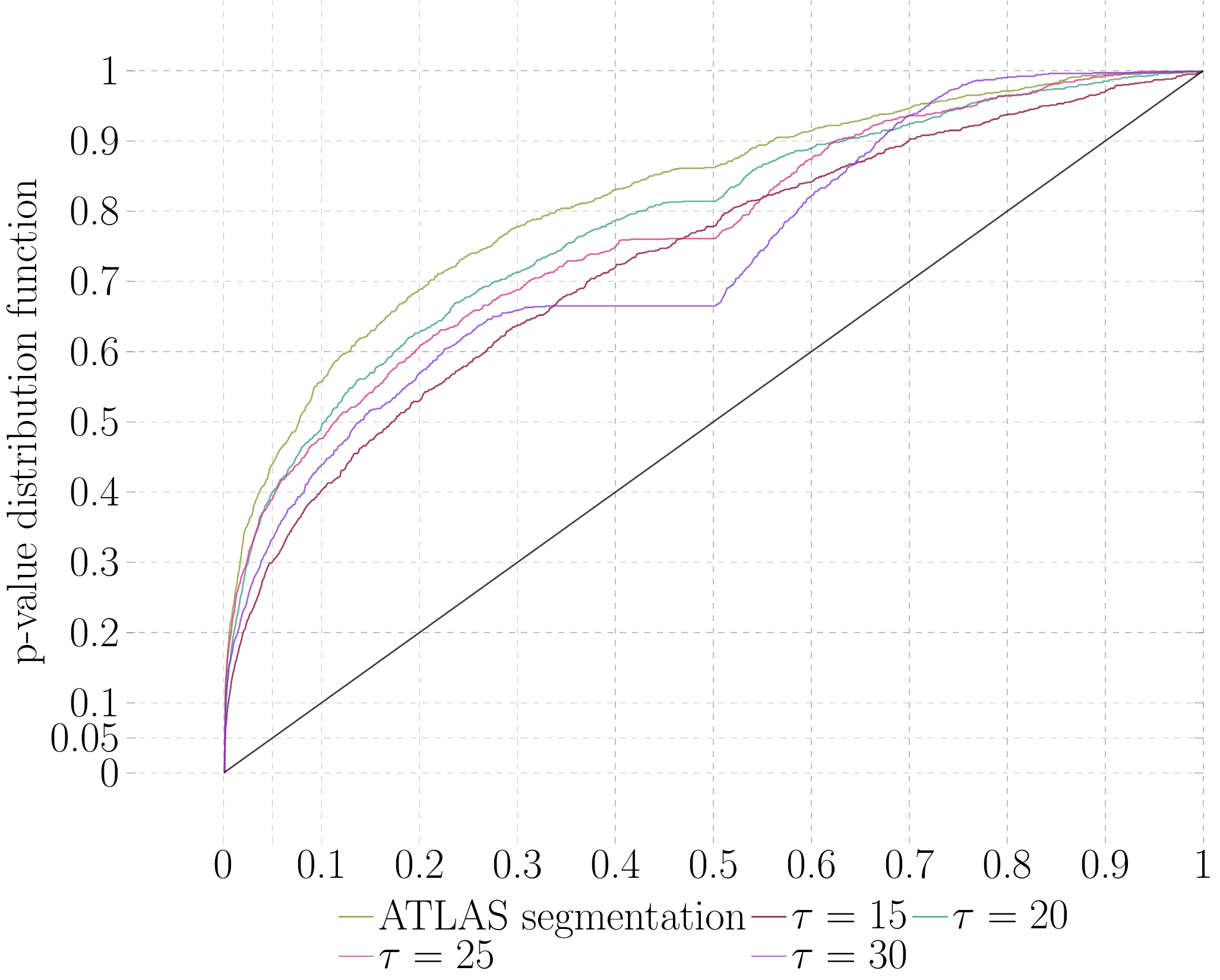} &
\includegraphics[width=.48\linewidth]{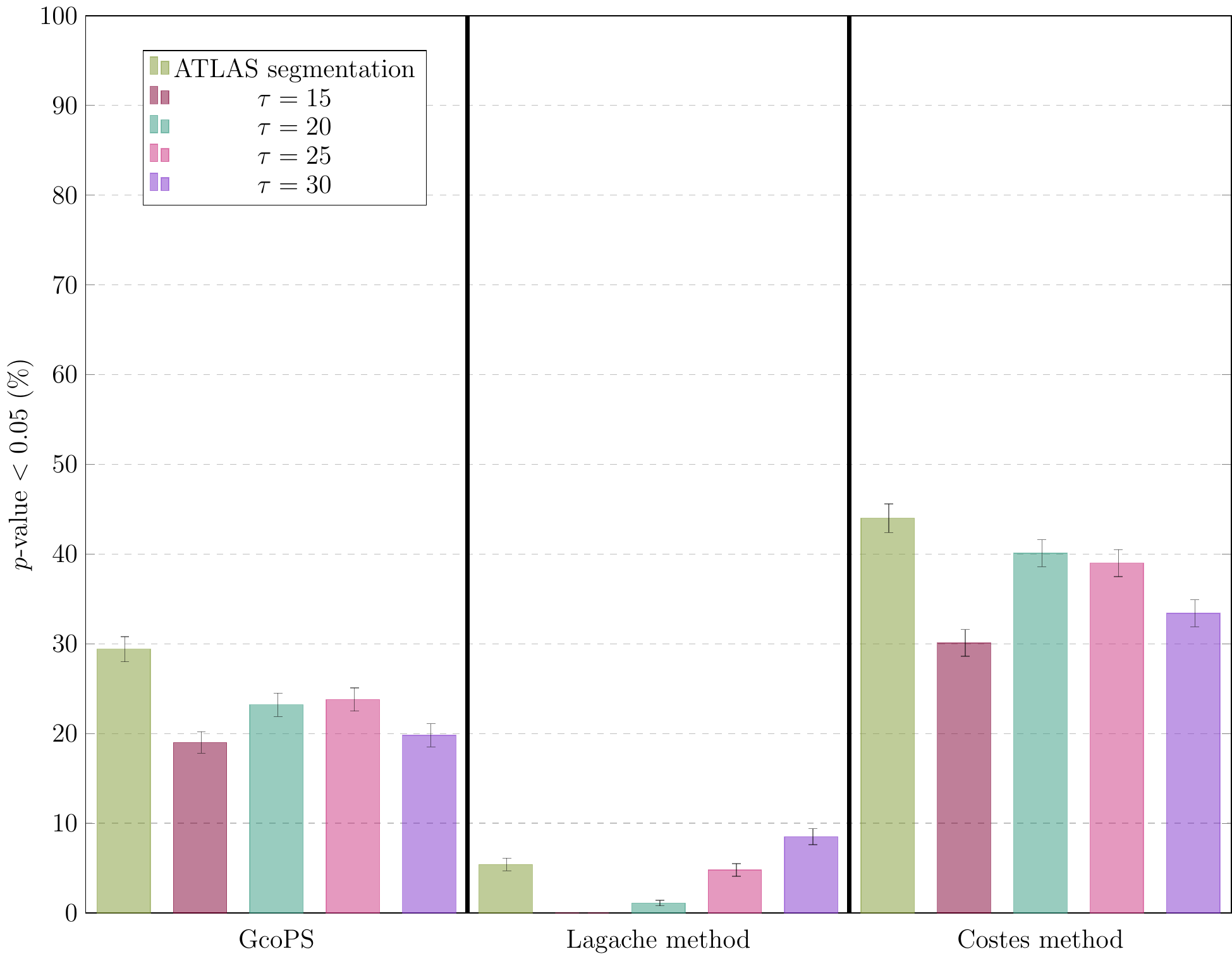} \\
\end{tabular}
\caption{Sensitivity to segmentation. 
Same plots as in Figure~\ref{fig:seg0}  but for 2.5\% forced neighbors.}
\label{fig:seg25}
\end{figure}

\begin{figure}
\begin{tabular}{cc}
{\bf  GcoPS} & {\bf  Lagache method} \\
\includegraphics[width=.48\linewidth]{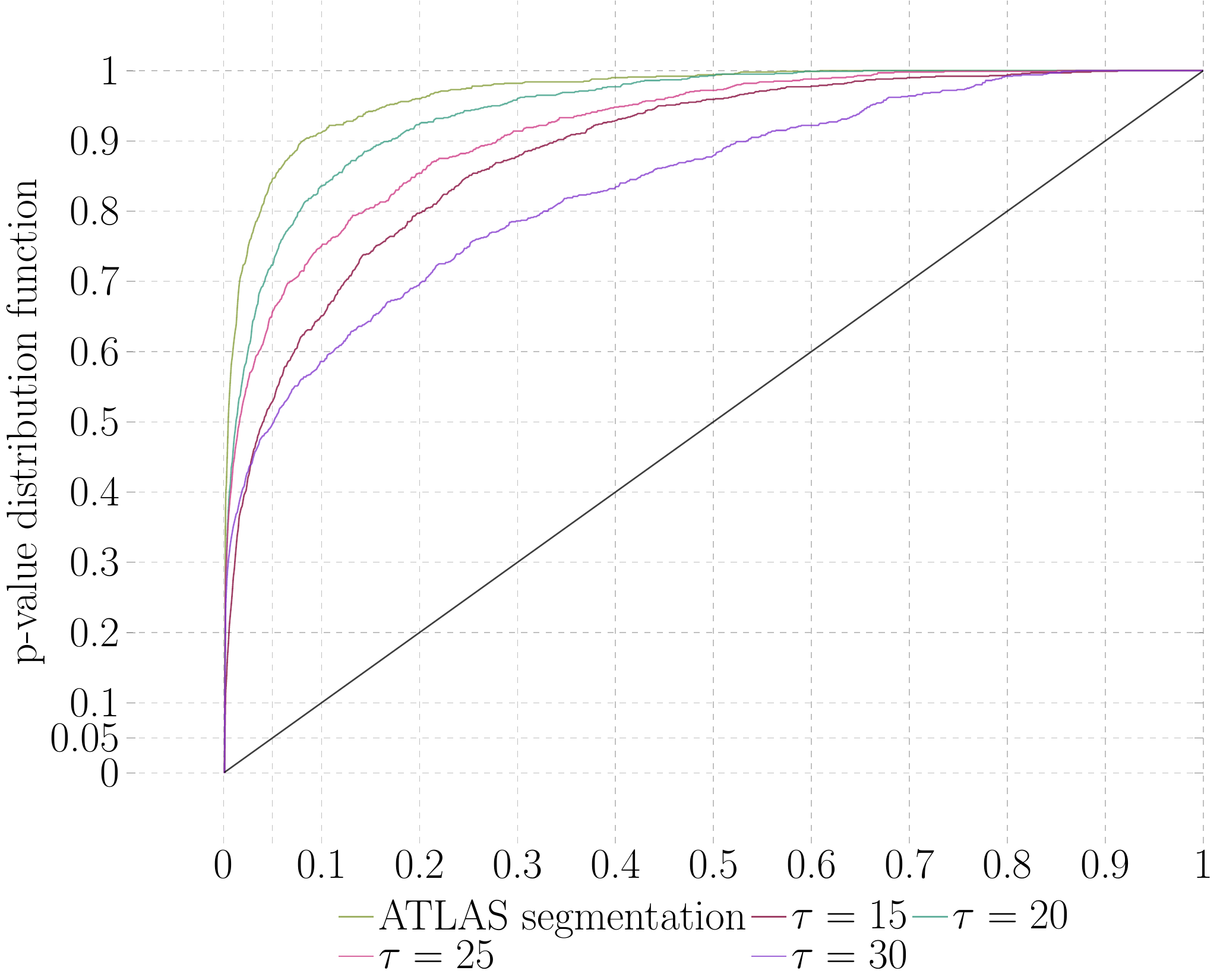} &
\includegraphics[width=.48\linewidth]{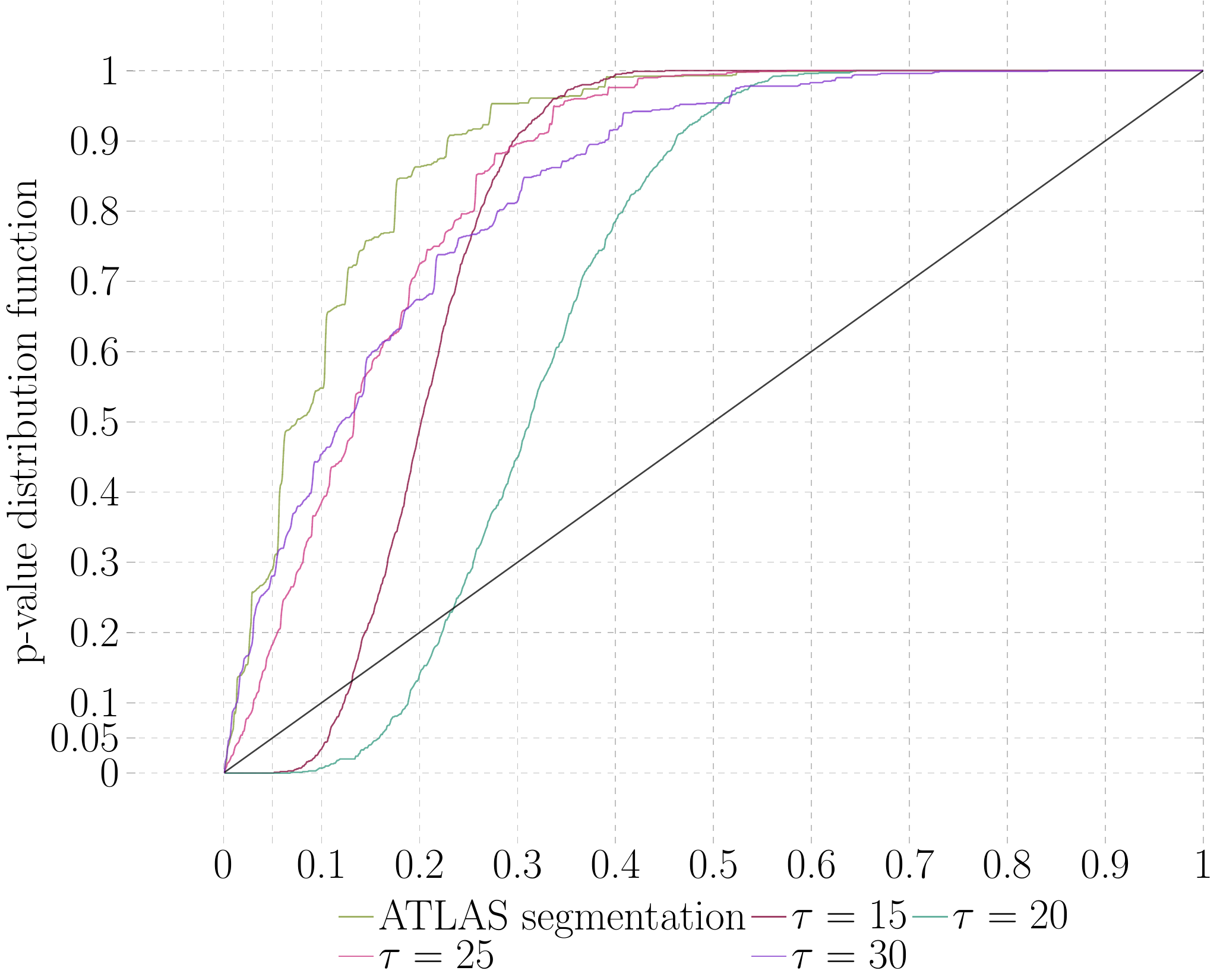}\\
$\,$ & \\
{\bf  Costes method} &  \\
\includegraphics[width=.48\linewidth]{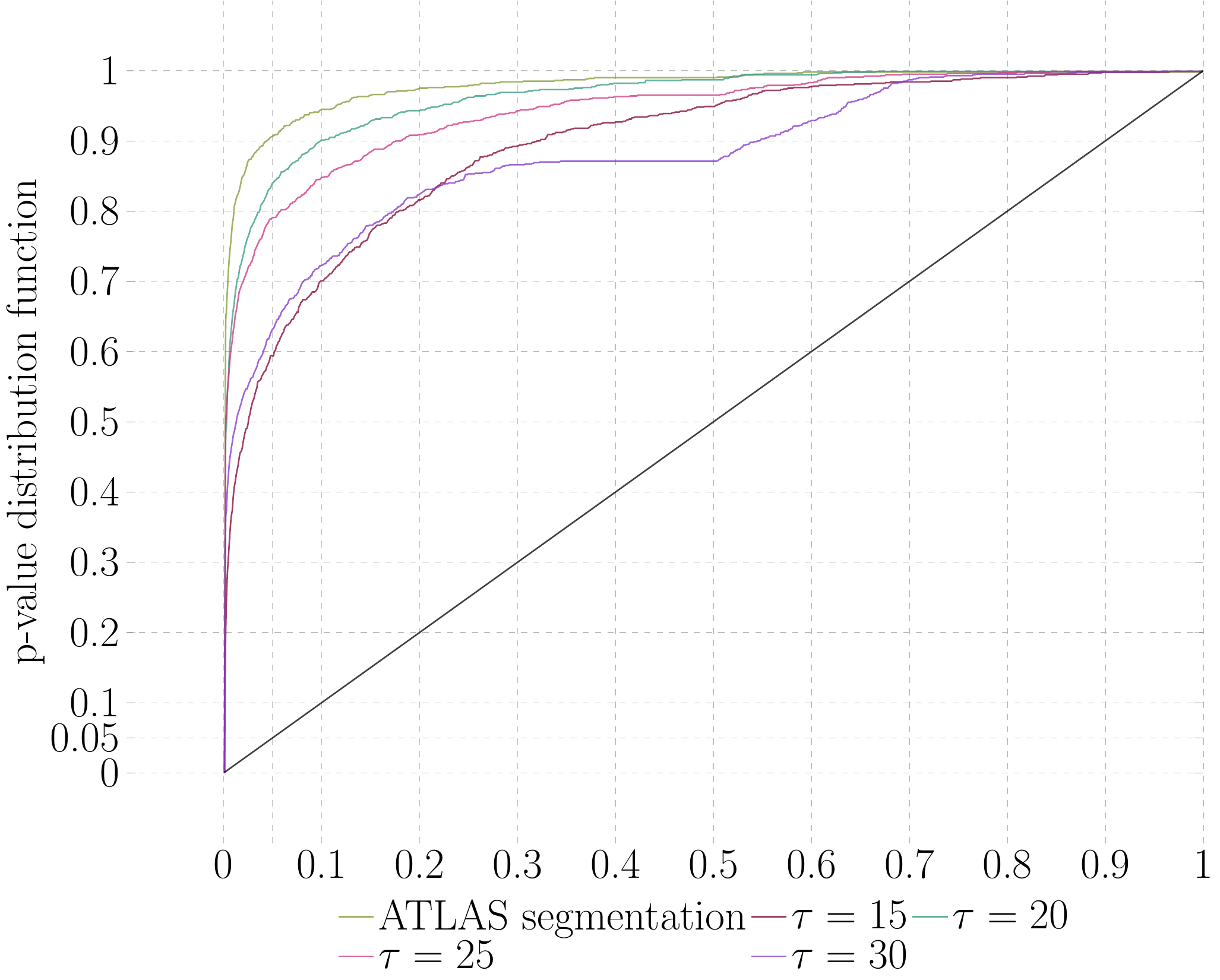} &
\includegraphics[width=.48\linewidth]{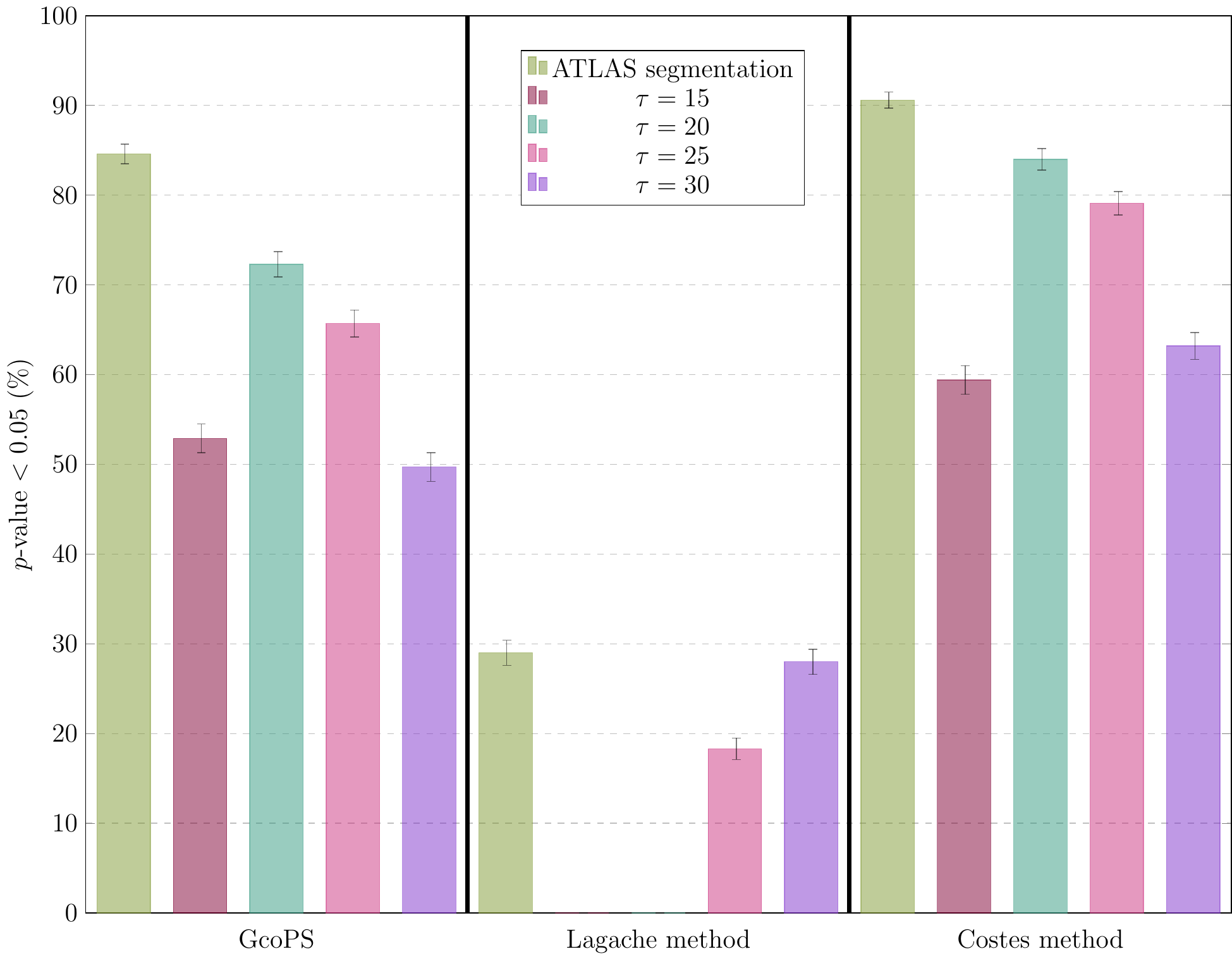} \\
\end{tabular}
\caption{Sensitivity to segmentation. 
Same plots as in Figure~\ref{fig:seg0}  but for 5\%  forced neighbors.}\label{fig:seg5}
\end{figure}

To evaluate the influence of the segmentation for the different methods, we consider the same simulated images with noise and shift as in the previous section but with four different thresholds of segmentation. We then apply the same methods as described in the previous section. As a reference, we also include the segmentation obtained with the ATLAS spot detection method \citep{Basset2015}, which is the segmentation method used in all experiments presented in the paper. 
An example of segmented image is given in Figure~\ref{fig:segexamples}. As a result, see Figures~\ref{fig:seg0}-\ref{fig:seg5}, the intensity-based method of  \cite{Costes2004} is not much affected by the pre-processing and is even slightly more sensitive than GcoPS when the  proportion of forced neighbors is 2.5\% and 5\%. However, this method clearly leads to too many false positive decisions when there is no colocalization, confirming a conclusion made in the previous section.  The object-based method of \cite{Lagache2015} is in turn clearly affected by over-segmentation, in  which case it completely fails to detect colocalization.  GcoPS is also less efficient when the segmentation is not well processed, but the results are overall still satisfying. In particular GcoPS  is very robust to pre-processing for images without colocalization, which is a safe guaranty against false positive decisions.

\subsection{Sensitivity to the shape and scale of objects}\label{sec:scale}
\begin{figure}
\begin{center}
\begin{tabular}{cc}
\begin{tabular}{c} 
\includegraphics[width=.2\linewidth,height=.2\linewidth]{GaussianLevels_spi8_rho0} \\
{ $\rho=0$}\\$\, $\\
\includegraphics[width=.2\linewidth,height=.2\linewidth]{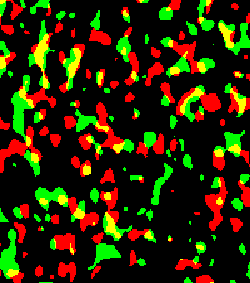}\\ 
{ $\rho=0.1$}\\$\, $\\
\includegraphics[width=.2\linewidth,height=.2\linewidth]{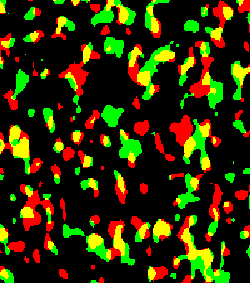}\\
{ $\rho=0.3$}
\end{tabular} &
\begin{tabular}{c} \includegraphics[width=0.5\linewidth]{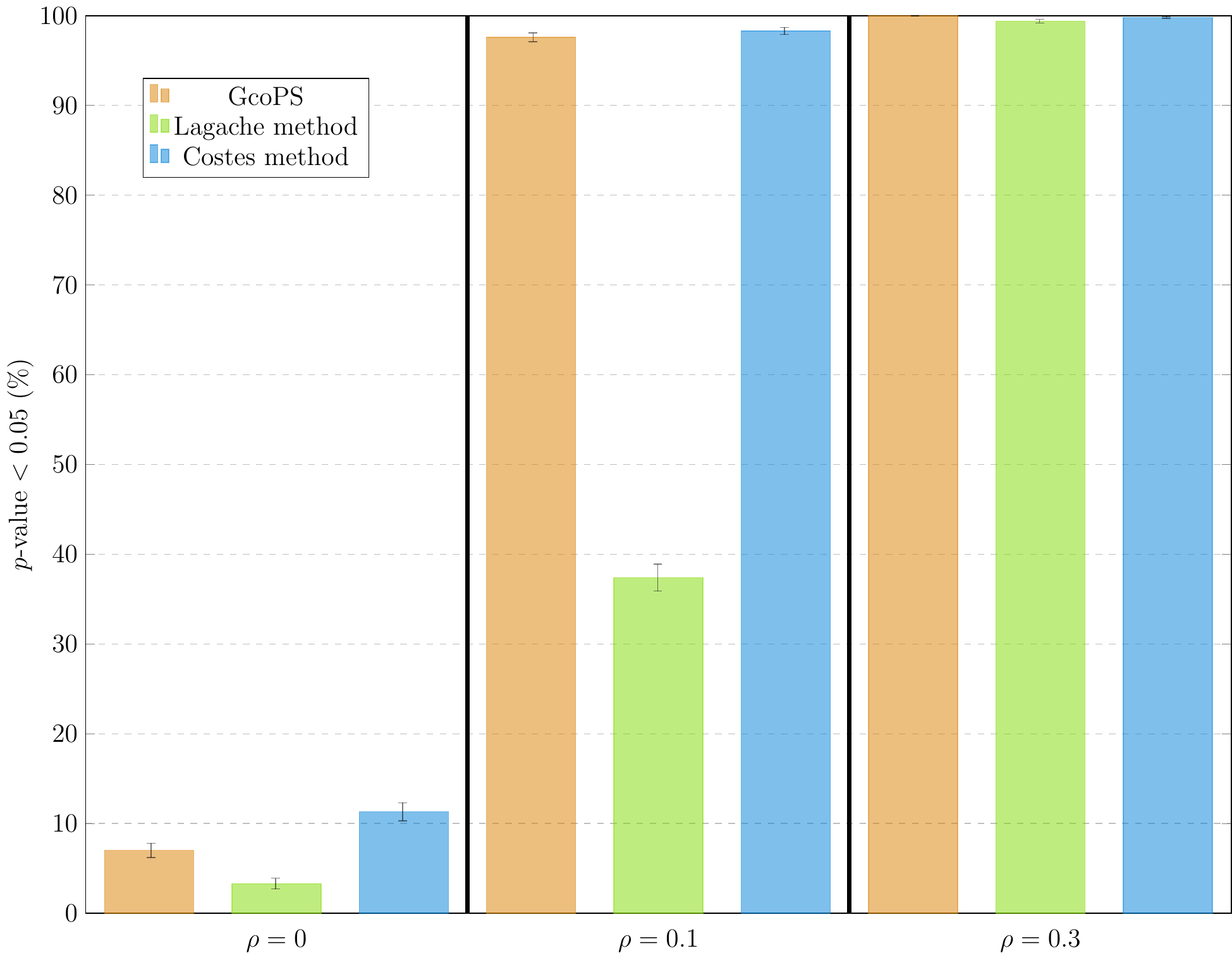} \\
\includegraphics[width=0.5\linewidth]{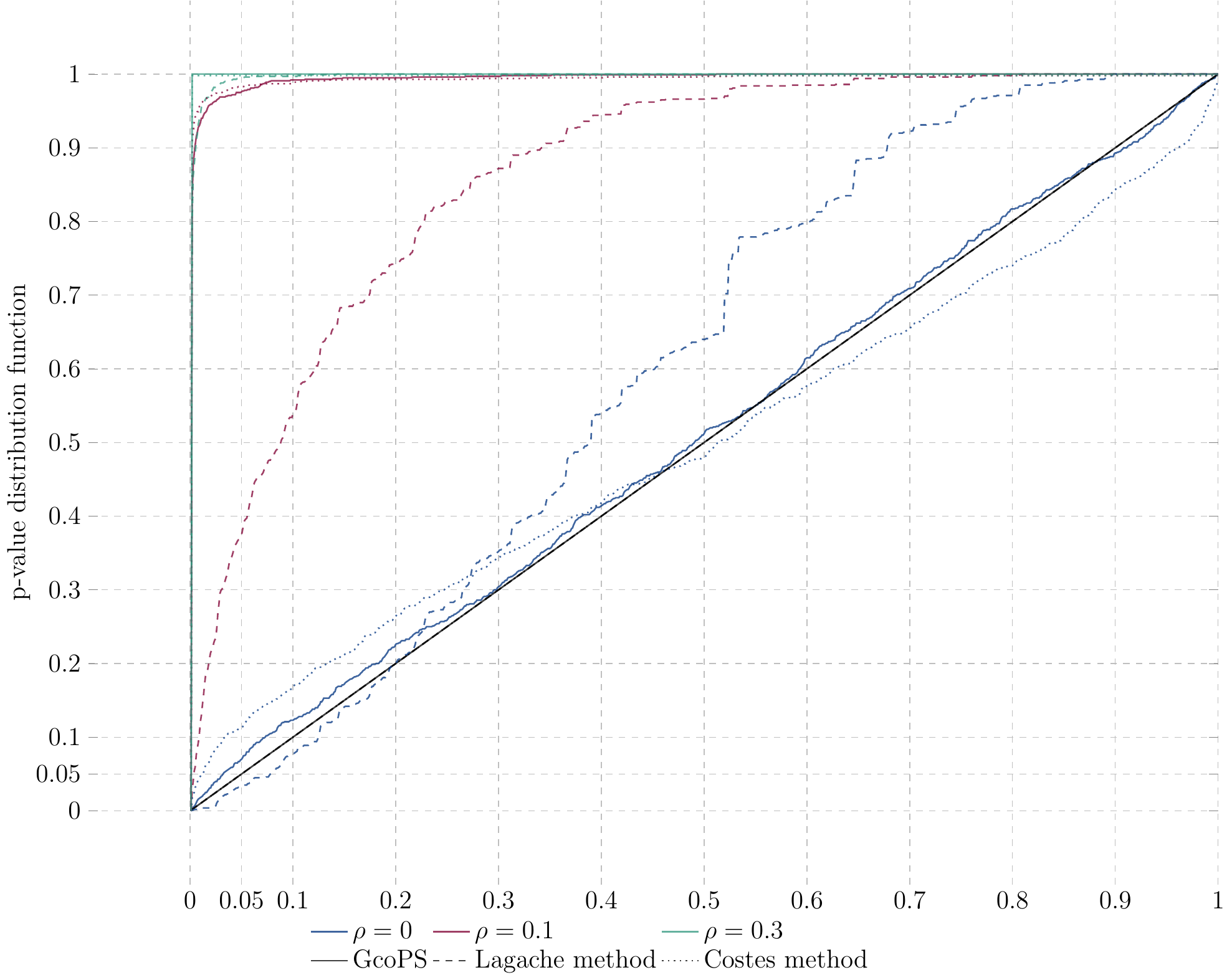} \end{tabular}
\end{tabular}
\end{center}
\caption{Sensitivity to the shape and scale of objects. 
Proportion of $p$-values lower than $0.05$ (top right) and  empirical distribution functions (bottom right) of $p$-values obtained with GcoPS,  the object-based method of \cite{Lagache2015} and the intensity-based method of  \cite{Costes2004} over 1000 simulated images obtained via Gaussian level sets with a correlation equal to 0, 0.1 and 0.3 between the two channels, resulting in non regularly shaped objects. An example of simulated images is shown at the left of the plot.}
\label{fig:scalemoderate}
\end{figure}

\begin{figure}
\begin{center}
\begin{tabular}{cc}
\begin{tabular}{c} 
\includegraphics[width=.2\linewidth,height=.2\linewidth]{GaussianLevels_spi20_rho0} \\
{ $\rho=0$}\\$\, $\\
\includegraphics[width=.2\linewidth,height=.2\linewidth]{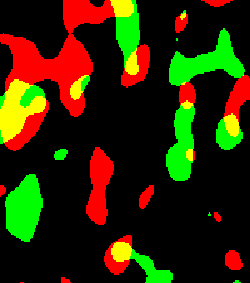}\\ 
{ $\rho=0.1$}\\$\, $\\
\includegraphics[width=.2\linewidth,height=.2\linewidth]{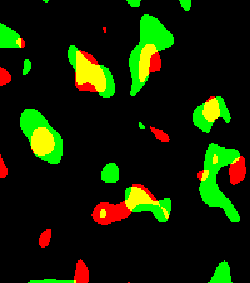}\\
{ $\rho=0.3$}
\end{tabular} &
\begin{tabular}{c} \includegraphics[width=0.5\linewidth]{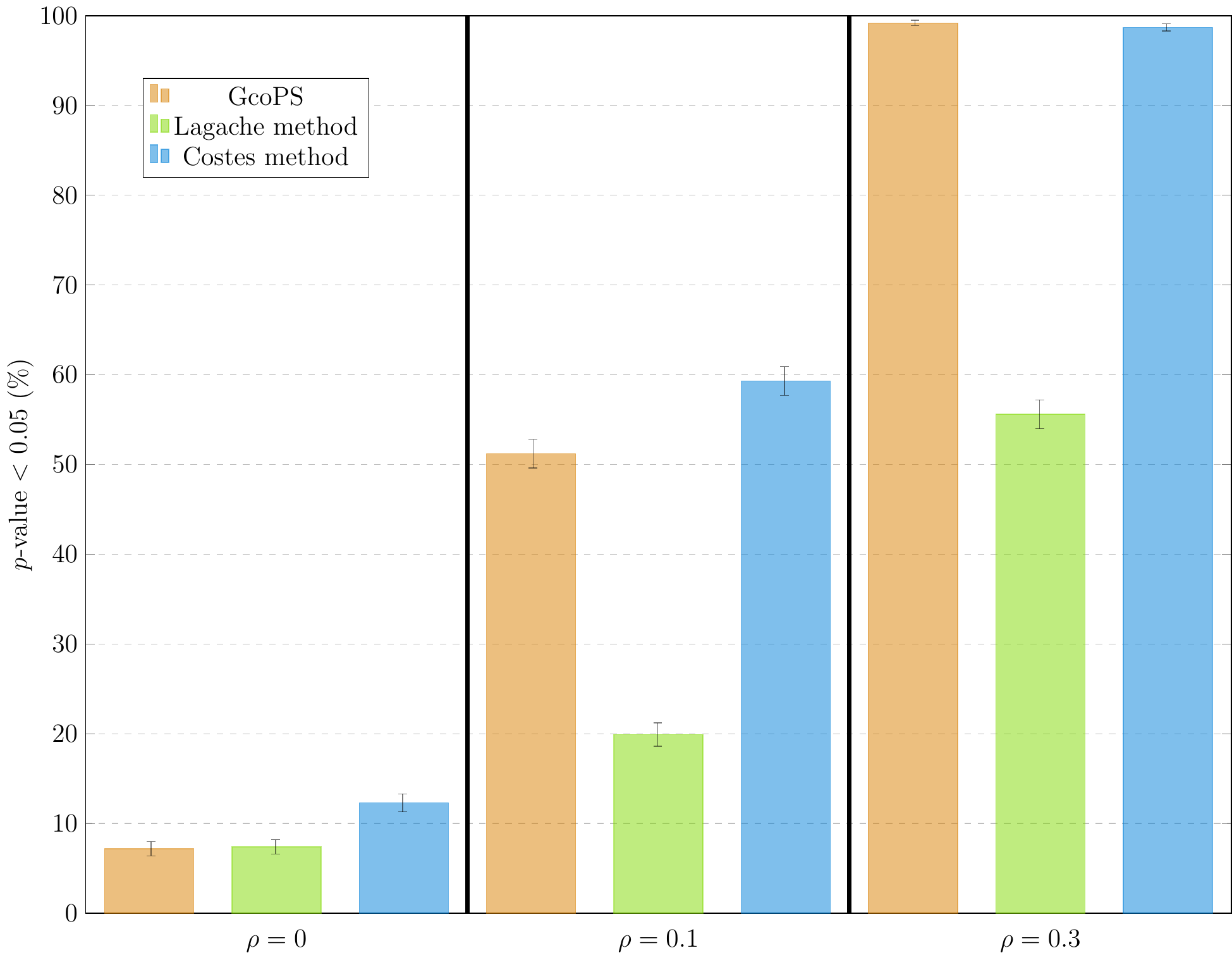} \\
\includegraphics[width=0.5\linewidth]{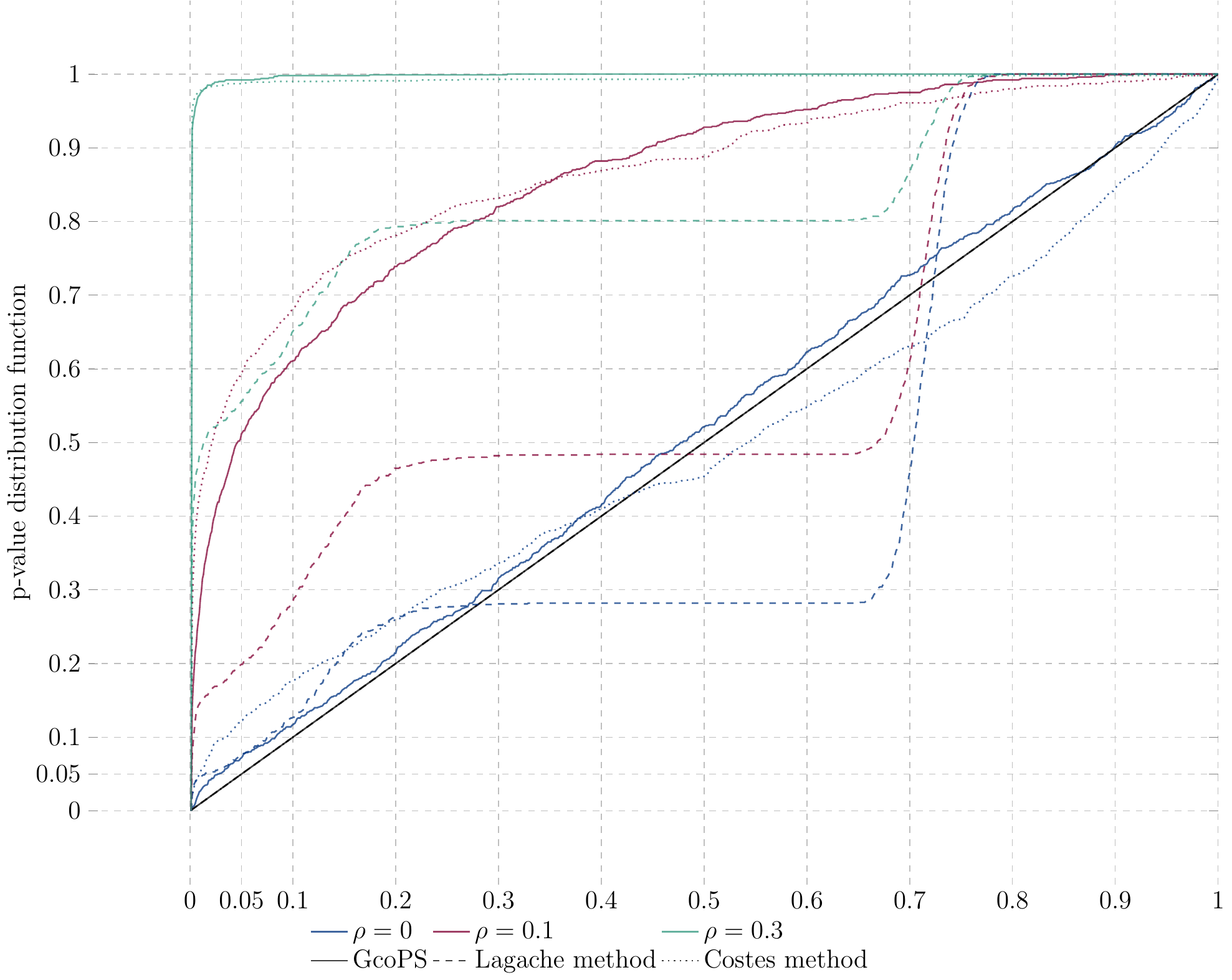} \end{tabular}
\end{tabular}
\end{center}
\caption{Sensitivity to the shape and scale of objects.  Same plots as in Figure~\ref{fig:scalemoderate} but the objects are larger.}
\label{fig:scalelarge}
\end{figure}

\begin{figure}
\begin{center}
\begin{tabular}{cc}
\begin{tabular}{c} 
\includegraphics[width=.2\linewidth,height=.2\linewidth]{GaussianLevels_spi50_rho0} \\
{$\rho=0$}\\$\, $\\
\includegraphics[width=.2\linewidth,height=.2\linewidth]{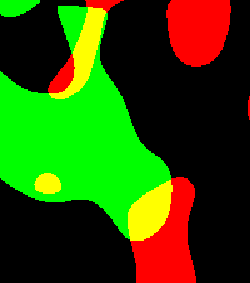}\\ 
{ $\rho=0.1$}\\$\, $\\
\includegraphics[width=.2\linewidth,height=.2\linewidth]{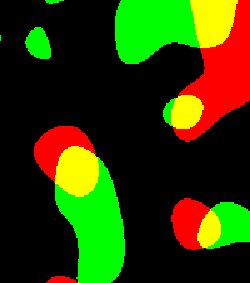}\\
{ $\rho=0.3$}
\end{tabular} &
\begin{tabular}{c} \includegraphics[width=0.5\linewidth]{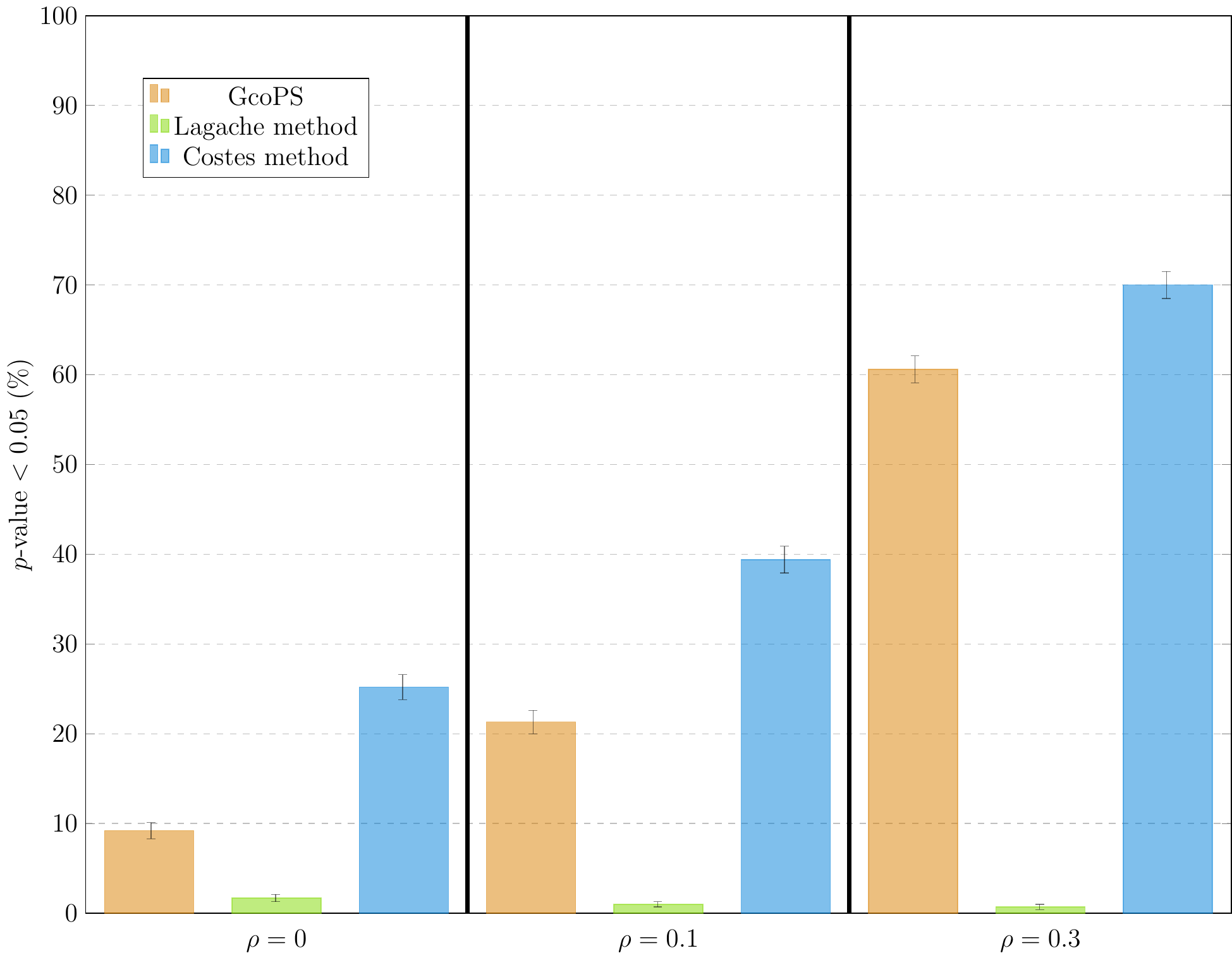} \\
\includegraphics[width=0.5\linewidth]{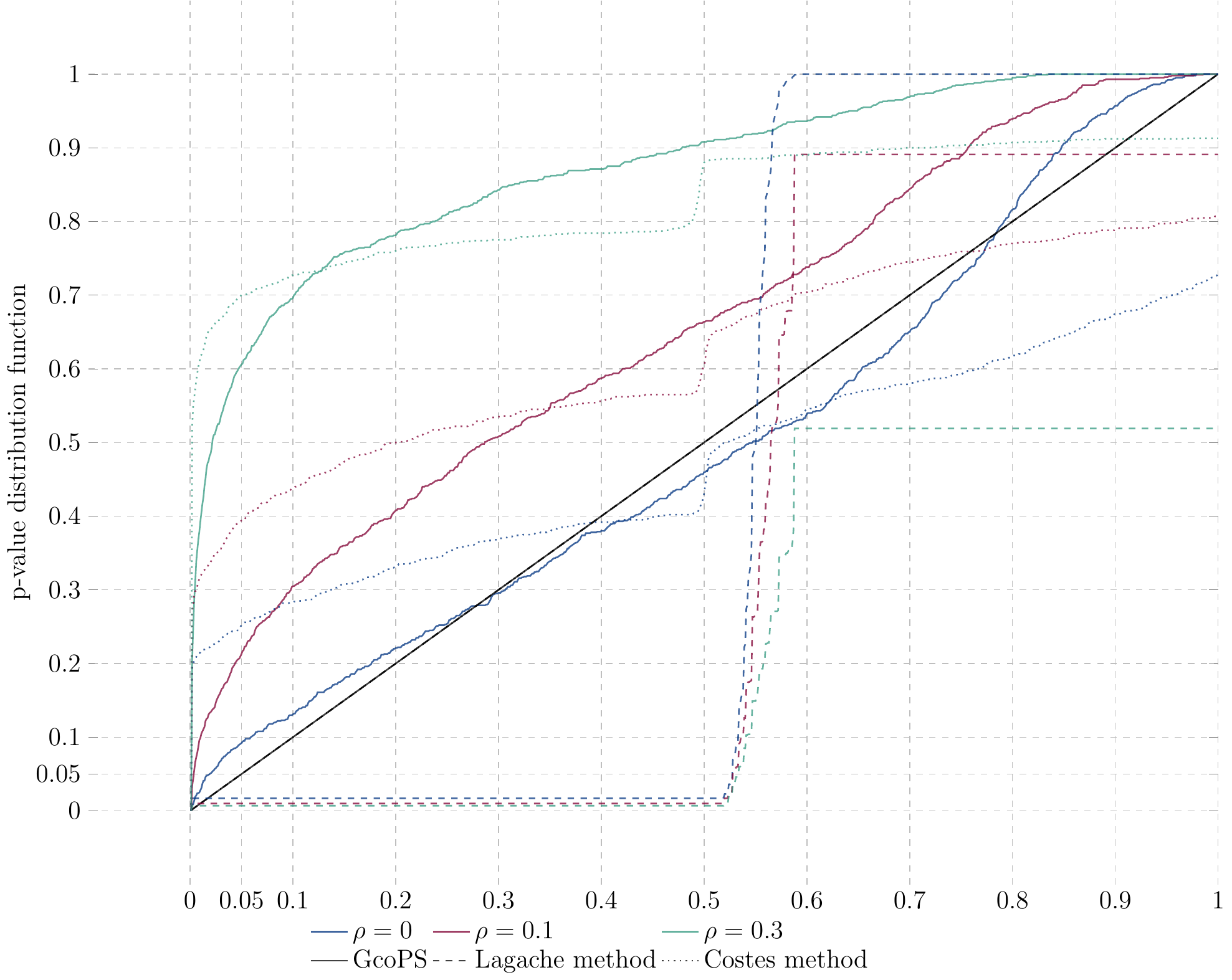} \end{tabular}
\end{tabular}
\end{center}
\caption[bof]{Sensitivity to the shape and scale of objects.  Same plots as in Figure~\ref{fig:scalemoderate} but the objects are even larger than in Figure~\ref{fig:scalelarge}.
}
\label{fig:scaleverylarge}
\end{figure}

To evaluate the sensitivity to the shape and the size (or scale) of objects, we have simulated Gaussian level sets with a correlation between the two channels equal to $0$ ({\it i.e.}, no colocalization), $0.1$ (slight colocalization) and $0.3$ (stronger colocalization) approximately. This method of simulation is detailed in Section~\ref{sec:gauss}. It allows to generate objects that exhibit non-regular shapes (different from balls), with a  typical scale that can be easily controlled. When the objects are small (not shown in our figures), the images are quite similar to the images generated in Section~\ref{sec:2D}, or in other words the objects are fairly assimilable to small balls, and the methods show similar efficiency  as described above. 
In presence of non regular objects, as shown in Figure~\ref{fig:scalemoderate}, the intensity-based method of  \cite{Costes2004} is once again too sensitive for colocalization while the object-based method of \cite{Lagache2015} is not sufficiently sensitive, especially when the correlation between the two channels is $0.1$. In contrast, the performance of GcoPS is not disrupted by the shape of objects.
In presence of larger objects, as depicted in Figure~\ref{fig:scalelarge}, the conclusions are similar. Finally, in the extreme case of very large objects, see Figure~\ref{fig:scaleverylarge}, the object-based method, that reduces each object to a single point, completely fails to detect colocalization, which is easily explained by the dramatic loss of information induced by this reduction. In this case, the Costes method is far too sensitive (about $25\%$ of false positive decisions in absence of colocalization), while GcoPS still performs well, though being a bit too sensitive when there is no colocalization. 
The robustness of GcoPS in presence of large objects shows that this method can be applied to small windows in the image, where the scaling makes objects larger, hence opening the possibility to effectively localize  the detection of colocalization.

\subsection{Performance  in presence of a different optical resolution in each image}\label{sec:resolution}
To generate images with a different optical resolution, we have simulated Gaussian level sets where the scale parameter ruling the typical size of objects is different in each image. The details are given in Section~\ref{sec:gauss}. The correlation between the two images is as in the previous section $\rho=0$, $\rho=0.1$ or $\rho=0.3$. 

\begin{figure}
\begin{center}
\begin{tabular}{cc}
\begin{tabular}{c} 
\includegraphics[width=.2\linewidth]{GaussianLevels_spi10-5_rho0} \\
{ $\rho=0$}\\$\, $\\
\includegraphics[width=.2\linewidth]{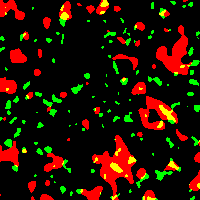}\\ 
{ $\rho=0.1$}\\$\, $\\
\includegraphics[width=.2\linewidth]{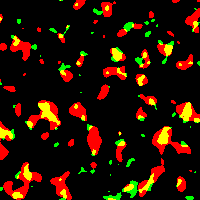}\\
{ $\rho=0.3$}
\end{tabular} &
\begin{tabular}{c} \includegraphics[width=0.5\linewidth]{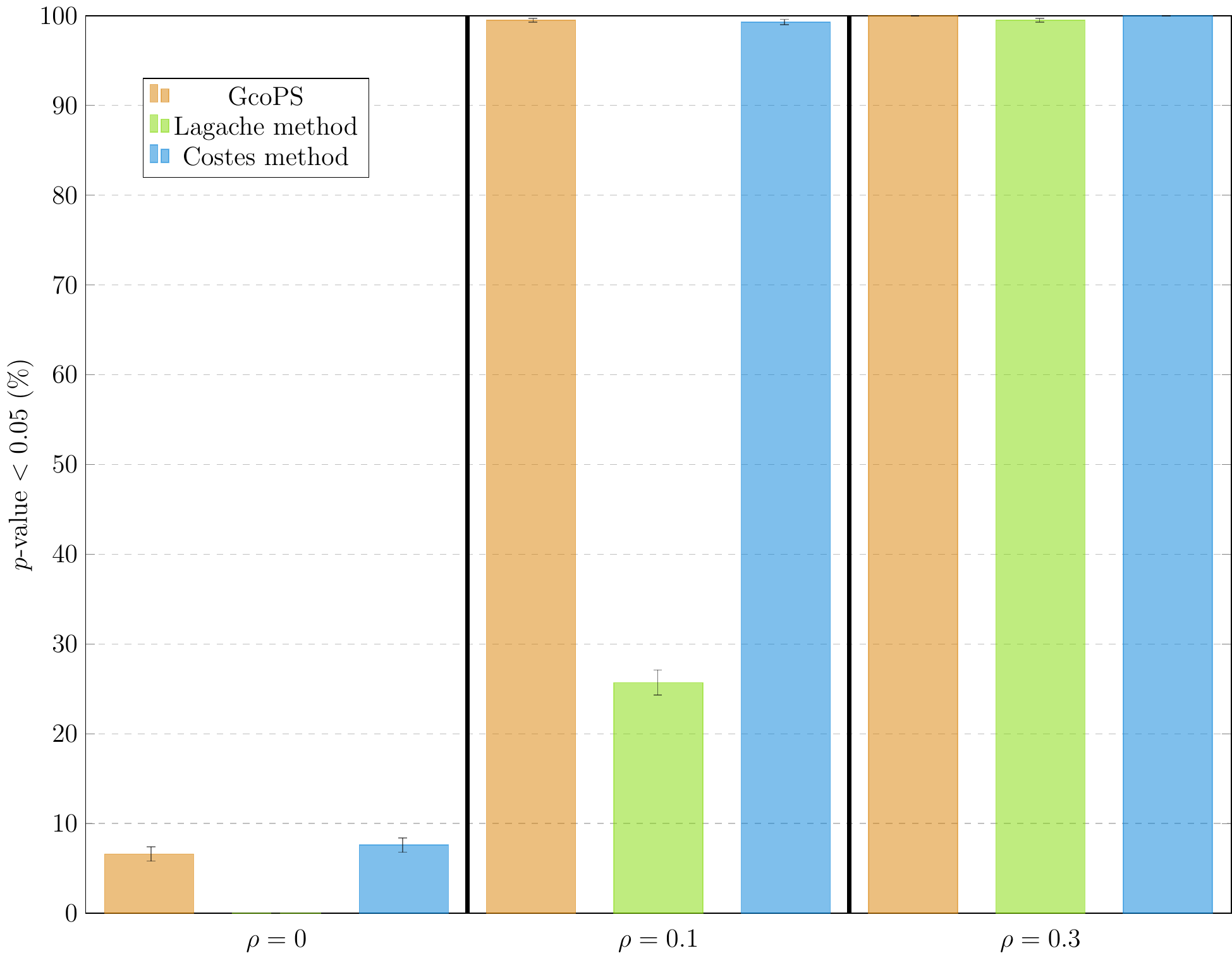} \\
\includegraphics[width=0.5\linewidth]{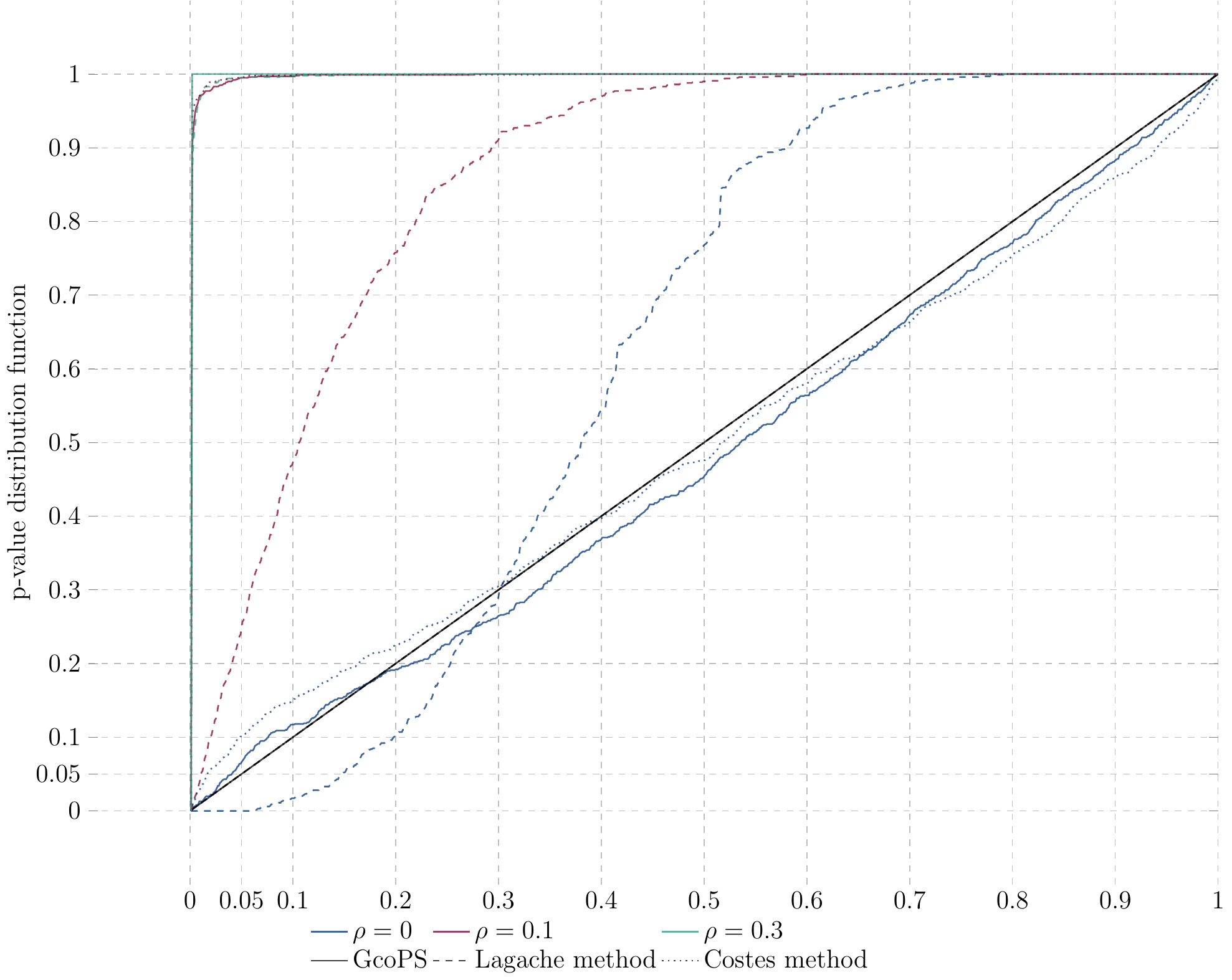} \end{tabular}
\end{tabular}
\end{center}
\caption{Sensitivity to the resolutions in each channel.
Proportion of $p$-values lower than $0.05$ (top right) and  empirical distribution functions (bottom right) of $p$-values obtained with GcoPS,  the object-based method of \cite{Lagache2015} and the intensity-based method of  \cite{Costes2004} over 1000 simulated images obtained via Gaussian level sets with a correlation equal to 0, 0.1 and 0.3 between the two channels, where the resolution is different in each channel. An example of simulated images is shown at the left of the plot.}
\label{fig:resolutiondifferent}
\end{figure}

\begin{figure}
\begin{center}
\begin{tabular}{cc}
\begin{tabular}{c} 
\includegraphics[width=.2\linewidth]{GaussianLevels_spi20-5_rho0} \\
{ $\rho=0$}\\$\, $\\
\includegraphics[width=.2\linewidth]{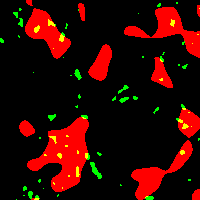}\\ 
{ $\rho=0.1$}\\$\, $\\
\includegraphics[width=.2\linewidth]{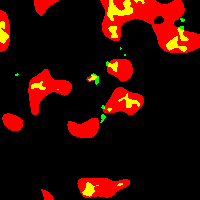}\\
{ $\rho=0.3$}
\end{tabular} &
\begin{tabular}{c} \includegraphics[width=0.5\linewidth]{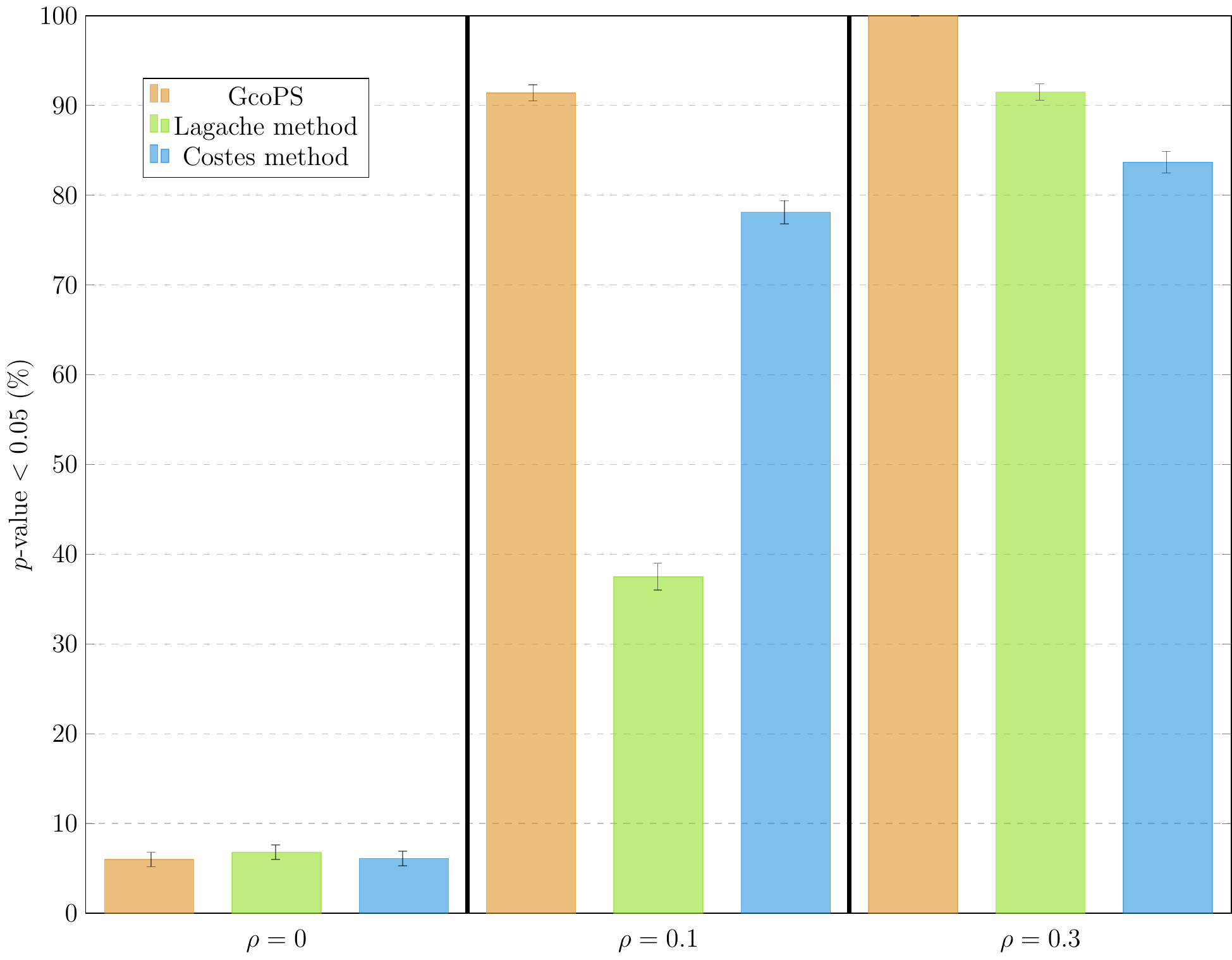} \\
\includegraphics[width=0.5\linewidth]{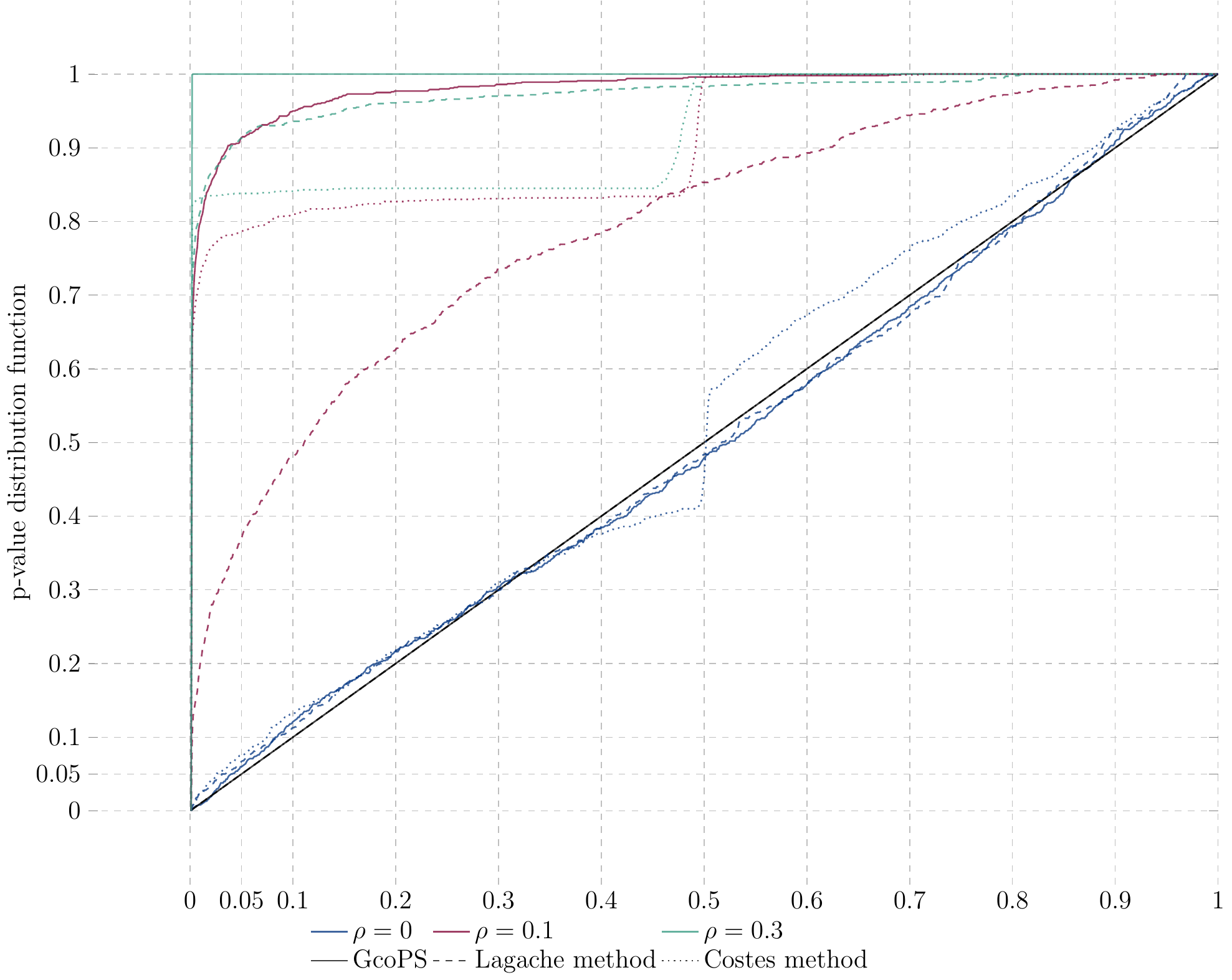} \end{tabular}
\end{tabular}
\end{center}
\caption[bof]{Sensitivity to the resolutions in each channel.
Same plots as in Figure~\ref{fig:resolutiondifferent} but the difference in resolution is larger. 
}
\label{fig:resolutionverydifferent}
\end{figure}

Figures~\ref{fig:resolutiondifferent} and \ref{fig:resolutionverydifferent} show situations with a  difference  in  optical resolution (moderate and strong respectively). The results demonstrate that the object-based method of \cite{Lagache2015}  is clearly outperformed by the other methods in presence of a difference of optical resolution. The intensity-based method of  \cite{Costes2004}  is more robust but GcoPS exhibits better efficiency.
This proves  that  GcoPS is able to process efficiently images for which a different microscopy technique is used for detecting each type of molecules.

\subsection{Evaluation on simulated 3D images}\label{sec:3D}
\begin{figure}
\begin{center}
\begin{tabular}{cc}
\begin{tabular}{c} 
\includegraphics[trim=2.8cm 1.55cm 2.8cm 1.55cm,clip,width=.25\linewidth]{GaussianLevels_spi8_3D_rho0} \\
{ $\rho=0$}\\$\, $\\
\includegraphics[trim=2.8cm 1.55cm 2.8cm 1.55cm,clip,width=.25\linewidth]{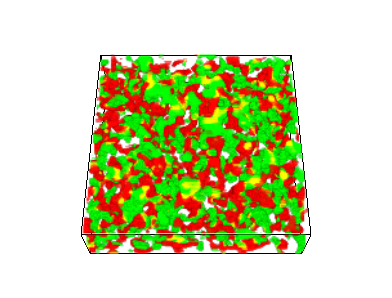}\\ 
{ $\rho=0.1$}\\$\, $\\
\includegraphics[trim=2.8cm 1.55cm 2.8cm 1.55cm,clip,width=.25\linewidth]{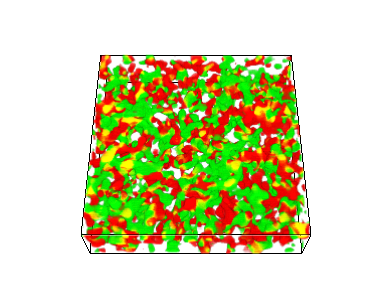}\\
{ $\rho=0.3$}
\end{tabular} &
\begin{tabular}{c} \includegraphics[width=0.5\linewidth]{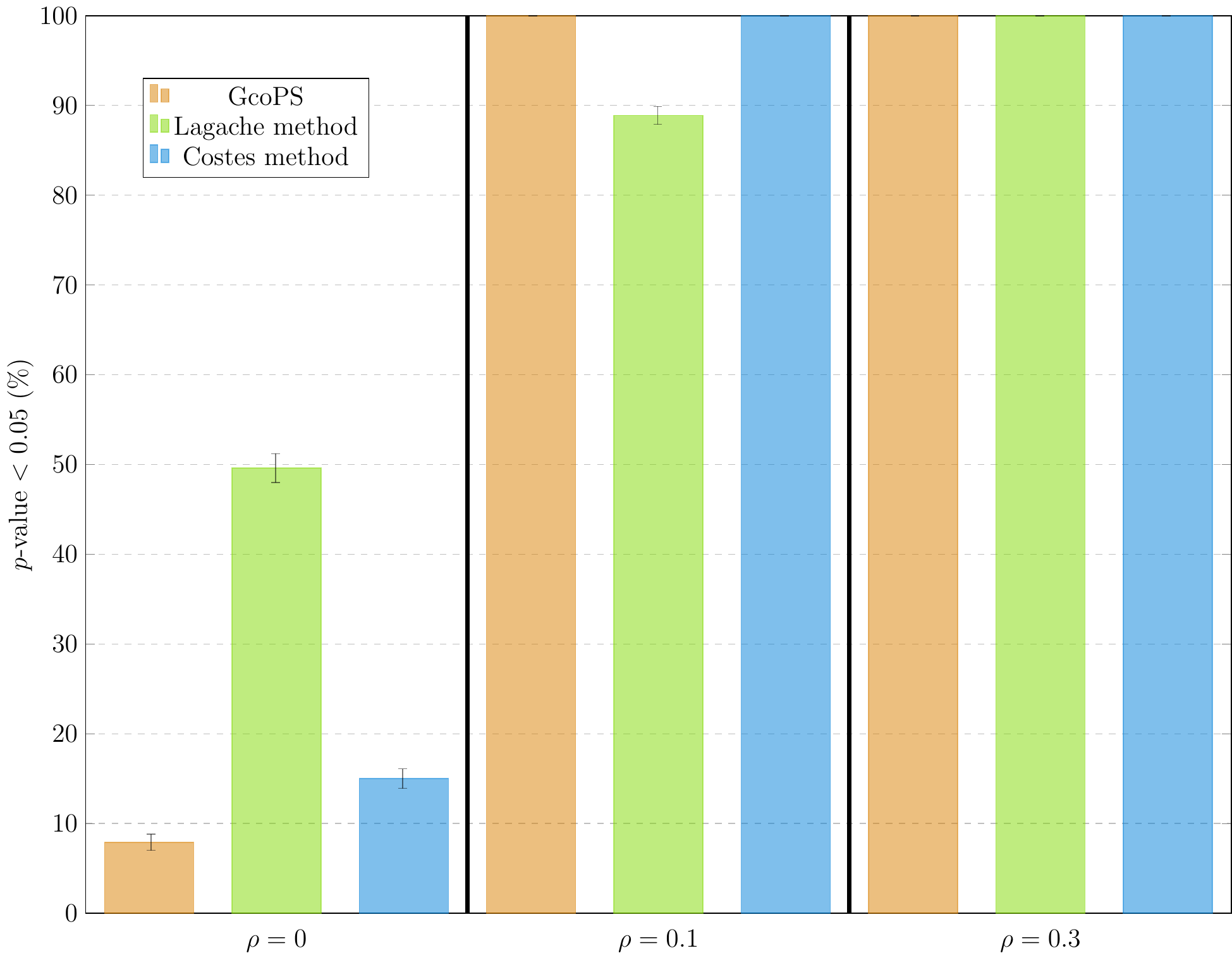} \\
\includegraphics[width=0.5\linewidth]{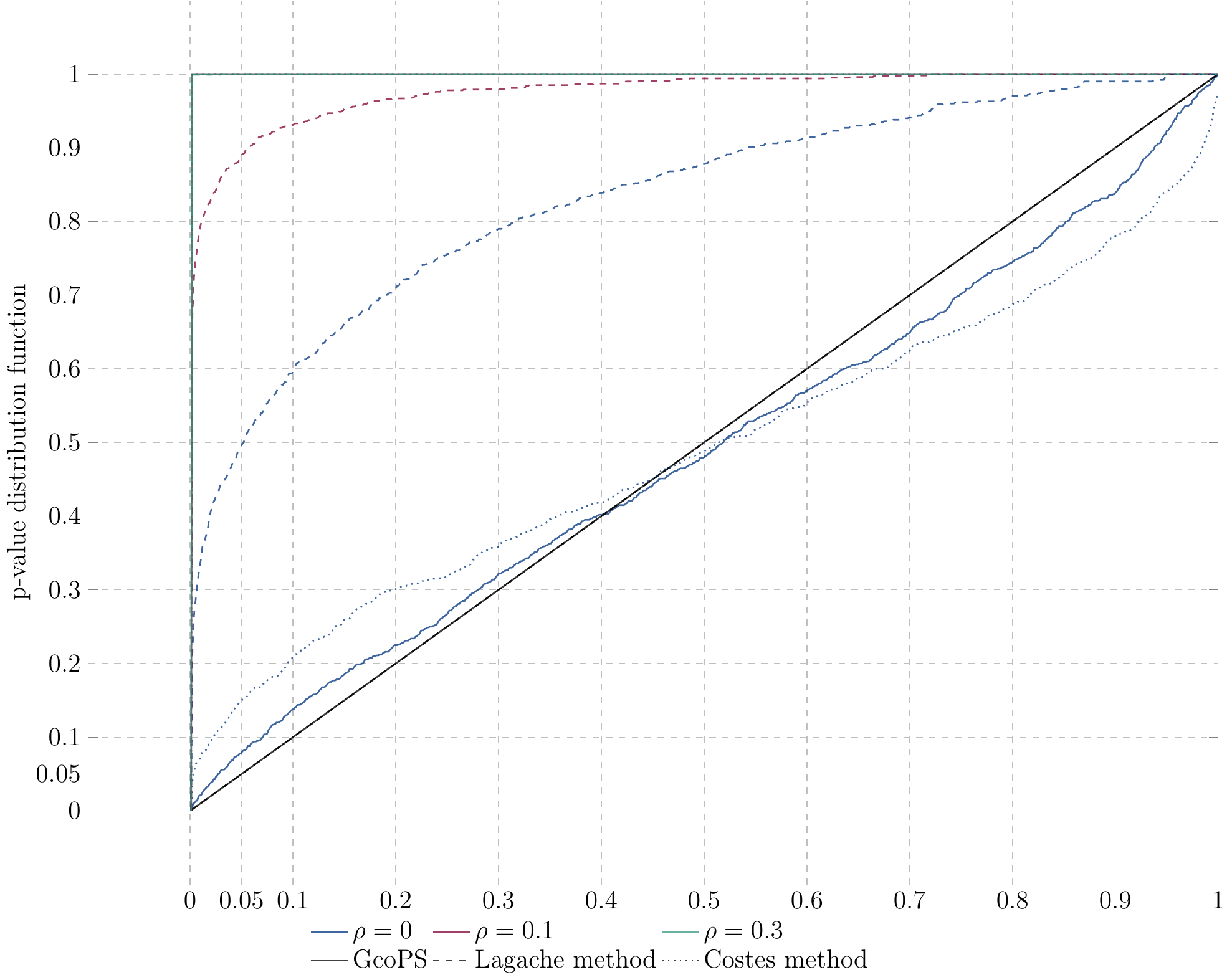} \end{tabular}
\end{tabular}
\end{center}
\caption{Evaluation on 3D images.
Proportion of $p$-values lower than $0.05$ (top right) and  empirical distribution functions (bottom right) of $p$-values obtained with GcoPS, the object-based method of \cite{Lagache2015} and the intensity-based method of  \cite{Costes2004} over 1000 simulated images obtained via 3D Gaussian level sets with a correlation equal to 0, 0.1 and 0.3 between the two channels. An example of simulated images is shown at the left of the plot.
}
\label{fig:3D}
\end{figure}

We have performed simulations of 3D objects using Gaussian level sets (see Section~\ref{sec:gauss} for details) to generate channels with a correlation equal to $0$, $0.1$ and $0.3$. The results are displayed in Figure~\ref{fig:3D}. The object-based method of \cite{Lagache2015} shows very unsatisfying results when there is no colocalization. The  intensity-based method of  \cite{Costes2004}  is in turn even more sensitive to colocalization in 3D than it is in 2D, leading to too many false positive conclusions when the two images are independent. The results obtained with GcoPS on 3D images  are in line with the results obtained with 2D images and demonstrate the better overall performance of this method. Note finally that we have also performed complementary simulations in 3D, not displayed here, to assess  the robustness of GcoPS against shape anisotropy (e.g. elongated shapes in 3D) and/or a low density of particles. The results demonstrate that the performance of GcoPS is not altered.

\vfill

\newpage

\section{Simulations by Gaussian level sets}\label{sec:gauss}
Let $X$, $Y$ and $\epsilon$ be three independent Gaussian random fields in $\mathbb R^d$ with isotropic covariance function  $$C(r)=\sigma^2 e^{-r^2/\alpha^2},\quad r\geq 0,$$
where $r$ denotes the radial distance between two points of the field, $\sigma^2$ is the variance and $\alpha>0$ is referred to as the scale parameter. These parameters  are denoted by $\sigma_X^2$, $\sigma_Y^2$, $\sigma_\epsilon^2$ and $\alpha_X$, $\alpha_Y$, $\alpha_\epsilon$, for $X$, $Y$ and $\epsilon$ respectively. Henceforth, for $\rho_0\in [-1,1]$, we set
$$\sigma^2_X=\sigma^2_Y=:\sigma_0^2\quad\text{and}\quad \sigma^2_\epsilon=\frac{\rho_0}{1-\rho_0} \sigma_0^2.$$
We define the  random fields
$$U=X+\epsilon \quad\text{and}\quad V=Y+\epsilon.$$
Note that $U$ and $V$ are Gaussian random fields with common variance $\sigma^2=\sigma_0^2/(1-\rho_0)$ and that $U$ and $V$ are correlated with correlation $\rho_0$.
We finally consider the random sets induced by the level sets of $U$ and $V$, for $\tau_1>0$ and $\tau_2>0$,
$$\Gamma_1=\{U>\tau_1 \sigma\} \quad\text{and}\quad \Gamma_2=\{V>\tau_2  \sigma\}.$$
We easily get the following properties :
\begin{itemize}
\item The random set $\Gamma_1$, respectively $\Gamma_2$, has a coverage rate equal to $p_1:=\mathbb P(o\in\Gamma_1)=1-\Phi(\tau_1)$ (for any $o\in\mathbb R^d$),  respectively $p_2:=1-\Phi(\tau_2)$, where $\Phi$ denotes the cumulative distribution function of a standard normal law. Recall that this coverage rate represents the proportion of $1$'s, in average,  generated by the binary field $\Gamma_1$ in a given domain.
\item  The random sets $\Gamma_1$ and $\Gamma_2$ are correlated with correlation 
 $$\rho=\left(\int_{\tau_1}^\infty\int_{\tau_2}^\infty f(u,v) {\rm d}u{\rm d}v - p_1 p_2\right)/\sqrt{p_1(1-p_1)p_2(1-p_2)},$$ 
where $f$ denotes the marginal probability density function of $(U,V)$ that is the density of a bivariate centered Gaussian random variable with covariance matrix $\sigma^2 \left[\begin{array}{cc} 1 & \rho_0\\ \rho_0 & 1\end{array}\right]$.
\end{itemize}

In order to generate two correlated binary images containing random spots in a given domain, say $\Omega_n$, we therefore simulate $\Gamma_1$ and $\Gamma_2$ in $\Omega_n$. The input parameters are first the scale parameters $\alpha_X$, $\alpha_Y$ and $\alpha_\epsilon$ that rule the size of the spots (the larger the scale parameters, the larger the spots), second the thresholds $\tau_1$ and $\tau_2$ that rule the density of spots in $\Omega_n$ (see the expression of $p_1$ and $p_2$), third $\rho_0$ and $\sigma_0$ that along with the thresholds influence the correlation $\rho$ between the two channels. 

Given the input parameters, the simulation is straightforward. It basically amounts to simulate  the Gaussian random fields $X$, $Y$ and $\epsilon$ on $\Omega_n$. We use at this step the \texttt{RandomFields} package \citep{RandomFields} of the free available software R \citep{R}. Then the random fields $U$ and $V$, and finally the binary images induced by $\Gamma_1$ and $\Gamma_2$,  are easily deduced.

In the simulations of Section~\ref{sec:scale},  we  used as input parameters $\sigma_0=1$, $\tau_1=\tau_2=1$, resulting in  a density of spots of $p_1=p_2\approx 16\%$, and $\rho_0=0, 0.2, 0.5$ leading to an actual correlation between the two channels of $\rho=0, 0.1, 0.3$ approximately. The domain of simulation was $\Omega_n=[0,250]^2$. As to the scale parameters (that have no impact on the values of $p_1$, $p_2$ and $\rho$), we chose $\alpha_X=\alpha_Y=\alpha_\epsilon=8, 20, 50$ resulting in small, large or very large spots. 
Concerning Section~\ref{sec:resolution}, the difference of  optical resolution in the two images can be controlled by different scale parameters and/or thresholds parameters in the two channels. We set $\alpha_X=5$, $\alpha_Y=\alpha_\epsilon=10$, $\tau_1=1.5$, $\tau_2=1$  (moderate difference of resolution) and $\alpha_X=5$, $\alpha_Y=\alpha_\epsilon=20$, $\tau_1=2$, $\tau_2=1$ (large difference). The value of $\rho_0$ in these simulations has been tuned to result in the same final correlation between the channels as in Section~\ref{sec:scale}, namely  $\rho=0, 0.1$ and $0.3$ approximately. Finally, in the 3D experiments of Section~\ref{sec:3D}, the domain of simulation is $\Omega_n=[0,250]^2\times[0,60]$ and the input parameters are exactly the same as in Section~\ref{sec:scale} with the choice $\alpha_X=\alpha_Y=\alpha_\epsilon=8$.

\section{Data preparation}
We refer to  \cite{Andreska2014} for the description of data shown in Figure~2 of the main manuscript, that are BDNF proteins and vGlut acquired with dSTORM.

For the set of experiments shown in Figure~1, wild type RPE1 cells were grown in Dulbecco's Modified Eagle Medium: Nutrient Mixture F-12 (DMEM/F12) supplemented with 10\% (vol/vol) FCS, in 6 well plates. RPE1 cells were transiently transfected with plasmids coding for Langerin-YFP and Langerin-mCherry or Rab11a-GFP and Langerin-mCherry using the following protocol: 2 $\mu$g of each DNAs, completed to 100 $\mu$L with DMEM/F12 (FCS free) were incubated for 5 min at room temperature. 6 $\mu$L of X-tremeGENE 9 DNA Transfection Reagent (Roche) completed to 100 $\mu$L with DMEM/F12 (FCS free), were added to the mix and incubated for further 15 min at room temperature. The transfection mix was then added to RPE1 cells grown one day before and incubated further at 37$^o$C overnight. Cells were then spread onto fibronectin Cytoo chips (Cytoo Cell Architect) for 4h at 37$^o$C with F-12 (with 10\% (vol/vol) FCS (feotal calf serum), 10 mM Hepes, 100 units/ml of penicillin and 100 $\mu$g/ml of Strep) before imaging. Cell adhesion on micropatterns both constrains the cells in terms of lateral movement and averages their size and shape (1100 $\mu$m2).

Live-cell imaging was performed using simultaneous dual color Total Internal Reflection Fluorescence (TIRF) microscopy. All imaging was performed in full conditioned medium at 37$^o$C and 5\% CO$_2$.  Simultaneous dual color TIRFM microscopy sequences were acquired on a Nikon TE2000 inverted microscope equipped with a x100 TIRF objective (NA=1.49), an azimuthal (spinning) TIRF module (Ilas2, Roper Scientifc), an image splitter (DV, Roper Scientific) installed in front of an EMCCD camera (Evolve, Photometrics) and a temperature controller (LIS). GFP and m-Cherry were excited with a 488 nm and a 561 nm laser, respectively (100mW). The system was driven by the Metamorph software (Molecular Devices). A range of angles corresponding to a set of penetration depths is defined for a given wavelength and optical index of the medium \citep{Boulanger2014}. We performed simultaneous double-fluorescence image acquisition using RPE1 cells double transfected with Langerin-YFP and Langerin-mCherry  or Rab11A-GFP  and Langerin-mCherry.  Image series corresponding to simultaneous two colors multi-angles TIRF image stacks were recorded at one stack of 12 angles every 360 ms during 14.76 s, with a 30 ms exposure time per frame. Three-dimensional reconstructions of the whole cells were performed on the first 300 nm in depth of the cells using a 30-nm axial pixel size, see \cite{Boulanger2014}.

\newpage

\section{Supplementary Figure}

\vfill

\begin{figure}[h]
\centering
 \includegraphics[scale=0.07]{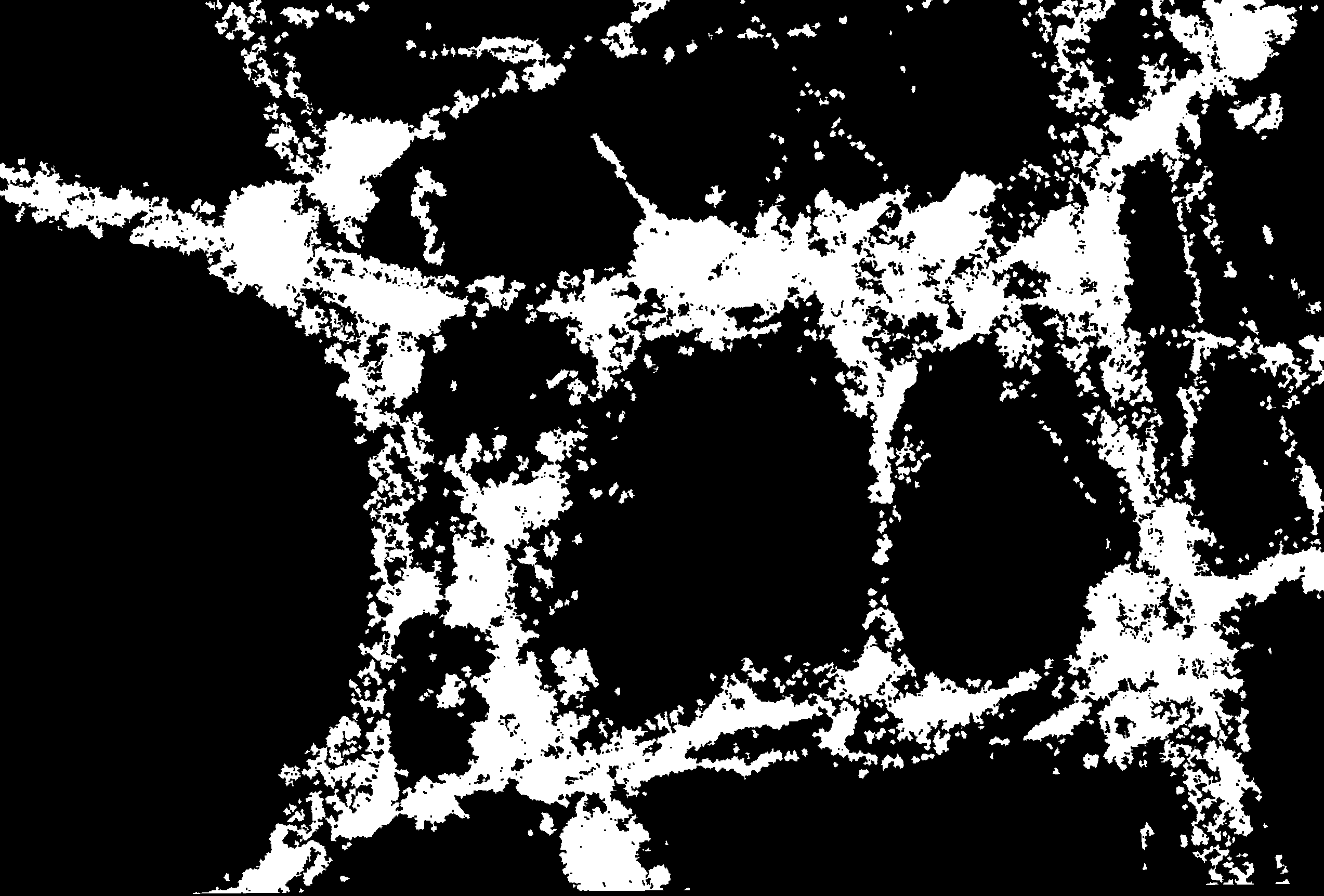} \\
 \includegraphics[scale=0.07]{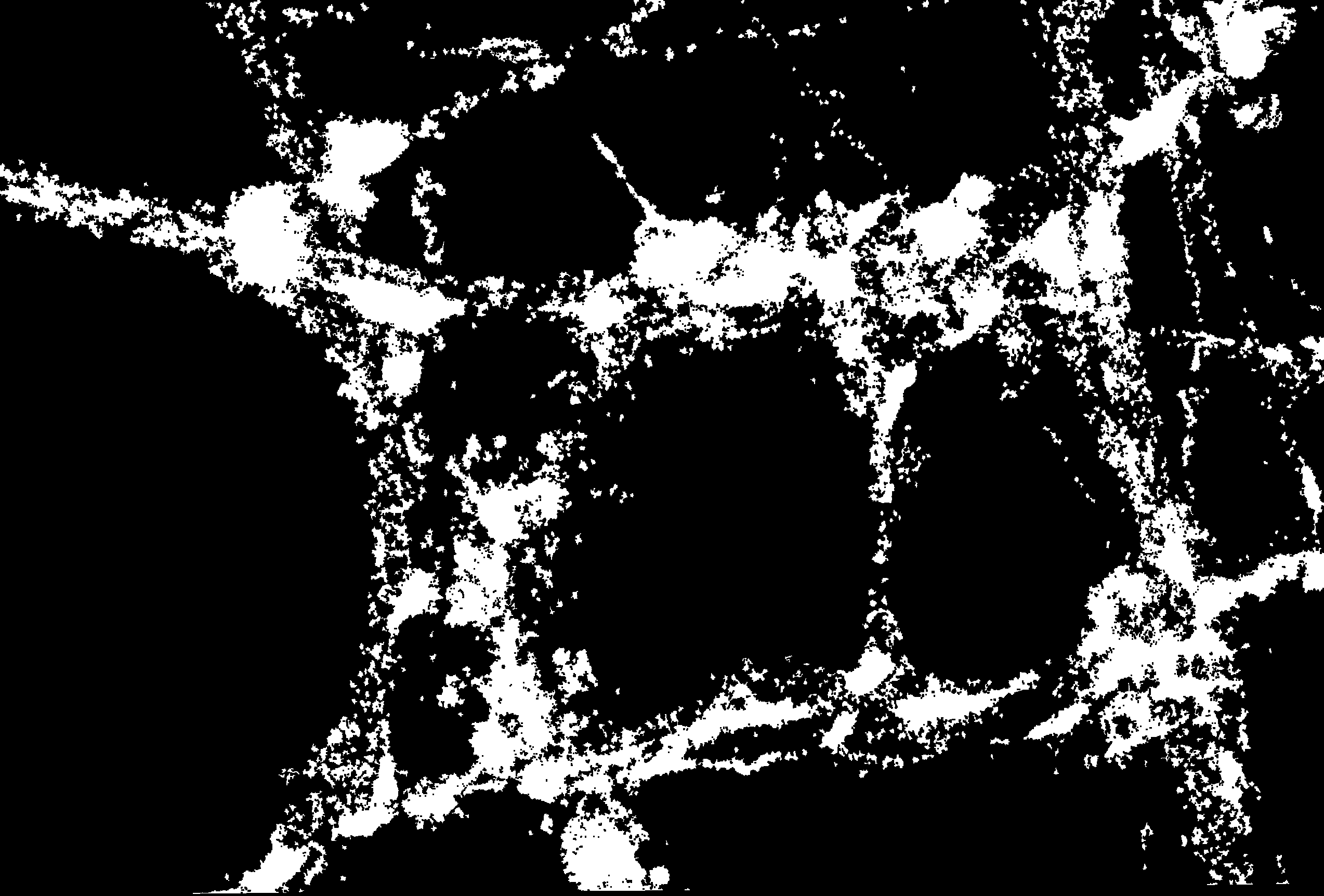}\\
 \includegraphics[scale=0.07]{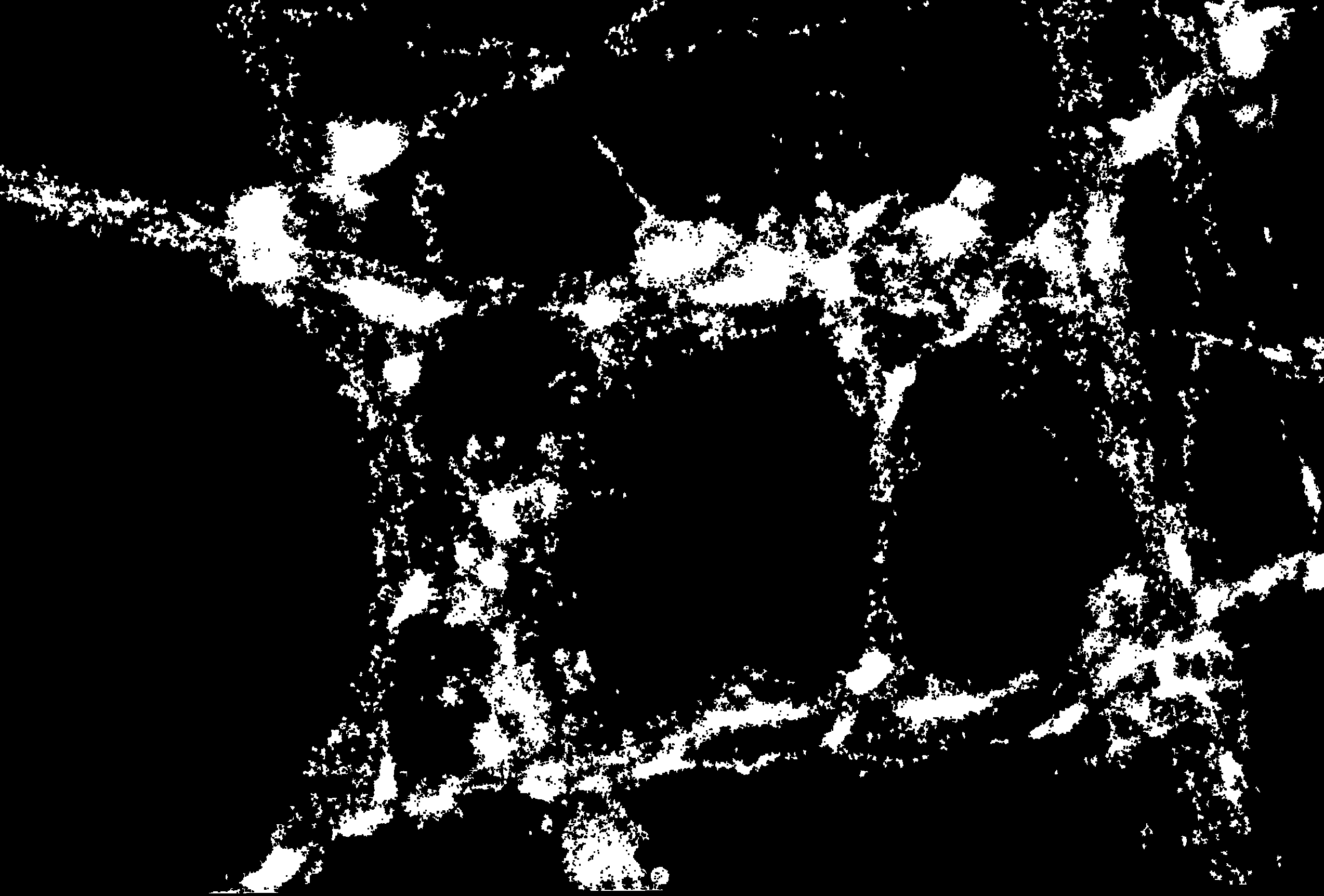} \\
\caption{vGlut segmentations. Three different segmentations of purple channel of image shown in  Figure~2 of the main manuscript.}
\end{figure}

\newpage

 \bibliographystyle{biom} 
\bibliography{references_gcops}

\label{lastpage}

\end{document}